\documentclass[10pt, aps, prd, twocolumn, superscriptaddress, nofootinbib]{revtex4-2}

\usepackage{amssymb,amsmath,mathrsfs,enumerate}
\usepackage{graphicx}
\usepackage{feynmp-auto}
\usepackage{mathtools}
\usepackage{caption}
\usepackage{subcaption}
\usepackage[dvipsnames]{xcolor}
\usepackage[colorlinks=true, linkcolor=teal, urlcolor=blue, citecolor=Mulberry]{hyperref}
\usepackage{orcidlink}
\usepackage{multirow}
\usepackage{booktabs}
\usepackage[normalem]{ulem}
\usepackage{microtype}  % nicer spacing (optional)
\usepackage{ragged2e}   % for \plotalign

\newcommand{\blfootnote}[1]{%
  \begingroup
  \renewcommand{\thefootnote}{}\footnote{#1}%
  \addtocounter{footnote}{-1}%
  \endgroup
}

\makeatletter
\patchcmd{\@makecaption}{\ignorespaces}{\justifying\ignorespaces}{}{}
\makeatother

\newcommand{\alpFluxSupernovaFig}{
\begin{figure*}[t]
    \centering
    \begin{subfigure}{0.325\textwidth}
        \centering
        \includegraphics[width=0.975\linewidth]{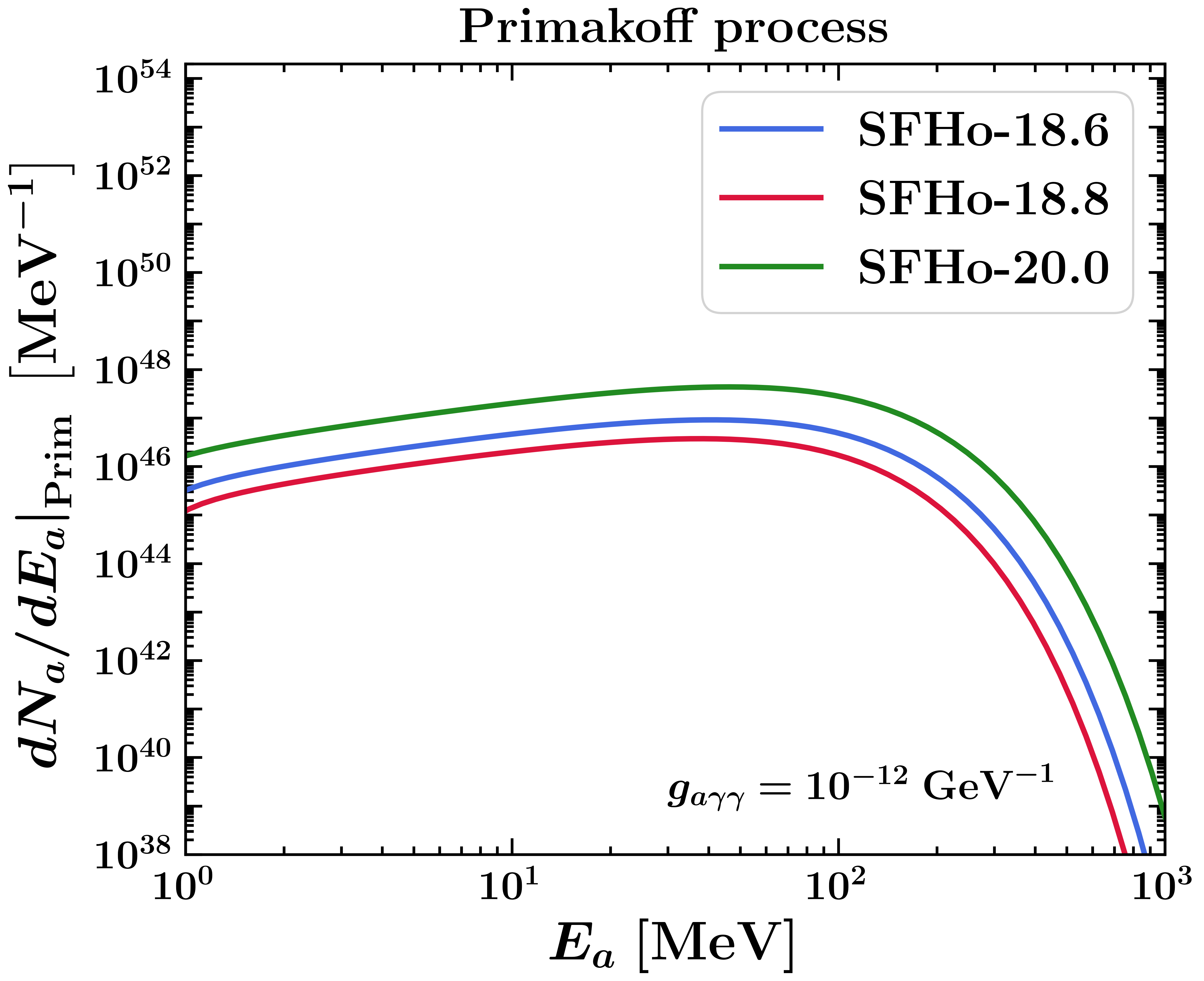}
        \caption{}
        \label{sf:alp_prim_flux}
    \end{subfigure}
    \begin{subfigure}{0.325\textwidth}
        \centering
        \includegraphics[width=0.975\linewidth]{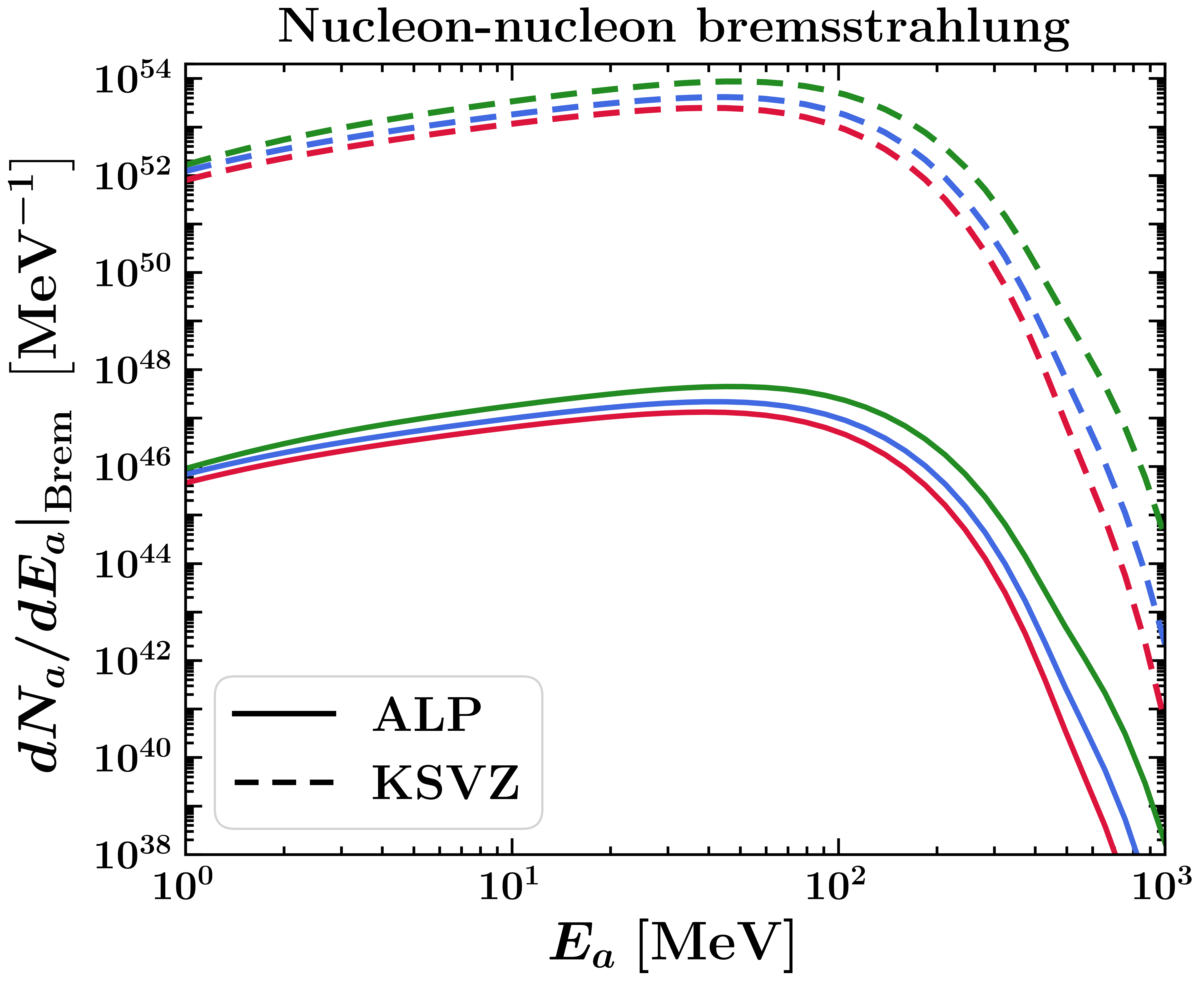}
        \caption{}
        \label{sf:alp_brem_flux}
    \end{subfigure}
    \begin{subfigure}{0.325\textwidth}
        \centering
        \includegraphics[width=0.975\linewidth]{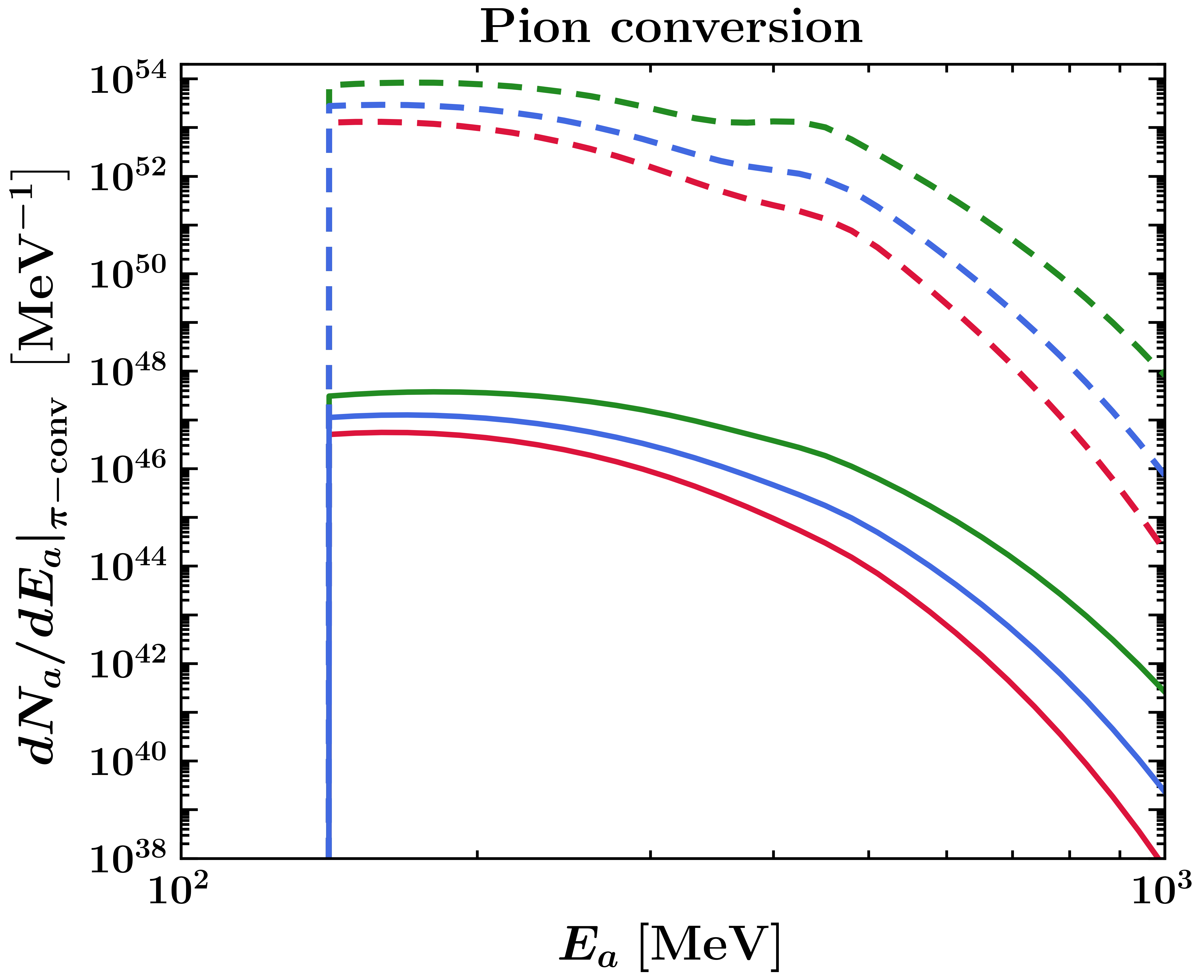}
        \caption{}
        \label{sf:alp_pion_conv_flux}
    \end{subfigure}
    \caption{\justifying Time-integrated axion spectra for the supernova progenitor profiles \textcolor[HTML]{4169E1}{SFHo-18.6 (blue)}, \textcolor[HTML]{DC143C}{SFHo-18.8 (red)}, and \textcolor[HTML]{228B22}{SFHo-20.0 (green)}. The spectra are shown for different production channels: the Primakoff process (left), nucleon–nucleon bremsstrahlung (middle), and pion conversion (right). Solid (dashed) lines correspond to the KSVZ (ALP) case with nucleon couplings $C_{app} \approx -0.47$, $C_{ann} \approx -0.02$ ($C_{app} \approx C_{ann} \approx 10^{-4}$) \cite{Manzari:2024jns}. The axion–photon coupling is fixed at $g_{a\gamma\gamma}=10^{-12} \, \mathrm{GeV}^{-1}$. All spectra are shown in the massless axion limit.}
    \label{fig:alp_flux_supernova}
\end{figure*}
}
\newcommand{\alpPhotonConv}{%
\begin{figure*}[t]
\centering

\begin{subfigure}{0.48\textwidth}
\centering
\includegraphics[width=\linewidth]{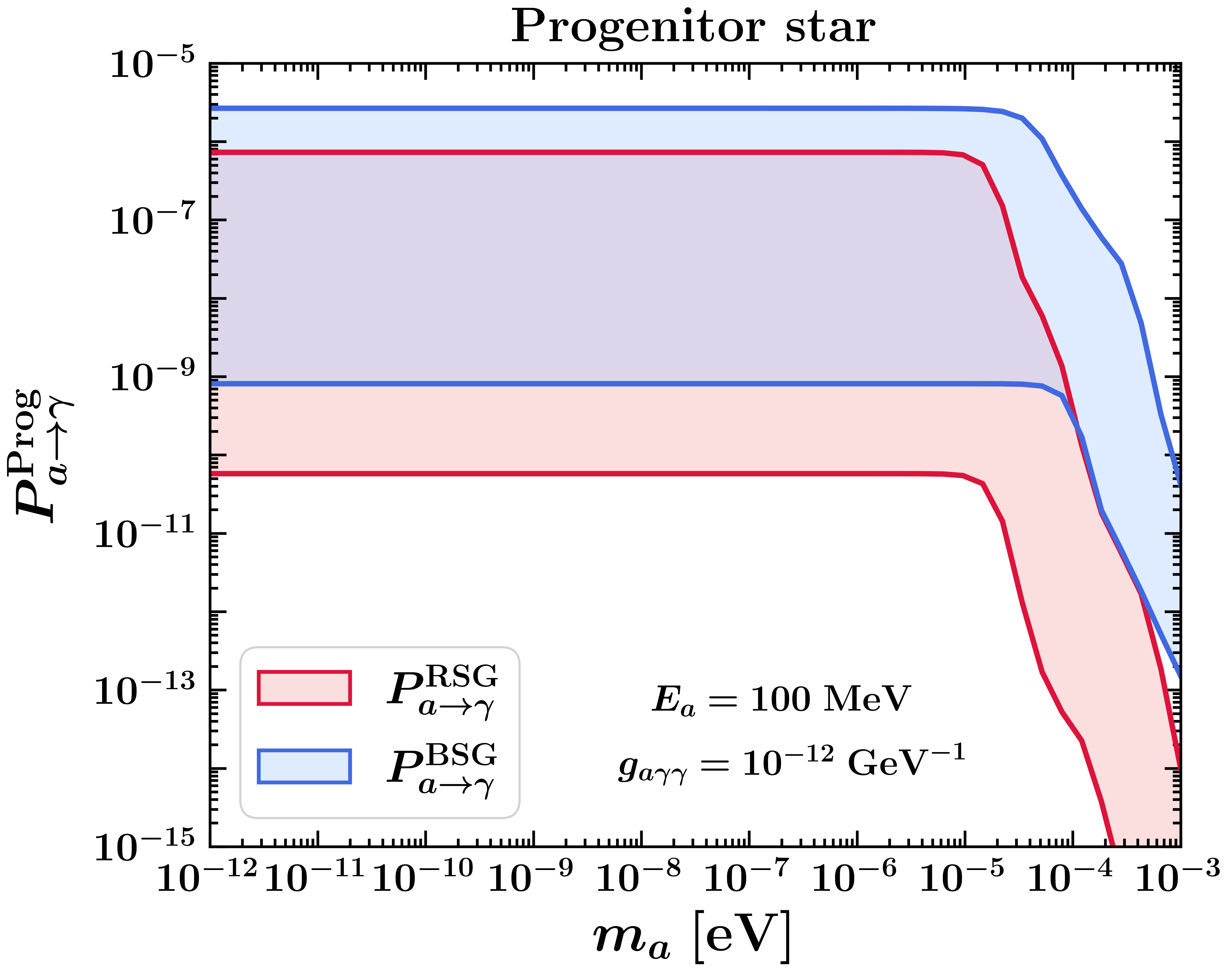}
\caption{}
\label{sf:patog_prog}
\end{subfigure}
\hfill
\begin{subfigure}{0.48\textwidth}
\centering
\includegraphics[width=\linewidth]{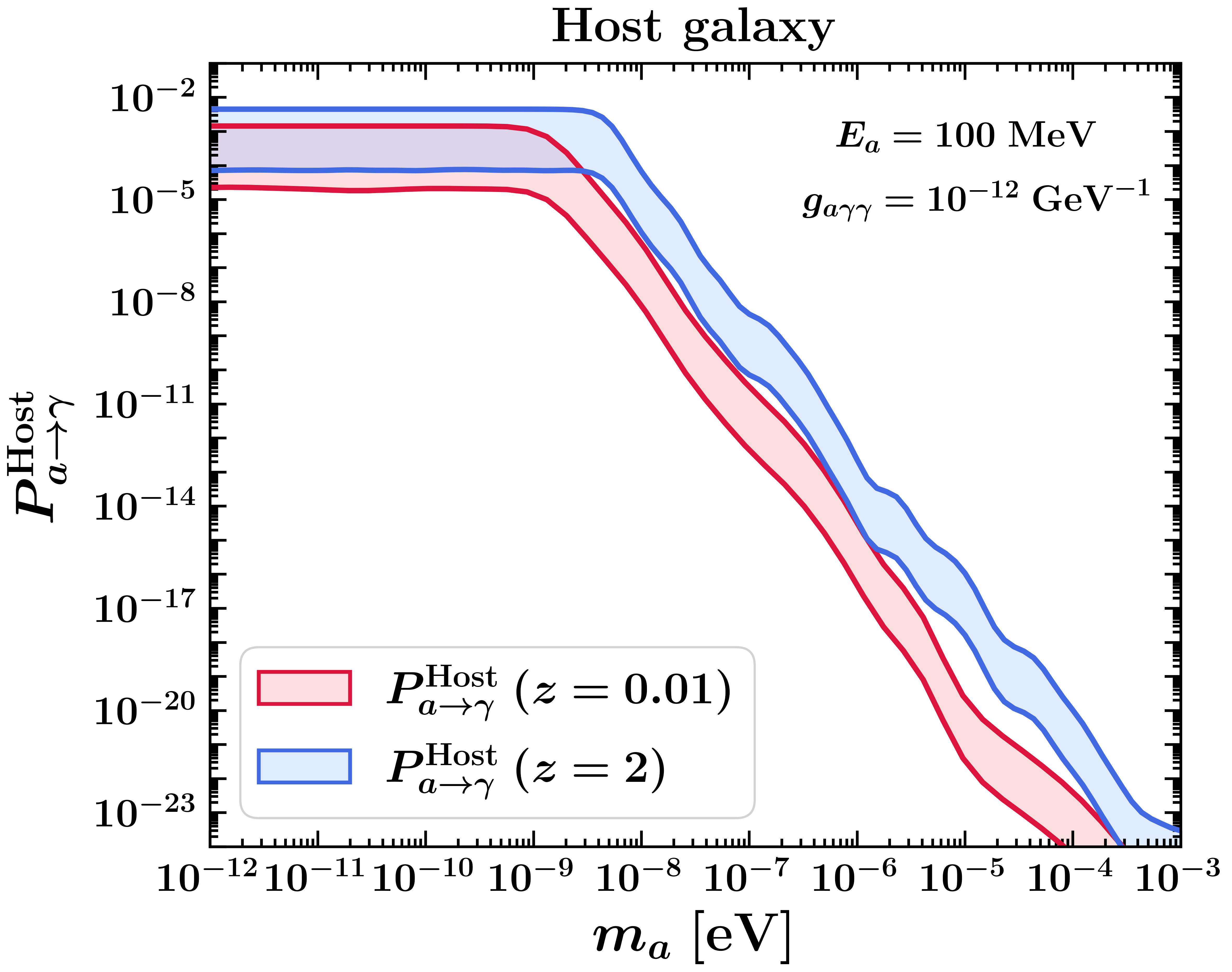}
\caption{}
\label{sf:patog_host}
\end{subfigure}

\medskip

\begin{subfigure}{0.48\textwidth}
\centering
\includegraphics[width=\linewidth]{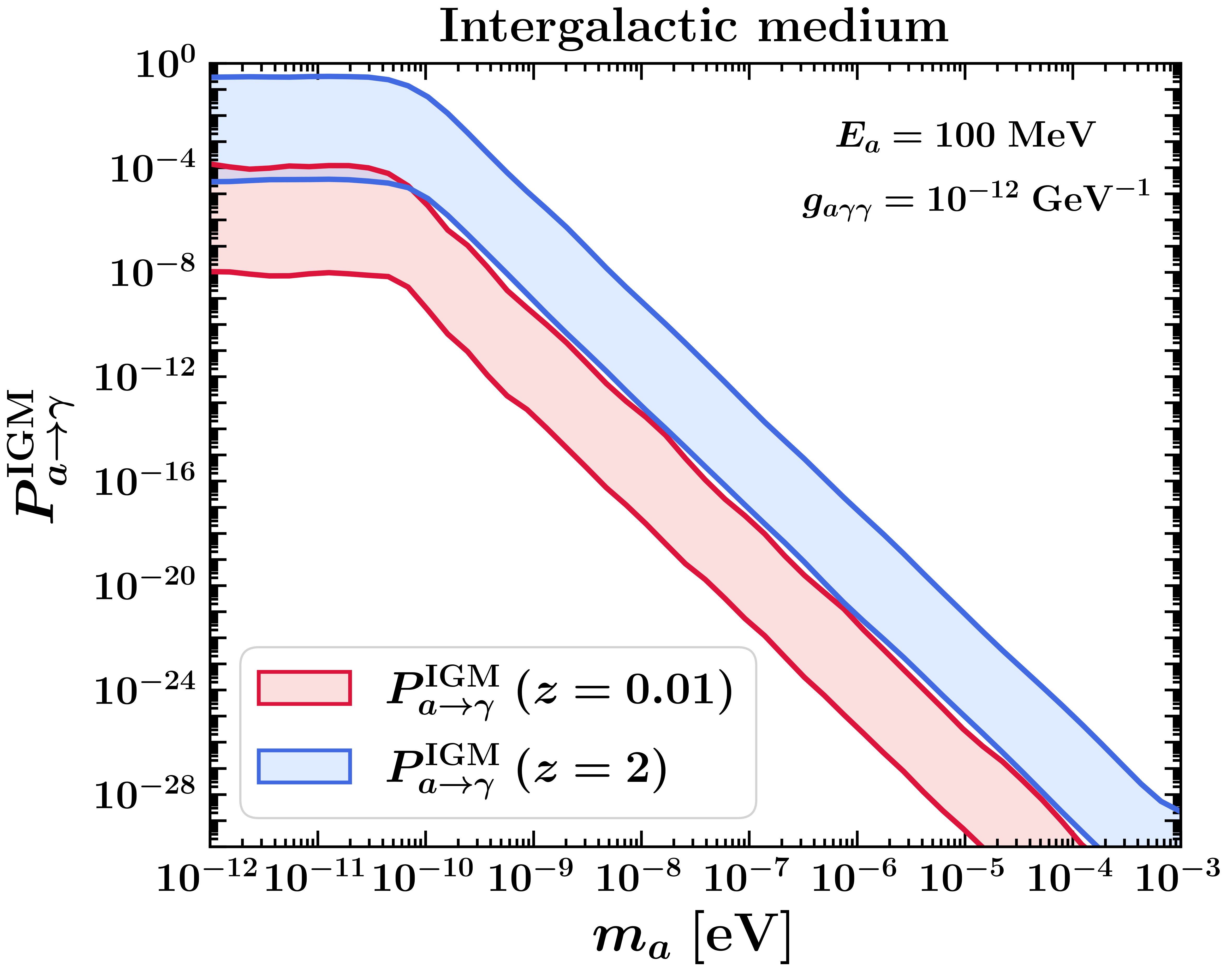}
\caption{}
\label{sf:patog_igm}
\end{subfigure}
\hfill
\begin{subfigure}{0.48\textwidth}
\centering
\includegraphics[width=\linewidth]{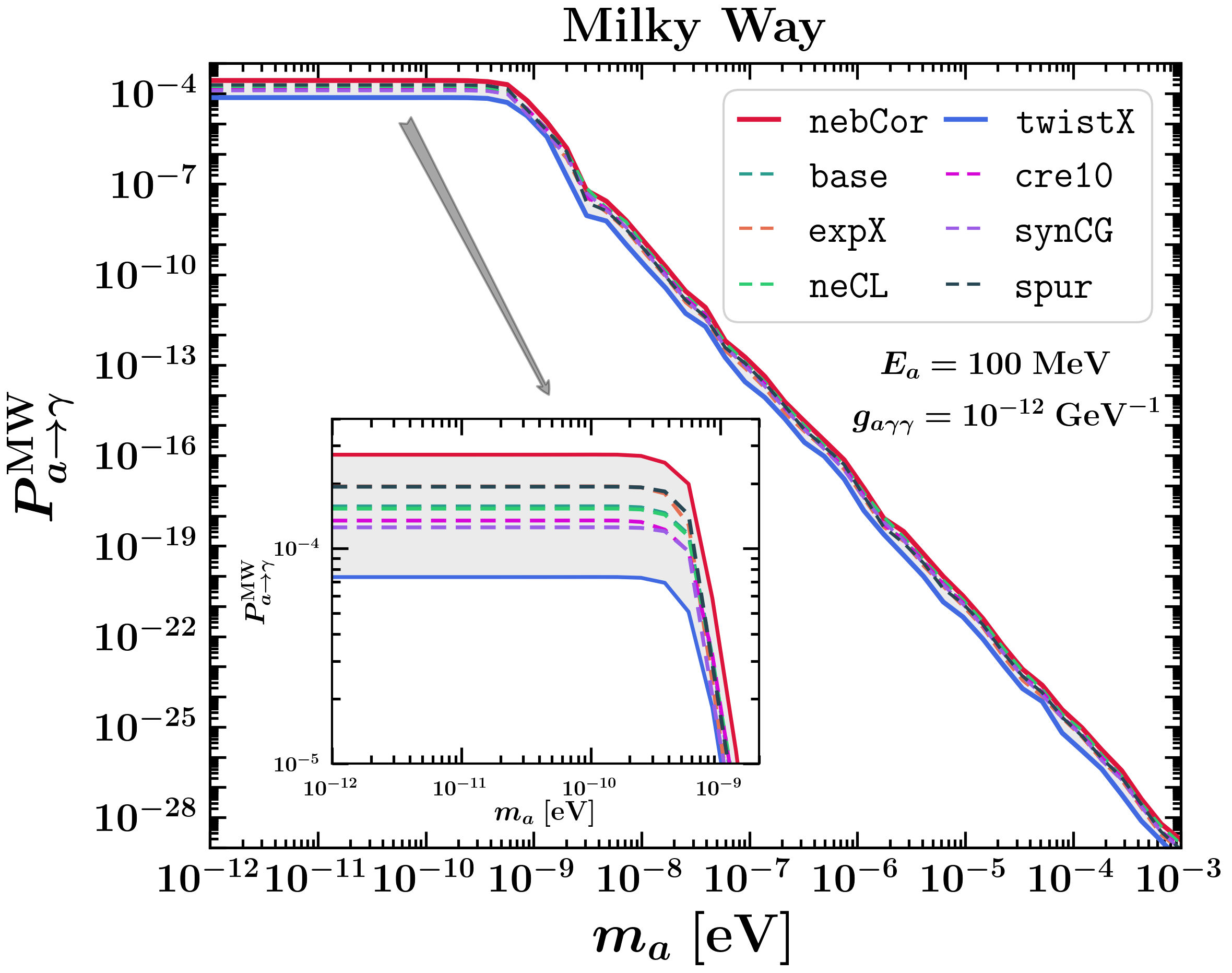}
\caption{}
\label{sf:patog_MW}
\end{subfigure}

\caption{\justifying Axion–photon conversion probabilities in various magnetic field environments: (a) Progenitor star (upper left), where \textcolor[HTML]{DC143C}{red} and \textcolor[HTML]{4169E1}{blue} bands indicate uncertainties from magnetic field modeling for \textcolor[HTML]{DC143C}{RSG} and \textcolor[HTML]{4169E1}{BSG}, respectively; (b) Host galaxy (upper right), showing results for \textcolor[HTML]{DC143C}{$z=0.01$ (red)} and \textcolor[HTML]{4169E1}{$z=5$ (blue)}. The bands represent the uncertainties arising from the host galaxy magnetic field configurations; (c) Intergalactic medium (lower left), with the shaded area represents uncertainties corresponding to the allowed range of IGM magnetic field strengths for sources at \textcolor[HTML]{DC143C}{$z=0.01$ (red)} and \textcolor[HTML]{4169E1}{$z=5$ (blue)}; (d) Galactic magnetic field (lower right), where eight curves represent UF23 models. The \textcolor[HTML]{DC143C}{red} (\textcolor[HTML]{4169E1}{blue}) line marks the \textcolor[HTML]{DC143C}{maximum} (\textcolor[HTML]{4169E1}{minimum}) conversion probability defining the uncertainty band. A zoomed version is shown to highlight the eight different UF23 models. The wiggles in some of the curves are numerical artifacts.}
\label{fig:patog_combined}

\end{figure*}
}
\newcommand{\sfrdFit}{
\begin{figure}[t]
    \centering
    \includegraphics[width=0.975\linewidth]{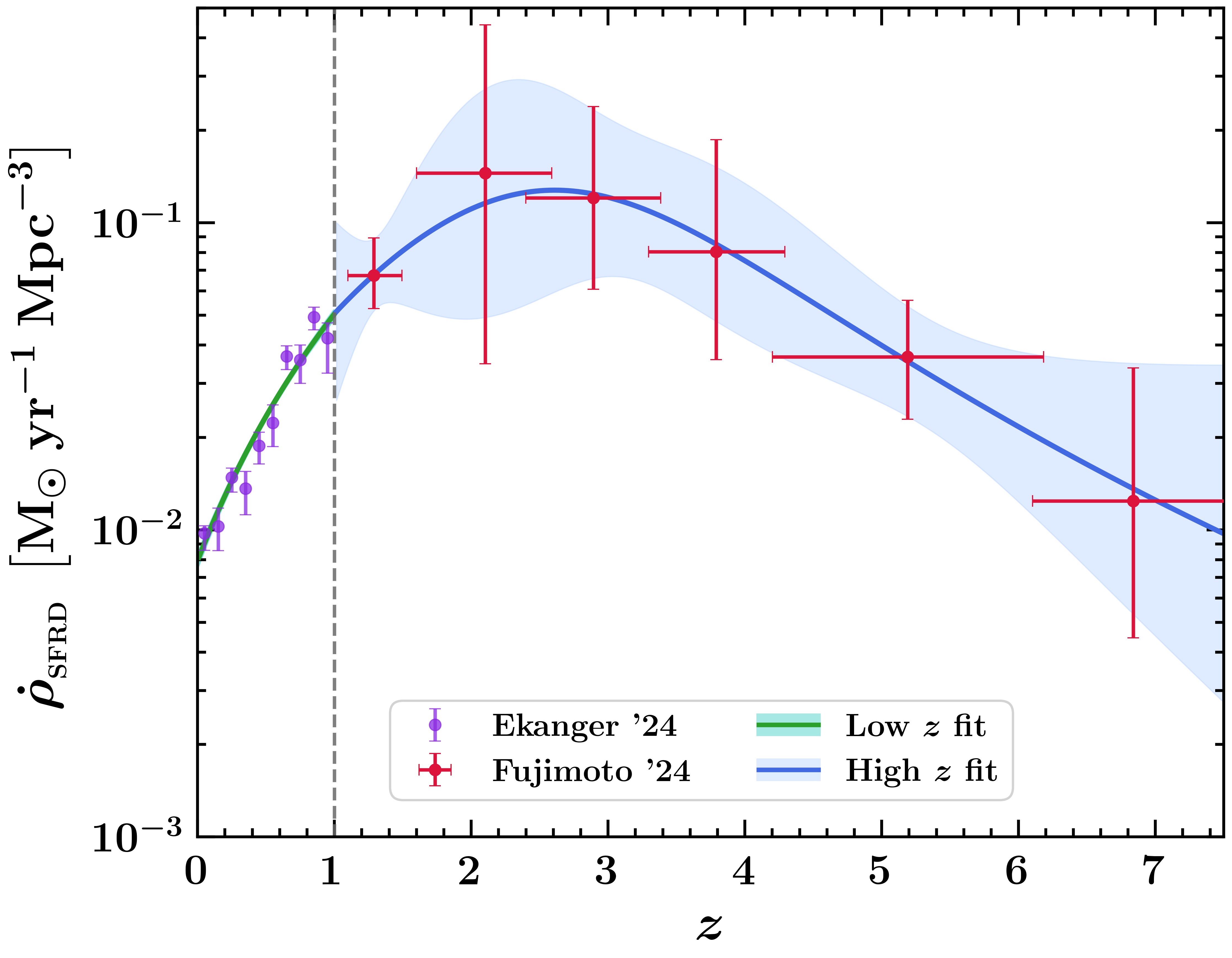}
    \caption{\justifying Fits to the cosmic SFRD in two regimes: a \textcolor[HTML]{228B22}{low redshift fit (green solid curve)} to the weighted SFRD compilation of Ref.\,\cite{Ekanger:2023qzw} (applied for $z \leq 1$), and a \textcolor[HTML]{4169E1}{high redshift fit (blue solid curve)} to the total (IR$+$UV, without dust correction) SFRD evolution from Ref.\,\cite{Fujimoto:2023vfa} (applied for $z > 1$). The \textcolor[HTML]{800080}{purple} and \textcolor[HTML]{DC143C}{red} data points with error bars correspond to the datasets used for the \textcolor[HTML]{800080}{low redshift} and \textcolor[HTML]{DC143C}{high redshift} fits, respectively. Shaded regions show the propagated $1\sigma$ fit uncertainties in each regime.}
    \label{fig:sfrd_fit}
\end{figure}
}
\newcommand{\gammaRayFlux}{
\begin{figure}[t]
    \centering
    \includegraphics[width=0.975\linewidth]{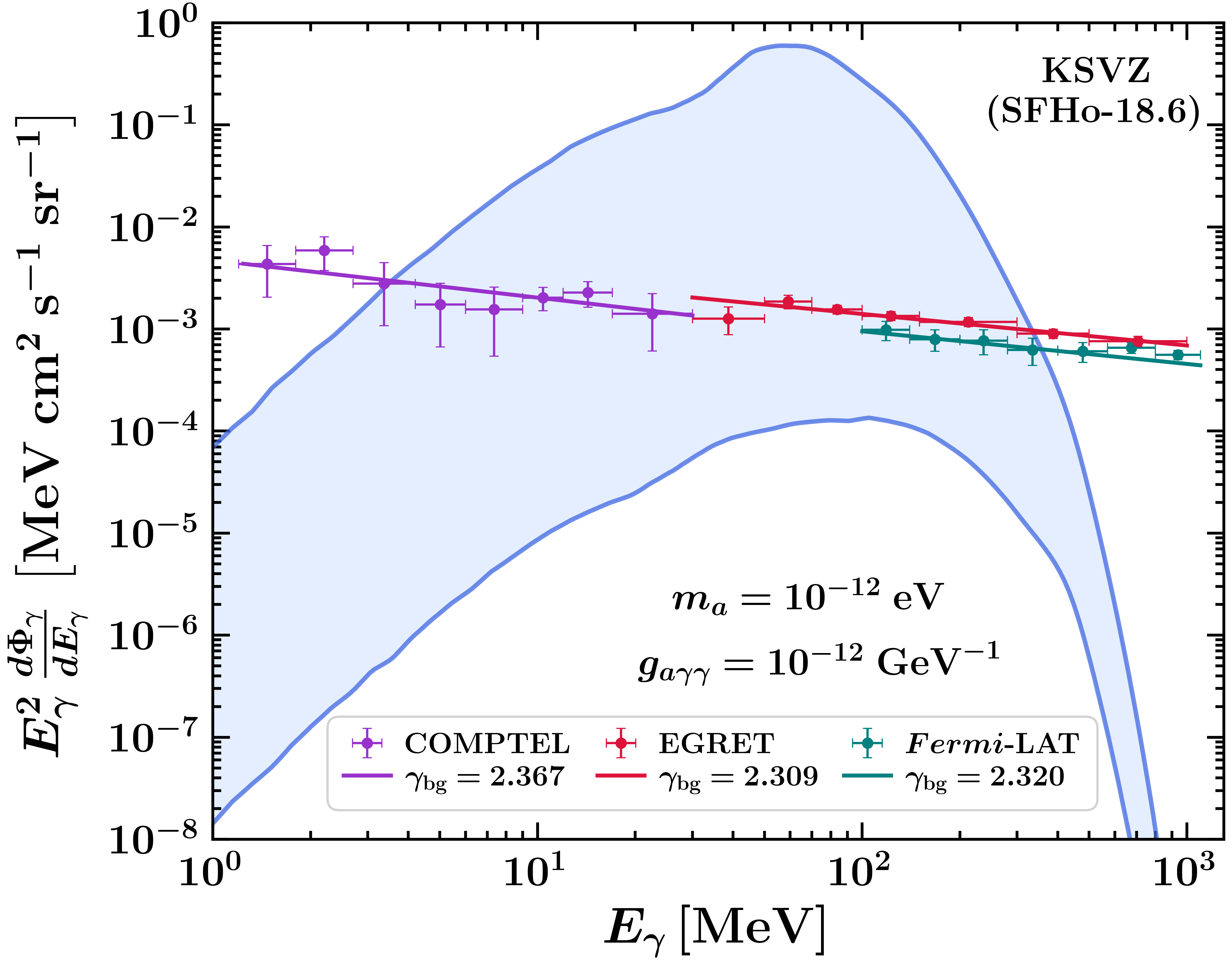}
    \caption{\justifying Diffuse axion-induced photon flux as a function of energy for the SFHo-18.6, together with the COMPTEL, EGRET, and \textit{Fermi}-LAT diffuse-background measurements. The photon flux is shown for $m_a=10^{-12}\,\rm eV$ and $g_{a\gamma\gamma}=10^{-12}\,\rm GeV$ in the KSVZ scenario. The \textcolor[HTML]{4169E1}{blue}-shaded region denotes the \textcolor[HTML]{4169E1}{uncertainty} enevelope obtained by propagating uncertainties in the CCSN rate and conversion probabilities across different astrophysical environments. The solid lines show the best-fit power-law background together with the data points of \textcolor[HTML]{800080}{COMPTEL (purple)}, \textcolor[HTML]{DC143C}{EGRET (red)}, and \textcolor{teal}{\textit{Fermi}-LAT(teal)}.}
    \label{fig:gamma_ray_flux}
\end{figure}
}
\newcommand{\chisqLimits}{
\begin{figure*}[t]
    \centering
    \begin{subfigure}{0.485\textwidth}
        \centering
        \includegraphics[width=0.975\linewidth]{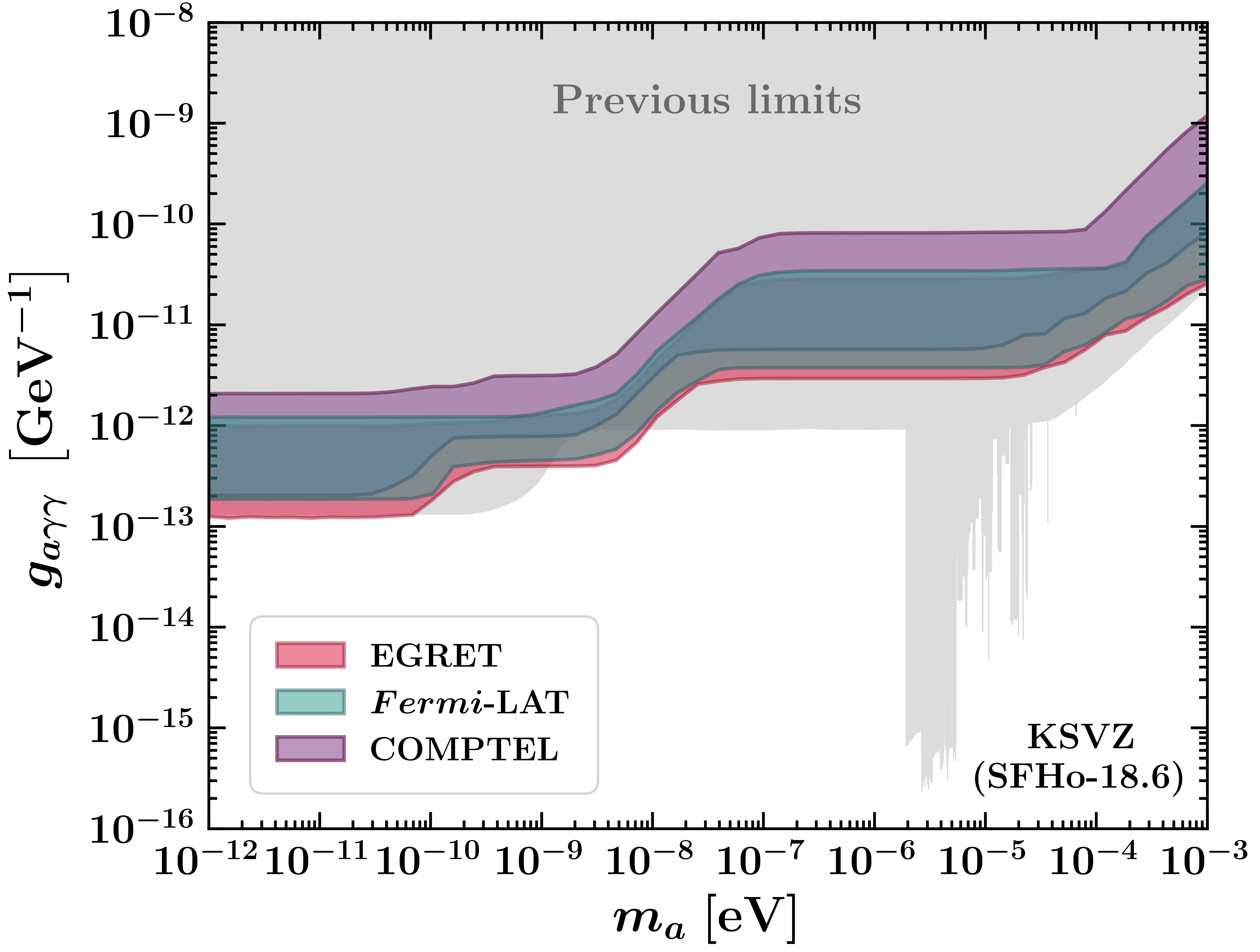}
        \caption{}
        \label{sf:chisq_limits_KSVZ}
    \end{subfigure}
    \begin{subfigure}{0.485\textwidth}
        \centering
        \includegraphics[width=0.975\linewidth]{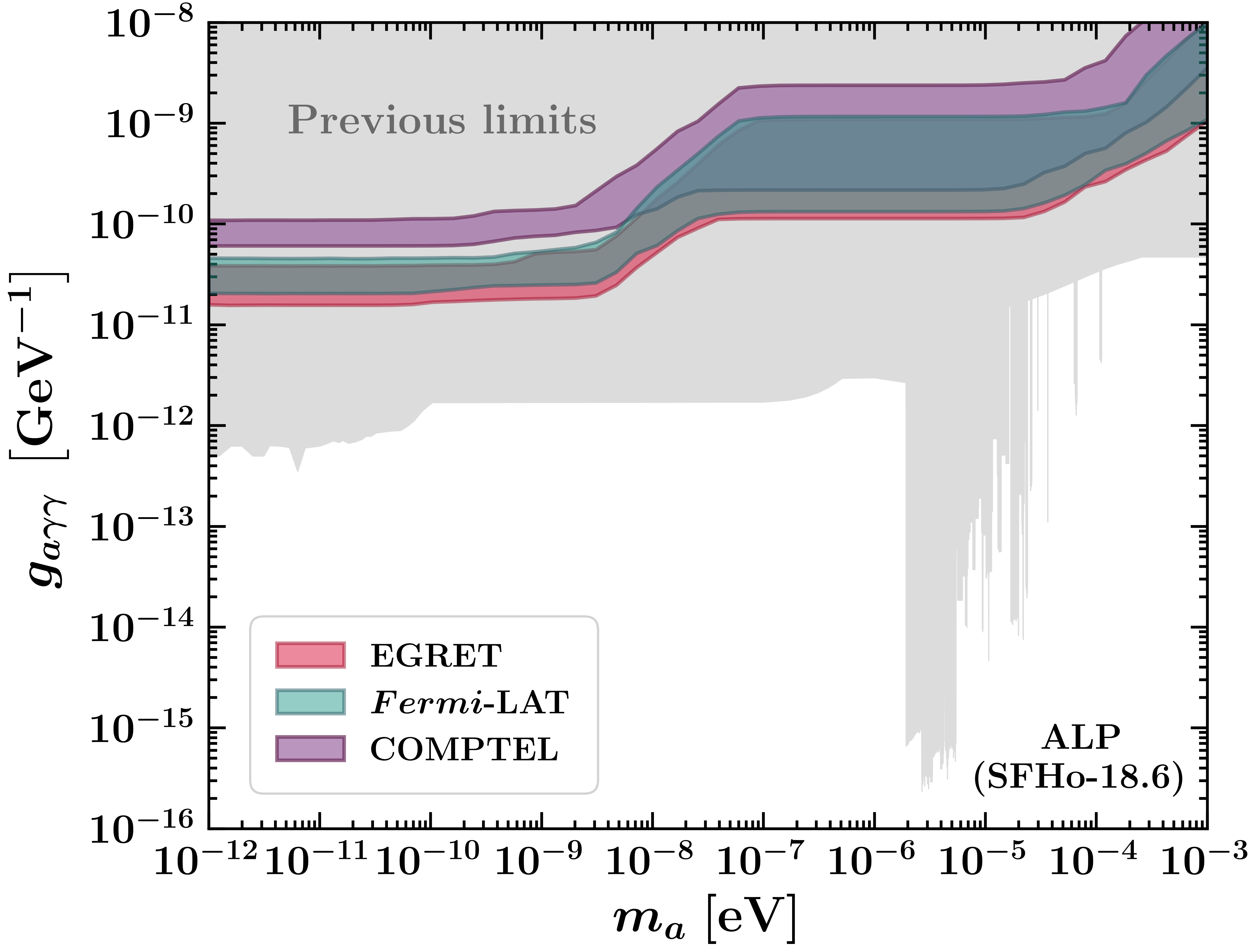}
        \caption{}
        \label{sf:chisq_limits_noUV}
    \end{subfigure}
    \caption{\justifying Constraints on $g_{a\gamma\gamma}$ with 95\% CL for SFHo-18.6 model in the KSVZ (left) and ALP (right) scenarios. The \textcolor[HTML]{800080}{purple}, \textcolor[HTML]{DC143C}{red}, and \textcolor{teal}{ teal} bands show the exclusion limits from \textcolor[HTML]{800080}{COMPTEL}, \textcolor[HTML]{DC143C}{EGRET}, and \textcolor{teal}{\textit{Fermi}-LAT}, respectively, incorporating astrophysical and conversion uncertainties. Existing constraints from astrophysical observations\,\cite{Wouters:2013hua, HESS:2013udx, Marsh:2017yvc, Kohri:2017ljt, Reynolds:2019uqt, Buen-Abad:2020zbd, Calore:2020tjw, Dessert:2020lil, Li:2020pcn, Meyer:2020vzy, Xiao:2020pra, Calore:2021hhn, Chan:2021gjl, Dessert:2021bkv, Li:2021gxs, Keller:2021zbl, Reynes:2021bpe, Dessert:2022yqq, Dolan:2022kul, Hoof:2022xbe, Jacobsen:2022swa, Foster:2022fxn, Noordhuis:2022ljw, Escudero:2023vgv, Battye:2023oac, Li:2024zst, Cyr:2024sbd, Manzari:2024jns, Ning:2024eky, Ruz:2024gkl, MAGIC:2024arq, Benabou:2025jcv}, as well as laboratory searches including haloscope experiments\,\cite{DePanfilis:1987dk, Wuensch:1989sa, Hagmann:1990tj, Hagmann:1996qd, 2010PhRvL.104d1301A, Brubaker:2016ktl, McAllister:2017lkb, Ouellet:2018beu, HAYSTAC:2018rwy, ADMX:2018gho, ADMX:2018ogs, ADMX:2019uok, Alesini:2019ajt, Lee:2020cfj, CAPP:2020utb, Alesini:2020vny, HAYSTAC:2020kwv, CAST:2020rlf, Gramolin:2020ict, Jeong:2020cwz, Devlin:2021fpq, Salemi:2021gck, Grenet:2021vbb, Thomson:2021zvq, ADMX:2021nhd, Adair:2022rtw, Kim:2022hmg, Yoon:2022gzp, Lee:2022mnc, Alesini:2022lnp, Quiskamp:2022pks, TASEH:2022vvu, Kim:2023vpo, Yang:2023yry, Thomson:2023moc, Quiskamp:2023ehr, QUAX:2023gop, HAYSTAC:2023cam, Heinze:2023nfb, Bae:2024kmy, Ahyoune:2024klt, QUAX:2024fut, Quiskamp:2024oet, HAYSTAC:2024jch, MADMAX:2024sxs, ADMX:2024xbv, Pandey:2024dcd, GigaBREAD:2025lzq, ADMX:2025vom} and helioscope experiments\,\cite{CAST:2007jps, Ehret:2010mh, Betz:2013dza, OSQAR:2015qdv, CAST:2017uph, CAST:2024eil}, are overlaid as a grey-shaded region for comparison.}
    \label{fig:chisq_limits}
\end{figure*}
}
\newcommand{\fisherLimits}{
\begin{figure*}[t]
    \centering
    \begin{subfigure}{0.485\textwidth}
        \centering
        \includegraphics[width=0.975\linewidth]{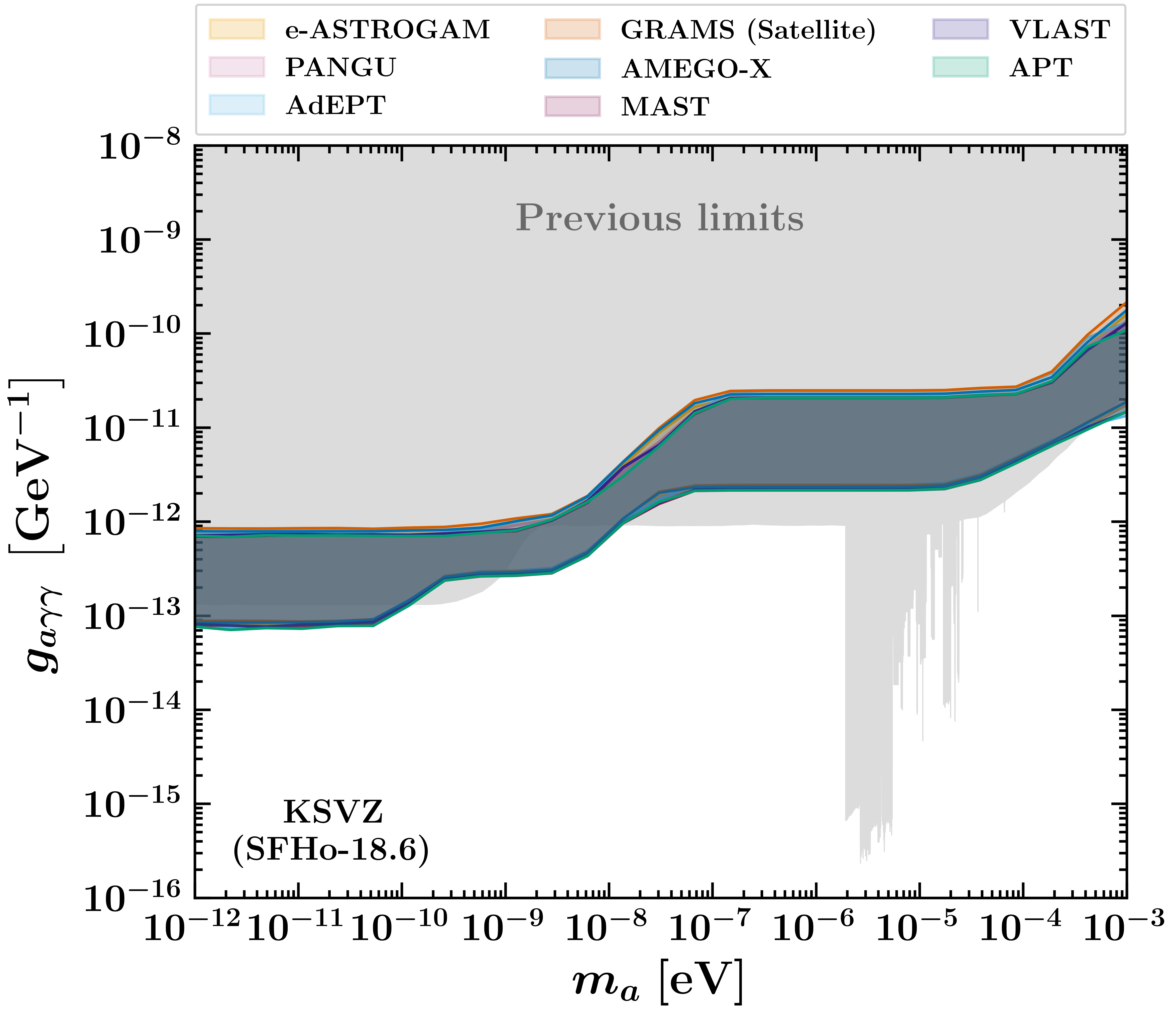}
        \label{sf:fisher_limits_KSVZ}
    \end{subfigure}
    \begin{subfigure}{0.485\textwidth}
        \centering
        \includegraphics[width=0.975\linewidth]{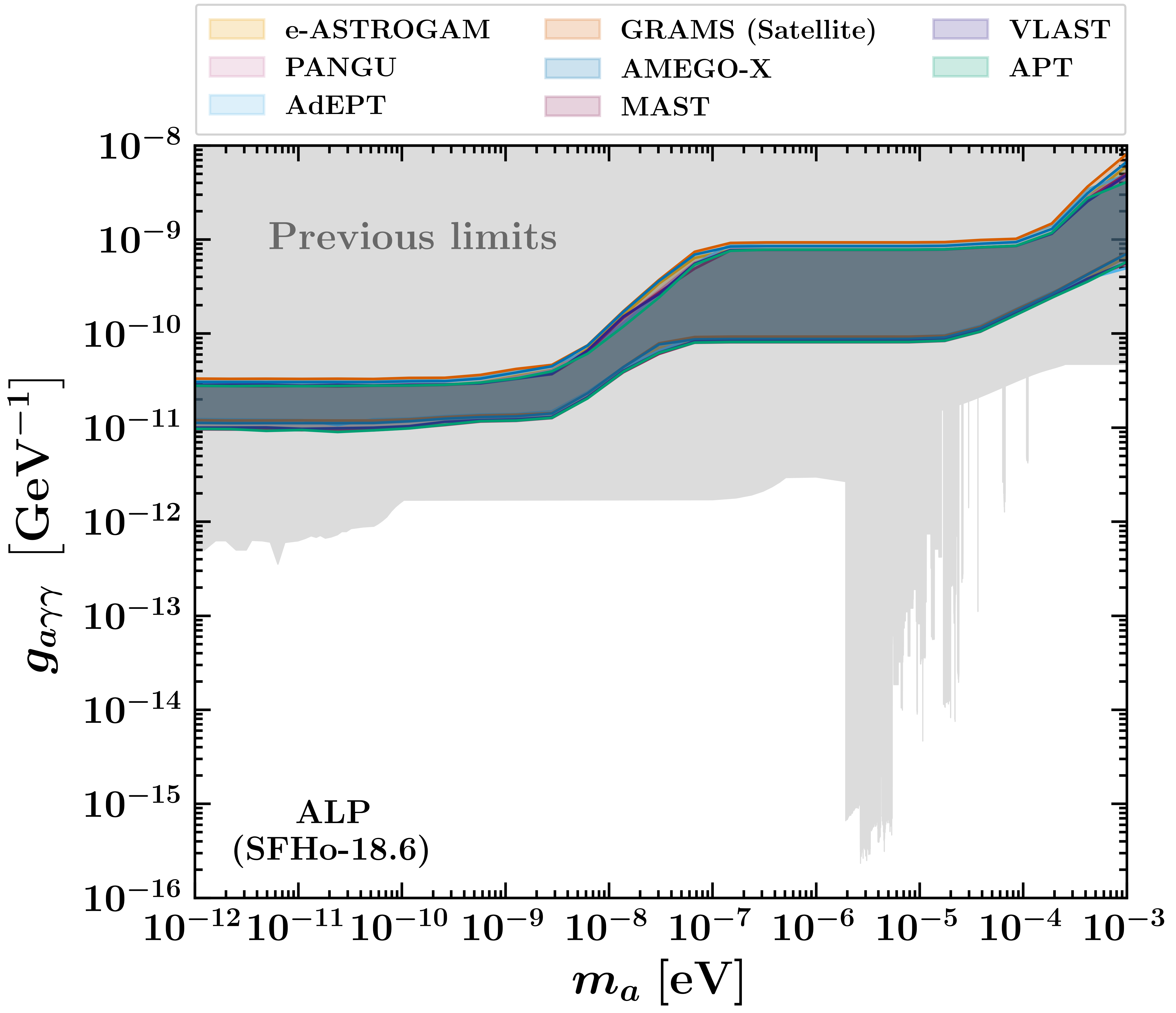}
        \label{sf:fisher_limits_noUV}
    \end{subfigure}
    \caption{\justifying Projected 95\% CL sensitivity on $g_{a\gamma\gamma}$ as a function of \ $m_a$ for the 
KSVZ (left panel) and ALP (right panel) models for SFHo-18.6 model using the Fisher forecast. The bands show sensitivities of the future telescopes: {\color[HTML]{0072B2}AMEGO-X (blue)}, 
{\color[HTML]{E69F00}\textit{e}-ASTROGAM (yellow)}, 
{\color[HTML]{009E73}APT (green)}, 
{\color[HTML]{D55E00}GRAMS Satellite (orange)}, 
{\color[HTML]{882255}MAST (dark red)}, 
{\color[HTML]{56B4E9}AdEPT (light blue)}, 
{\color[HTML]{CC79A7}PANGU (pink)}, and 
{\color[HTML]{332288}VLAST (dark blue)}.The width of each band reflects astrophysical uncertainties. Existing constraints from astrophysical observations\,\cite{Wouters:2013hua, HESS:2013udx, Marsh:2017yvc, Kohri:2017ljt, Reynolds:2019uqt, Buen-Abad:2020zbd, Calore:2020tjw, Dessert:2020lil, Li:2020pcn, Meyer:2020vzy, Xiao:2020pra, Calore:2021hhn, Chan:2021gjl, Dessert:2021bkv, Li:2021gxs, Keller:2021zbl,Reynes:2021bpe, Dessert:2022yqq, Dolan:2022kul, Hoof:2022xbe, Jacobsen:2022swa, Foster:2022fxn,Noordhuis:2022ljw, Escudero:2023vgv,Battye:2023oac, Li:2024zst, Cyr:2024sbd, Manzari:2024jns, Ning:2024eky, Ruz:2024gkl, MAGIC:2024arq, Benabou:2025jcv}, laboratory searches (haloscope\,\cite{DePanfilis:1987dk, Wuensch:1989sa, Hagmann:1990tj, Hagmann:1996qd, 2010PhRvL.104d1301A, Brubaker:2016ktl, McAllister:2017lkb, Ouellet:2018beu, HAYSTAC:2018rwy, ADMX:2018gho, ADMX:2018ogs, ADMX:2019uok, Alesini:2019ajt, Lee:2020cfj, CAPP:2020utb, Alesini:2020vny, HAYSTAC:2020kwv, CAST:2020rlf, Gramolin:2020ict, Jeong:2020cwz, Devlin:2021fpq, Salemi:2021gck, Grenet:2021vbb, Thomson:2021zvq, ADMX:2021nhd, Adair:2022rtw, Kim:2022hmg, Yoon:2022gzp, Lee:2022mnc, Alesini:2022lnp, Quiskamp:2022pks, TASEH:2022vvu, Kim:2023vpo, Yang:2023yry, Thomson:2023moc, Quiskamp:2023ehr, QUAX:2023gop, HAYSTAC:2023cam, Heinze:2023nfb, Bae:2024kmy, Ahyoune:2024klt, QUAX:2024fut, Quiskamp:2024oet, HAYSTAC:2024jch, MADMAX:2024sxs, ADMX:2024xbv, Pandey:2024dcd, GigaBREAD:2025lzq, ADMX:2025vom} and helioscope\,\cite{CAST:2007jps,Ehret:2010mh,Betz:2013dza, OSQAR:2015qdv, CAST:2017uph,CAST:2024eil}) are overlaid as a grey-shaded region for comparison. All forecasts assume an observation time of $T_{\rm obs}=10^{7}\,\mathrm{s}$ 
and high-latitude sky coverage excluding $|b|<20^\circ$.}
    \label{fig:fisher_limits}
\end{figure*}
}
\newcommand{\patogMWROIavg}{
\begin{figure}[!htbp]
    \centering
    \includegraphics[width=0.96\linewidth]{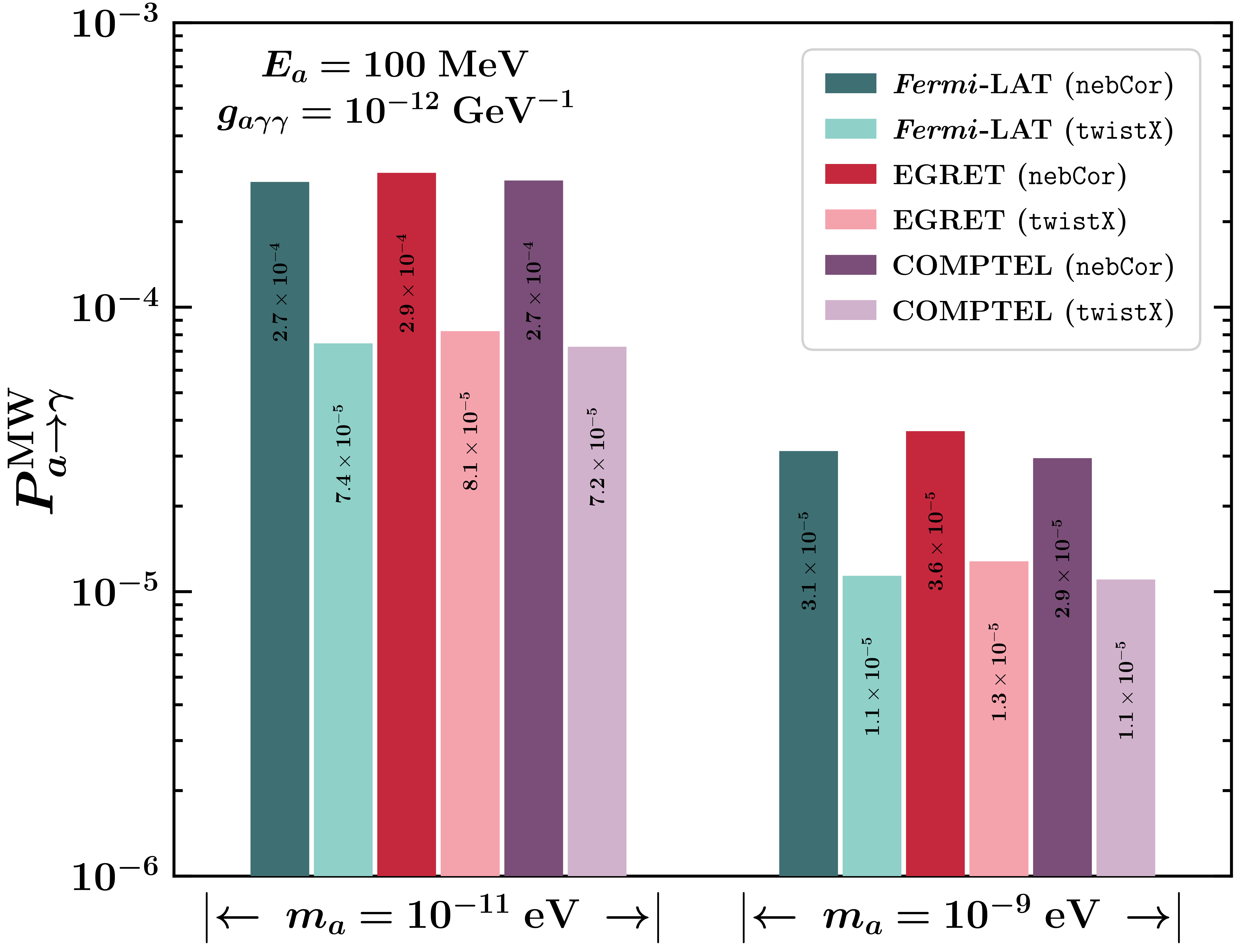}
    \caption{\justifying Comparison of the ROI-averaged MW axion-photon conversion probability $P_{a\to\gamma}^{\rm MW}$ across the telescopes \textcolor[HTML]{800080}{COMPTEL (purple)}, \textcolor[HTML]{DC143C}{EGRET (red)} and \textcolor{teal}{\textit{Fermi}-LAT (teal)},  evaluated at $E_\gamma = 100~{\rm MeV}$ and $g_{a\gamma\gamma} = 10^{-12}~{\rm GeV}^{-1}$ for two representative axion masses. Dark (light) bars correspond to the \texttt{nebCor} (\texttt{twistX}) variant of the UF23 GMF model \,\cite{Unger:2023lob}, bracketing the maximal and minimal conversion scenarios.}
    \label{fig:roi_comparison}
\end{figure}
}
\newcommand{\chisqLimitsOther}{
\begin{figure*}[!htbp]
    \centering
    \begin{subfigure}{0.485\textwidth}
        \centering
        \includegraphics[width=0.975\linewidth]{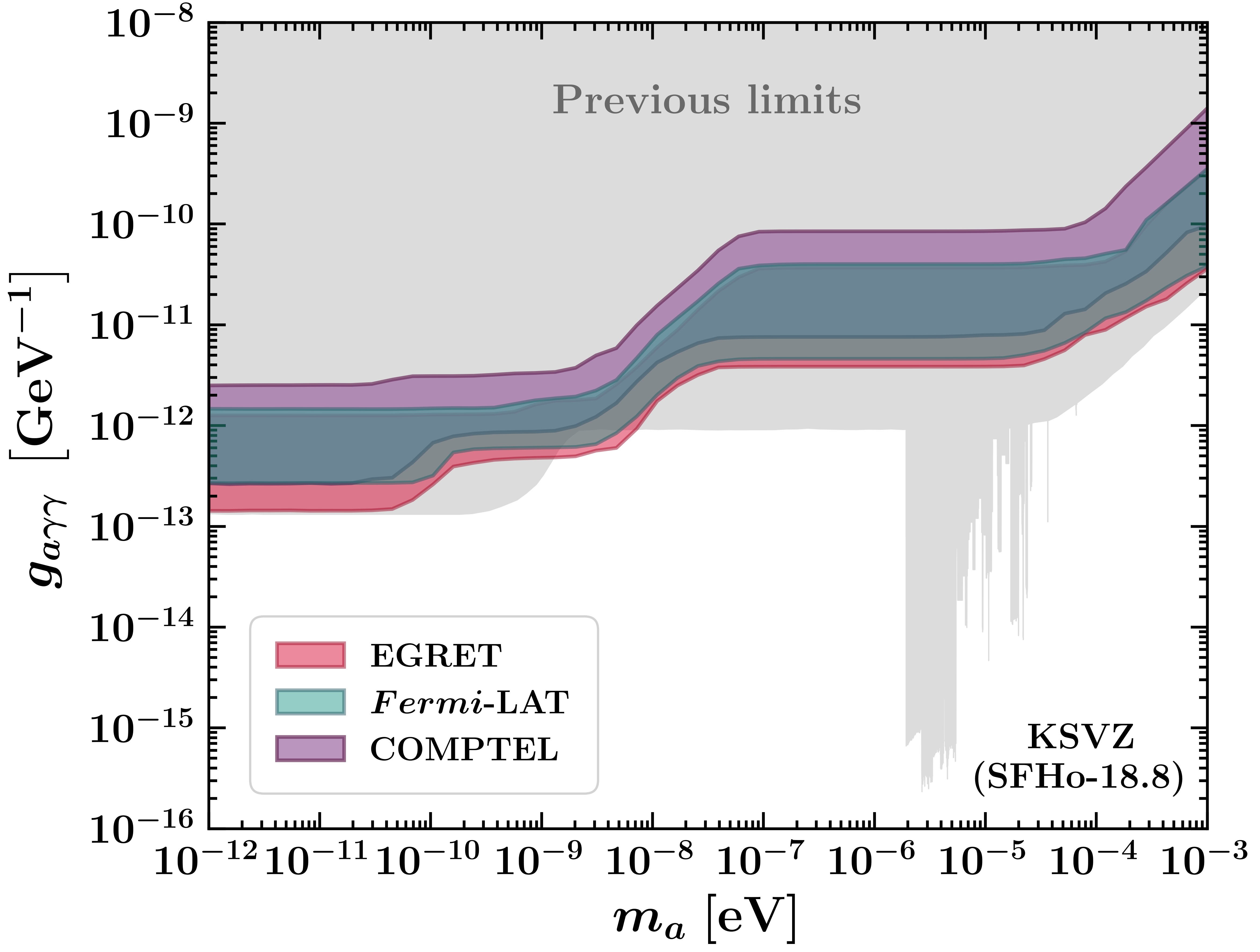}
        \caption{}
        \label{sf:chisq_limits_KSVZ_18p8}
    \end{subfigure}
    \begin{subfigure}{0.485\textwidth}
        \centering
        \includegraphics[width=0.975\linewidth]{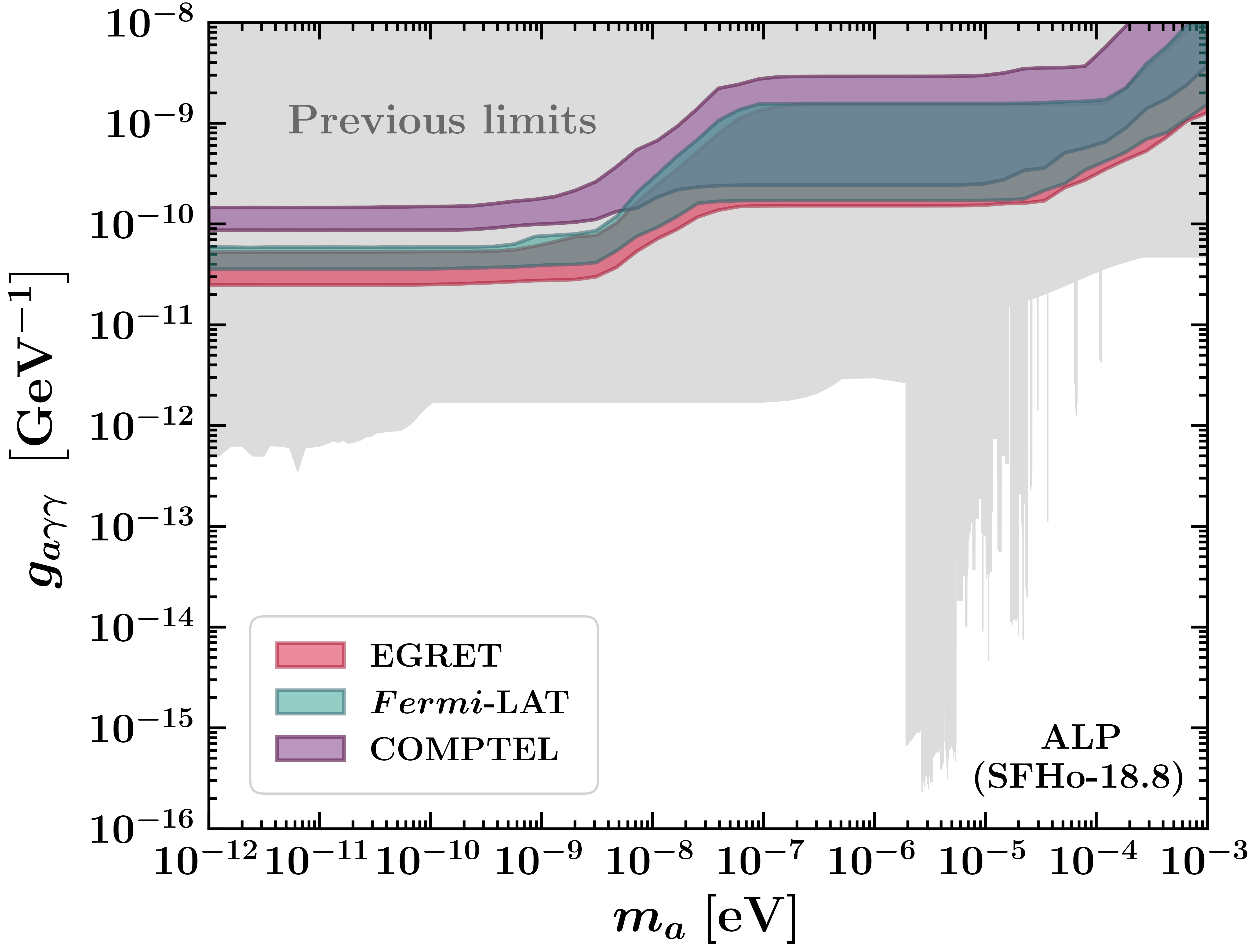}
        \caption{}
        \label{sf:chisq_limits_noUV_18p8}
    \end{subfigure}
    \begin{subfigure}{0.485\textwidth}
        \centering
        \includegraphics[width=0.975\linewidth]{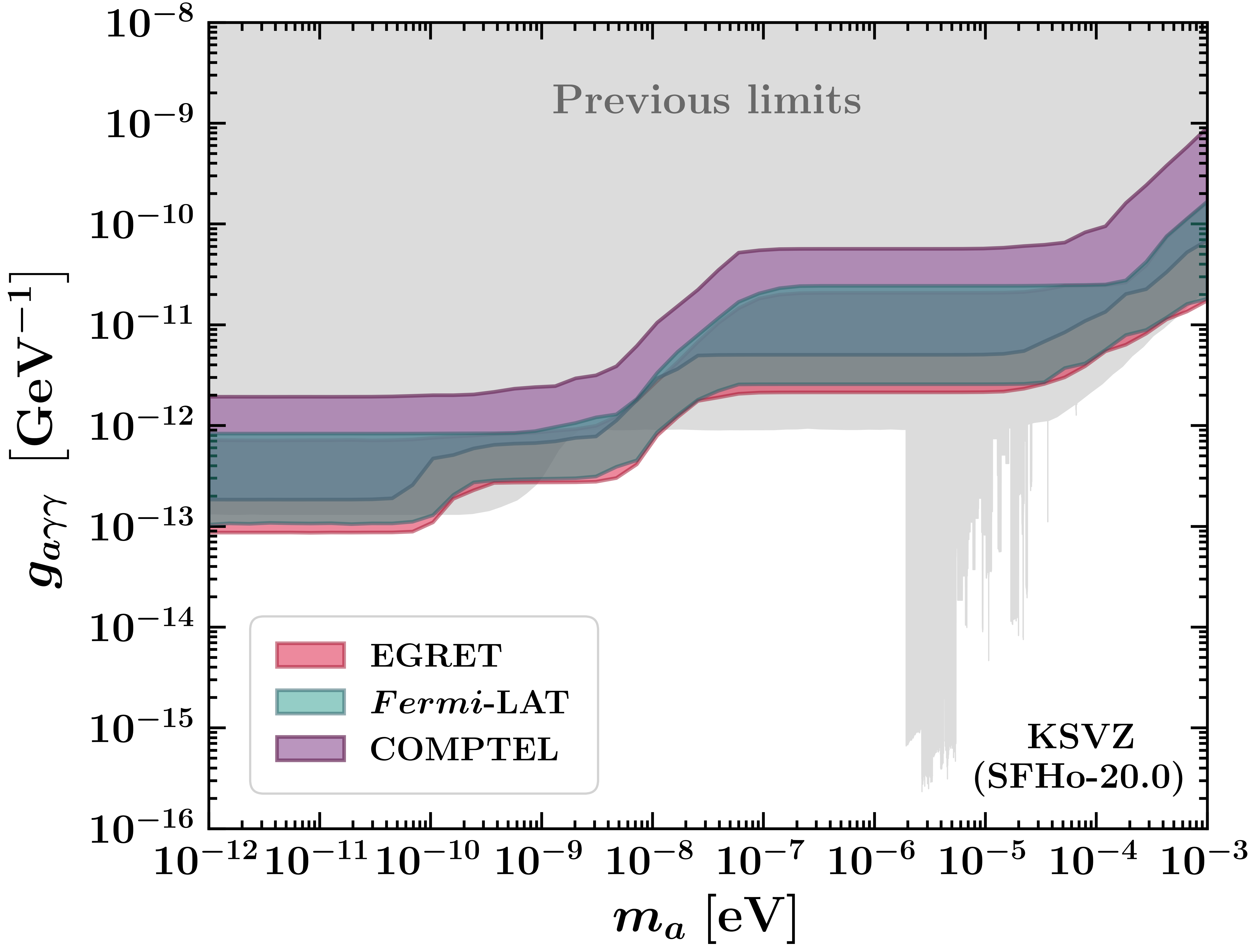}
        \caption{}
        \label{sf:chisq_limits_KSVZ_20p0}
    \end{subfigure}
    \begin{subfigure}{0.485\textwidth}
        \centering
        \includegraphics[width=0.975\linewidth]{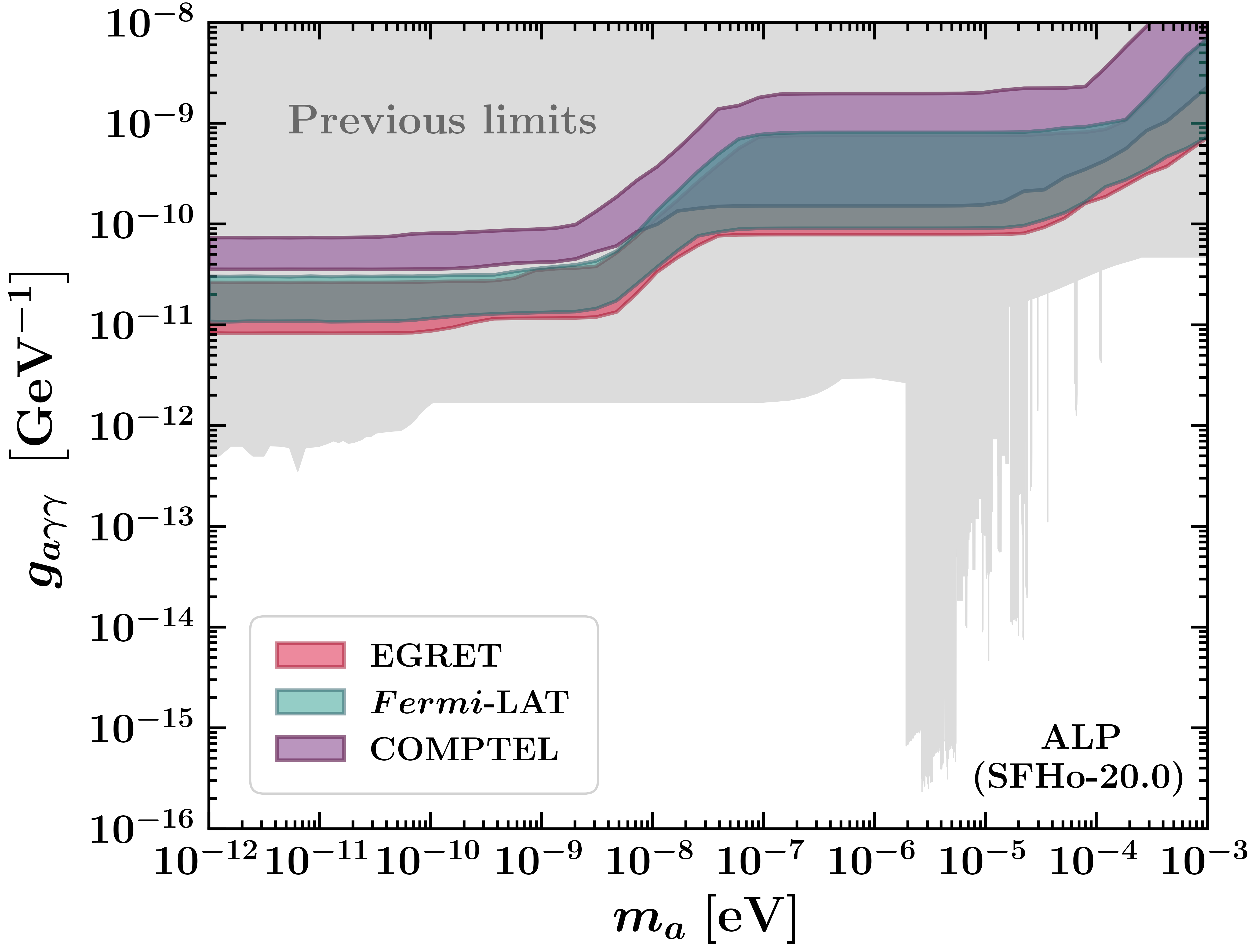}
        \caption{}
        \label{sf:chisq_limits_noUV_20p0}
    \end{subfigure}
    \caption{\justifying Same as Fig.~\ref{fig:chisq_limits} but for SFHo-18.8 (top) and SFHo-20.0 (bottom) CCSN profiles.}
    \label{fig:chisq_limits_other_models}
\end{figure*}
}
\newcommand{\FisherLimitsOther}{
\begin{figure*}[!htbp]
    \centering
    \begin{subfigure}{0.485\textwidth}
        \centering
        \includegraphics[width=0.975\linewidth]{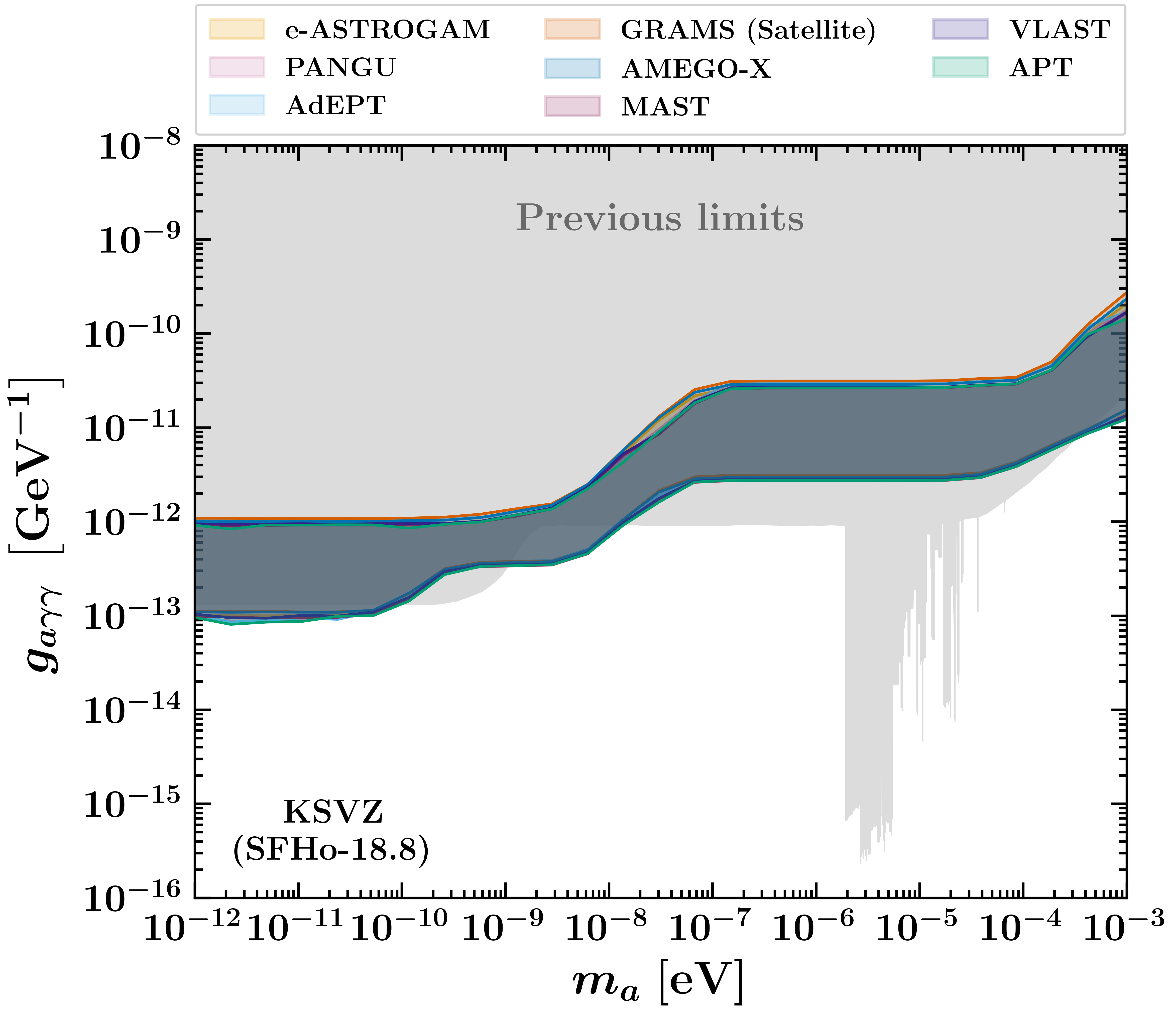}
        \caption{}
        \label{sf:fisher_limits_KSVZ_18p8}
    \end{subfigure}
    \begin{subfigure}{0.485\textwidth}
        \centering
        \includegraphics[width=0.975\linewidth]{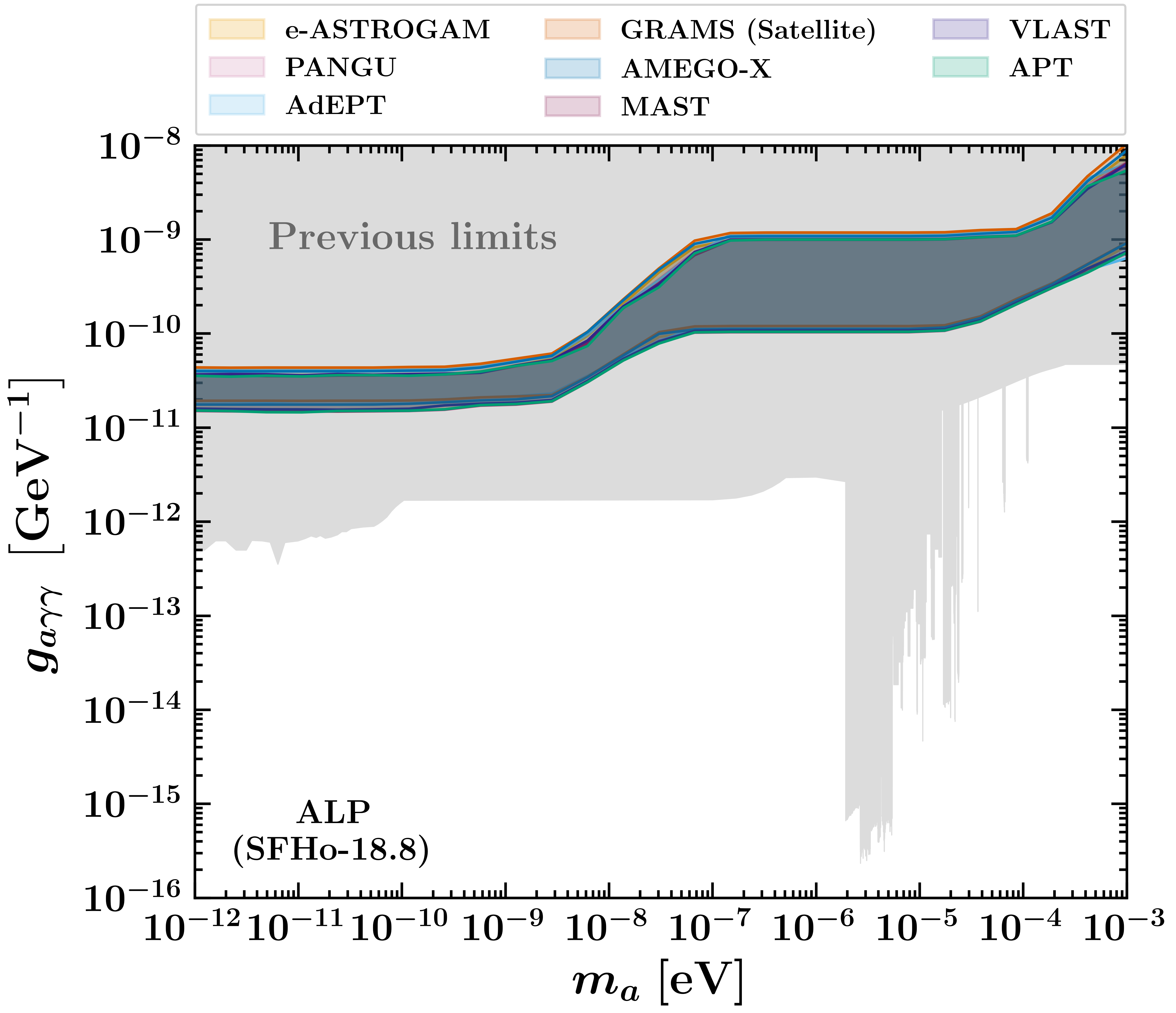}
        \caption{}
        \label{sf:fisher_limits_noUV_18p8}
    \end{subfigure}
    \begin{subfigure}{0.485\textwidth}
        \centering
        \includegraphics[width=0.975\linewidth]{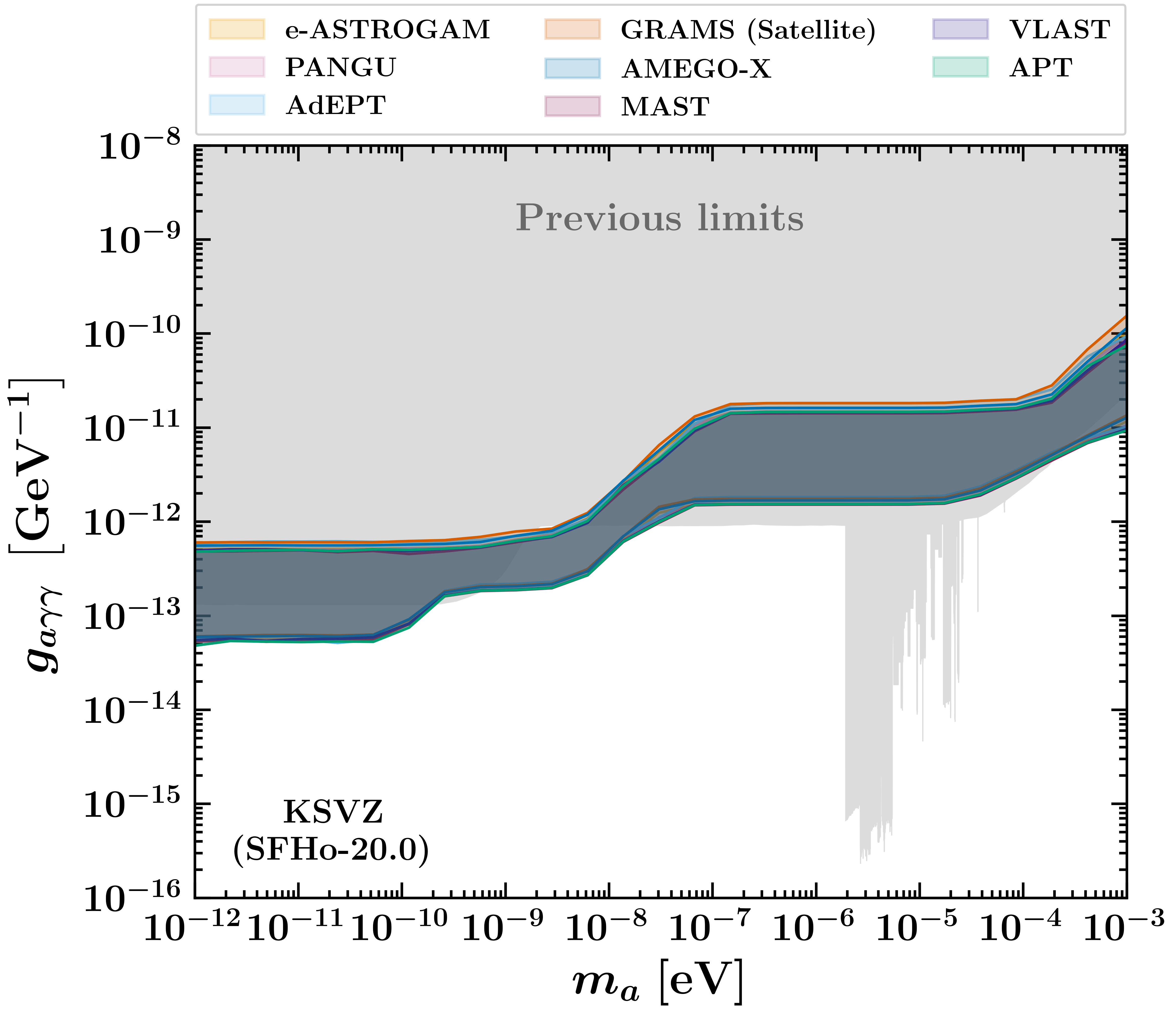}
        \caption{}
        \label{sf:fisher_limits_KSVZ_20p0}
    \end{subfigure}
    \begin{subfigure}{0.485\textwidth}
        \centering
        \includegraphics[width=0.975\linewidth]{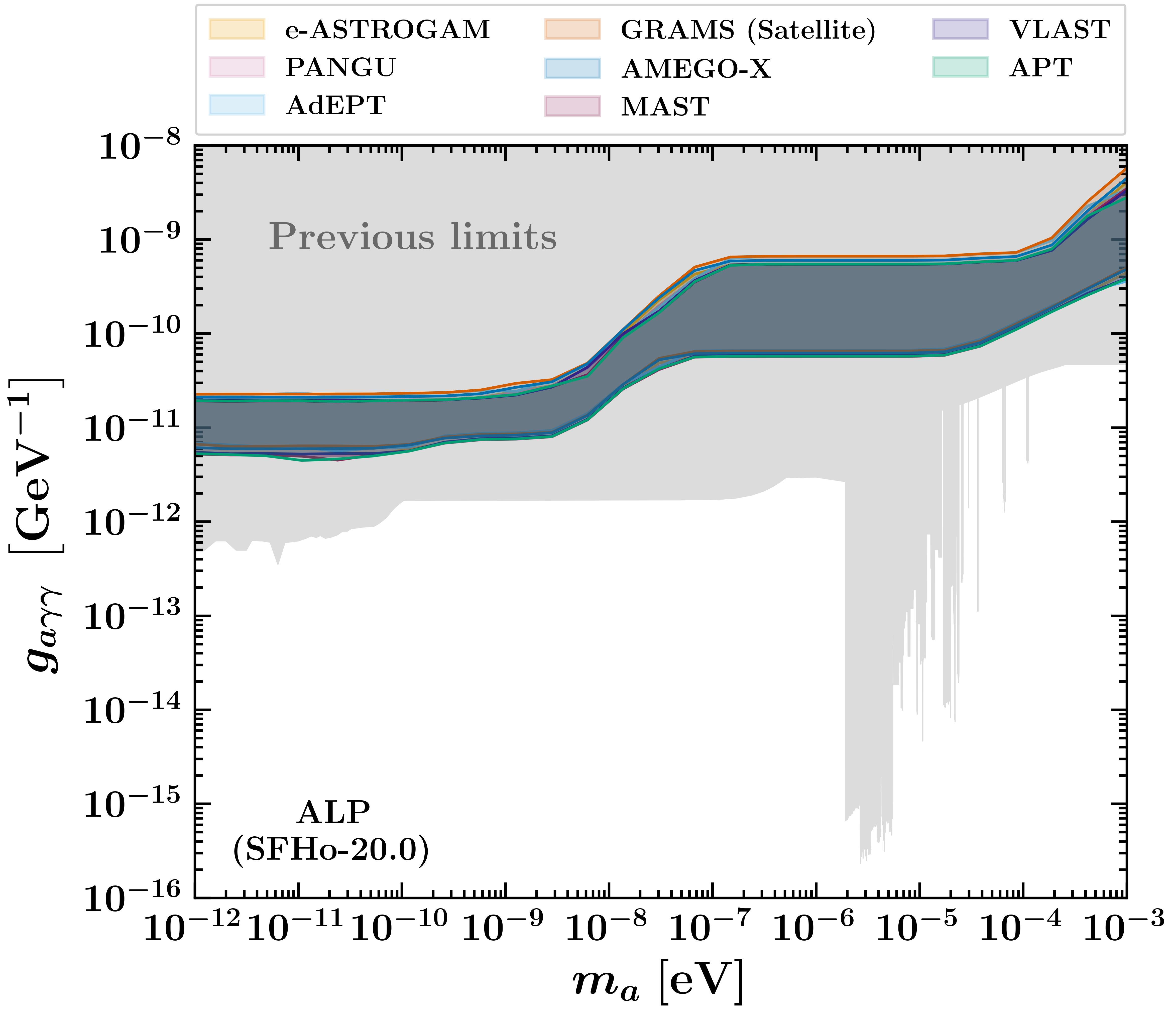}
        \caption{}
        \label{sf:fisher_limits_noUV_20p0}
    \end{subfigure}
    \caption{\justifying Same as Fig.~\ref{fig:fisher_limits} but for SFHo-18.8 (top) and SFHo-20.0 (bottom) CCSN profiles.}
    \label{fig:fisher_limits_other_models}
\end{figure*}
}
\newcommand{\FisherCorner}{%
\begin{figure*}[!htbp]
    \centering
    \includegraphics[width=0.9\linewidth]{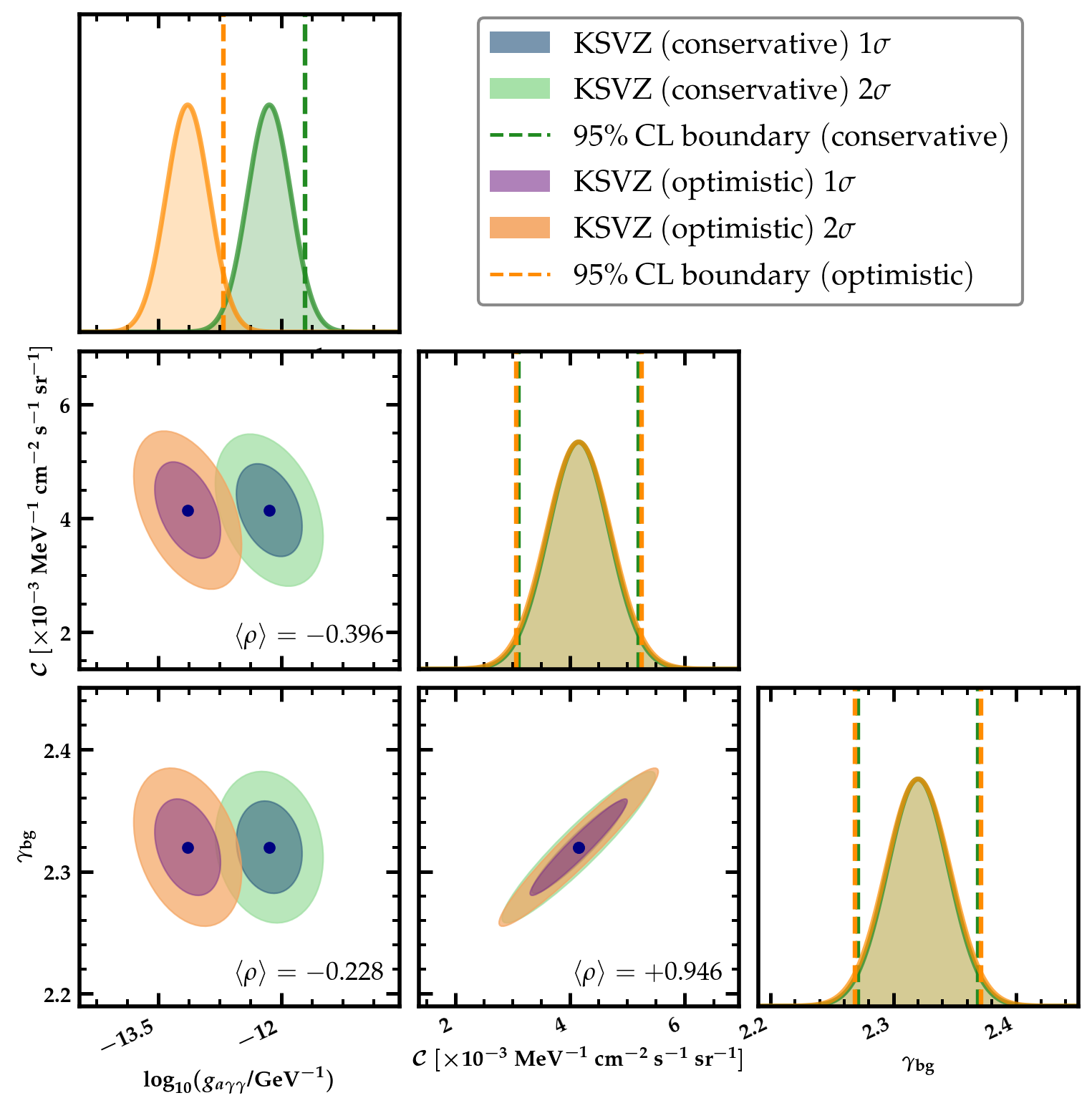}
    \caption{\justifying Fisher forecast corner plot for the APT experiment at axion mass 
$m_a = 10^{-12}\,\mathrm{eV}$ (KSVZ model) and observation time 
$T_{\rm obs}=10^{7}\,\mathrm{s}$ for the SFHo--18.6 CCSN model. 
Diagonal panels show the marginalized posterior distributions for 
$\log_{10}(g_{a\gamma\gamma}/\mathrm{GeV}^{-1})$, the background 
amplitude $\mathcal{C}$, and the spectral index $\gamma_{\rm bg}$, 
with dashed vertical lines indicating the $95\%$ CL boundaries. 
\textcolor[HTML]{FF8C00}{Orange} shading corresponds to the 
\textcolor[HTML]{FF8C00}{optimistic}  scenario, 
while \textcolor[HTML]{228B22}{green} shading corresponds to the 
\textcolor[HTML]{228B22}{conservative} scenario of axion--photon coupling. Off-diagonal panels 
display the \textcolor[HTML]{7B2D8B}{$1\sigma$} (dark) and 
\textcolor[HTML]{F4A460}{$2\sigma$} (light) Fisher confidence ellipses 
for the optimistic case and the \textcolor[HTML]{1F4E79}{$1\sigma$} 
(dark) and \textcolor[HTML]{9CDE9F}{$2\sigma$} (light) ellipses for 
the conservative case in each parameter plane. The dark blue dot marks 
the fiducial parameter values, and the mean Pearson correlation 
coefficient $\langle\rho\rangle$ is quoted in each panel.}
    \label{fig:fisher_corner_ksvz}
\end{figure*}
}
\begin{document}
\title{Lights, Camera, Axion: Tracing Axions from Supernovae in the Diffuse $\gamma$-ray Sky}

\author{Brijesh Kanodia \orcidlink{0000-0001-8999-8708}}
\email{brijeshk@iisc.ac.in}
\affiliation{\mbox{Department of Physics, Indian Institute of Science, C. V. Raman Avenue, Bengaluru 560012, India}}
\affiliation{\mbox{Centre for High Energy Physics, Indian Institute of Science, C. V. Raman Avenue, Bengaluru 560012, India}}

\author{Debajit Bose \orcidlink{0000-0001-8594-8885}}
\email{debajitbose550@gmail.com}
\affiliation{\mbox{Centre for High Energy Physics, Indian Institute of Science, C. V. Raman Avenue, Bengaluru 560012, India}}

\author{Subhadip Bouri \orcidlink{0000-0002-4971-8916}}
\email{subhadipb@iisc.ac.in}
\affiliation{\mbox{Department of Physics, Indian Institute of Science, C. V. Raman Avenue, Bengaluru 560012, India}}
\affiliation{\mbox{Centre for High Energy Physics, Indian Institute of Science, C. V. Raman Avenue, Bengaluru 560012, India}}

\author{Ranjan Laha \orcidlink{0000-0001-7104-5730}}
\email{ranjanlaha@iisc.ac.in}
\affiliation{\mbox{Centre for High Energy Physics, Indian Institute of Science, C. V. Raman Avenue, Bengaluru 560012, India}}

\date{\today}
%
%
%%%%%%%%%%%%%%%%%%%%%%%%%%%%%%%%%%%%%%%%%%%%%%%%%%%%%%%%%%%%%%%%%%%%%%%%%%%
%%%%%%%%%%%%%%%%%%%%%%%%       ABSTRACT       

\begin{abstract}

Axions produced copiously in core-collapse supernovae can convert into photons as they propagate through various astrophysical magnetic fields. The cumulative emission from the cosmic population of supernovae can therefore generate a diffuse gamma-ray signal through axion–photon conversion. In this work, we develop a comprehensive framework to compute the diffuse gamma-ray flux by modeling axion production in supernovae and, \textit{for the first time}, consistently accounting for their conversion into photons across all relevant magnetic field environments — progenitor, host galaxy, intergalactic medium, and the Milky Way — together with an updated cosmic star formation rate. Using measurements of the diffuse gamma-ray sky from COMPTEL, EGRET, and \textit{Fermi}-LAT, we derive competitive constraints on the axion–photon coupling over a wide range of axion masses. We further forecast the sensitivity of upcoming MeV gamma-ray telescopes to this diffuse signal using a Fisher forecast analysis.

\end{abstract}
%%%%%%%%%%%%%%%%%%%%%%%%%%%%%%%%%%%%%%%%%%%%%%%%%%%%%%%%%%%%%%%%%%%%%%%%%%%
%%%%%%%%%%%%%%%%%%%%%%%%%%%%%%%%%%%%%%%%%%%%%%%%%%%%%%%%%%%%%%%%%%%%%%%%%%%
%
%
\maketitle
\blfootnote{B.K., D.B., and S.B. contributed equally to this work.}

%
%
%%%%%%%%%%%%%%%%%%%%%%%%%%%%%%%%%%%%%%%%%%%%%%%%%%%%%%%%%%%%%%%%%%%%%%%%%%%
%%%%%%%%%%%%%%%%%%%%%%%%       INTRO      %%%%%%%%%%%%%%%%%%%%%%%%%%%%%%%%%
\section{Introduction}
\label{sec:intro}

The Standard Model (SM) of particle physics provides a remarkably successful description of fundamental interactions at the smallest length scales. Despite this success, several observations including neutrino oscillations, dark matter, and the strong CP problem indicate the existence of physics beyond the SM. One class of well-motivated extensions of the SM introduces pseudo-Nambu-Goldstone bosons, known as QCD axions, as a solution to the strong CP problem\,\cite{Peccei:1977hh, Peccei:1977ur, Weinberg:1977ma, Wilczek:1977pj, Dine:1981rt, Kim:2008hd, Marsh:2015xka}. Particles with properties similar to those of axions are predicted in the compactification of extra dimensions in string theory\,\cite{Svrcek:2006yi, Conlon:2006tq, Arvanitaki:2009fg, Cicoli:2012sz, DiLuzio:2020wdo, Gendler:2024adn, Cicoli:2026fqp}; these particles are collectively known as axion-like particles (ALPs). Since ALPs are not necessarily tied to the strong sector, they span a much broader parameter space compared to the QCD axion \footnote{Hereafter, the term ``axion'' is used in a broader sense to encompass ALPs and not necessarily tied to the strong sector.}. Axions can also serve as viable dark matter candidates through the misalignment mechanism\,\cite{Preskill:1982cy, Abbott:1982af, Arias:2012az, Co:2019jts, Eroncel:2022vjg}. One of the most interesting features of axions is their dimension-5 coupling to SM photons, which enables the conversion between axions and photons in the presence of external electric or magnetic fields. Through this coupling, loop diagrams allow axions to couple to leptons and hadrons; however, there are models in which axions have tree level couplings to leptons and hadrons\,\cite{Kim:1979if, Shifman:1979if, DiLuzio:2020wdo}.

One of the most important consequences of axion interactions with SM particles is its potential production in stellar environments, which can modify stellar properties. Core-collapse supernova (CCSN) which host the densest and hottest interiors among astrophysical objects, provide especially favorable conditions for copious axion production. If the interactions of such axions are in a favorable range, the produced axions can escape from the supernova\footnote{We use ``supernova" and ``CCSN" interchangeably throughout the text.} interior, thus carrying away energy and leading to enhanced cooling of the supernova. Observations of electron anti-neutrinos from Supernova 1987A (SN~1987A), comprising about 25 detected events\,\cite{Kamiokande-II:1987idp, Bionta:1987qt, Hirata:1988ad, IMB:1988suc, Alekseev:1988gp}, place constraints on axion couplings\,\cite{Turner:1987by, Burrows:1988ah, Raffelt:2006cw, Carenza:2019pxu, Carenza:2020cis, Carenza:2021pcm, Lella:2023bfb, Carenza:2023lci}, as any additional energy loss would modify the observed neutrino signal. An alternative way to probe axion couplings arises from the conversion of axions into photons in the presence of astrophysical magnetic fields. As axions escape from the dense supernova interior and propagate towards us, they can convert into photons in the magnetic fields of the intervening media. Given the typical supernova temperature of $\sim 40 \, {\rm MeV}$ (see Fig.~1 of Ref.\,\cite{Bollig:2020xdr}), the produced axions lie predominantly in the MeV energy range. Consequently, the resulting photons from axion conversion also fall in the MeV regime, making them accessible to terrestrial $\gamma$-ray experiments. Coincidentally, the Solar Maximum Mission (SMM) telescope had partial observational overlap in the direction of SN~1987A during the period when neutrinos from the event were detected. The non-observation of gamma rays by SMM from the direction of SN~1987A has been used to place constraints on axion-photon coupling\,\cite{Raffelt:1990yz,Raffelt:1996wa,Grifols:1996id,Brockway:1996yr,Payez:2014xsa,Hoof:2022xbe, Manzari:2024jns,Caputo:2024oqc, Candon:2025fnb}.

With the improved sensitivity of gamma-ray telescopes, a future Galactic CCSN can substantially strengthen the existing limits on axion parameters\,\cite{Payez:2014xsa, Meyer:2016wrm, Calore:2023srn, Lella:2024hfk, Manzari:2024jns, Carenza:2024ehj, Xie:2025igf}. However, Galactic supernovae are rare and unpredictable, which makes targeted observations difficult. Extending the search to extra-galactic supernova events increases the rate but makes individual gamma-ray detections challenging. An interesting alternative is to consider the cumulative axion emission from all past CCSNe throughout the Universe. These axions can convert into photons as they traverse astrophysical magnetic fields and contribute to the diffuse extra-galactic gamma-ray background\,\cite{Raffelt:2011ft, Calore:2020tjw, Calore:2021hhn, Lella:2024dmx, Eby:2024mhd, Candon:2025fnb}. Considering axion–photon conversion only within the Milky Way (MW) magnetic field leads to a strong suppression of the flux for axion masses $m_a \gtrsim \mathcal{O}(10^{-9}) \, \rm eV$, beyond which the progenitor magnetic fields must also be considered.

In this article, we develop a comprehensive framework to compute the diffuse extra-galactic gamma-ray flux by modeling the magnetic fields of progenitor stars, the host galaxy, the intergalactic medium (IGM), and the MW, considering various uncertainties in each component. We also adopt updated measurements of the star formation rate density (SFRD) across low and high redshifts to derive the cosmic CCSN rate. After performing a detailed calculation of the resulting diffuse gamma-ray flux, we derive constraints on the axion–photon coupling from Imaging Compton Telescope (COMPTEL)\,\cite{1998PhDT.........3K, 2000AIPC..510..467W}, Energetic Gamma Ray Experiment Telescope (EGRET)\,\cite{EGRET:1997qcq, Strong:2004ry,Thompson:2008rw}, and Fermi Large Area Telescope (\textit{Fermi}-LAT)\,\cite{Fermi-LAT:2009ihh,Fermi-LAT:2014ryh} observations. Finally, we forecast the sensitivity of upcoming gamma-ray missions to this diffuse signal and estimate their projected reach for the axion–photon coupling using a Fisher forecast analysis. To estimate future sensitivities, we consider the following upcoming telescopes:\,\,All-sky Medium Energy Gamma-ray Observatory eXplorer (AMEGO-X)\,\cite{Caputo:2022xpx}, e-ASTROGAM\,\cite{eASTROGAM:2016bph}, Massive Argon Space Telescope (MAST)\,\cite{Dzhatdoev:2019kay}, Gamma-Ray and AntiMatter Survey (GRAMS)\,\cite{GRAMS:2021tax, Aramaki:2019bpi}, Advanced Energetic Pair Telescope (AdEPT)\,\cite{Hunter:2013wla}, Pair-production Gamma-ray Unit (PANGU)\,\cite{Wu:2014tya}, Advanced Particle-astrophysics Telescope (APT)\,\cite{APT:2021lhj}, and Very Large Area gamma-ray Space Telescope (VLAST)\,\cite{Pan:2024adp}.
%
%
%%%%%%%%%%%%%%%%%%%%%%%%%%%%%%%%%%%%%%%%%%%%%%%%%%%%%%%%%%%%%%%%%%%%%%%%%%%
%%%%%%%%%%%%%%%%%%%%%%%%%%%%%%%%%%%%%%%%%%%%%%%%%%%%%%%%%%%%%%%%%%%%%%%%%%%

%
%
%%%%%%%%%%%%%%%%%%%%%%%%% FIGURE: SN ALP FLux %%%%%%%%%%%%%%%%%%%%%%%%%%%%%
\alpFluxSupernovaFig
%%%%%%%%%%%%%%%%%%%%%%%%%%%%%%%%%%%%%%%%%%%%%%%%%%%%%%%%%%%%%%%%%%%%%%%%%%%
%
%
%%%%%%%%%%%%%%%%%%%%%%%%%%%%%%%%%%%%%%%%%%%%%%%%%%%%%%%%%%%%%%%%%%%%%%%%%%%
%%%%%%%%%%%%%%%%%%%%%%%%       ALP FLUX      %%%%%%%%%%%%%%%%%%%%%%%%%%%%%%
\section{Axion flux from Supernova}
\label{sec:alp_flux}

The effective Lagrangian between the axion field $(a)$ and the photon field is\,\cite{Raffelt:1987im}
\begin{equation}\label{eq:L_agg}
    \mathcal{L}_{a \gamma \gamma} = - \dfrac{1}{4} \, g_{a \gamma \gamma} \, a \, F^{\mu \nu} \, \tilde{F}_{\mu \nu} \, ,
\end{equation}
where $F_{\mu \nu}$ and $\tilde{F}_{\mu \nu}$ denote the electromagnetic field strength tensor and its dual, respectively. The effective axion–photon coupling can be written as $g_{a \gamma \gamma} = C_{a \gamma \gamma} \, \alpha_{\rm EM} / (2 \pi f_a)$, with $\alpha_{\rm EM}$ being the fine-structure constant, $f_a$ the axion decay constant, and we take the coefficient to be $C_{a \gamma \gamma} = 1$. The axion can also couple to quarks. However, below the QCD phase transition scale, the more relevant interactions are with hadrons. The interaction Lagrangian between axions and nucleons is\,\cite{Carena:1988kr, Chang:1993gm, DiLuzio:2020wdo}
\begin{equation}\label{eq:L_aNN}
    \mathcal{L}_{aNN} = \dfrac{\partial^\mu a}{2 f_a} \left( C_{app} \, \bar{p} \gamma_\mu \gamma_5 p \, + \, C_{ann} \, \bar{n} \gamma_\mu \gamma_5 n  \right) \, ,
\end{equation}
where $C_{app}$ and $C_{ann}$ denote the axion couplings to protons and neutrons, respectively. Axions can also couple to pions, both directly and through interactions mediated by the $\Delta$-baryon, and the corresponding Lagrangian is\,\cite{Chang:1993gm, DiLuzio:2020wdo, Carenza:2023lci}
\begin{equation}\label{eq:L_apiN}
\begin{aligned}
    \mathcal{L}_{a \pi N} & = \dfrac{\partial^\mu a}{2 f_a} \left[ i \, \dfrac{C_{a \pi N}}{f_\pi} \left(\pi^+ \bar{p} \gamma_\mu n \, - \, \pi^- \bar{n} \gamma_\mu p  \right) \right.
    \\ & \left. + C_{a N \Delta} \left(\bar{p} \, \Delta_\mu^+ \, + \, \overline{\Delta_\mu^+} \, p \, + \, \bar{n} \, \Delta_\mu^0 \, + \, \overline{\Delta_\mu^0} \, n \right) \right] \, ,
\end{aligned}
\end{equation}
where $f_{\pi} = 92.4 \, {\rm MeV}$, $C_{a \pi N} = (C_{app} - C_{ann})/\sqrt{2} g_A$ with $g_A = 1.28$, and $C_{a N \Delta} = - \sqrt{3}/2 (C_{app} - C_{ann})$\,\cite{ParticleDataGroup:2024cfk, Carenza:2023lci}. 

In general, axions may not have UV contributions to the nucleon couplings. However, in such cases, nucleon interactions are generated by renormalization group running. In this work, we consider two scenarios: (i) nucleon operators generated by loop diagrams, for which $C_{app} \approx C_{ann} \approx 10^{-4}$\,\cite{Manzari:2024jns} (hereafter denoted as ``ALP”); and (ii) KSVZ axions\,\cite{Kim:1979if, Shifman:1979if}, for which non-zero UV couplings to quarks are present, i.e., $C_{app} \approx -0.47$, $C_{ann} \approx -0.02$\,\cite{Manzari:2024jns} (hereafter denoted as ``KSVZ”). To calculate the axion flux produced in the supernova, the radial profiles of the CCSN are required. We adopt profiles based on the simulations of\,\cite{Bollig:2020xdr}, with data obtained from the Garching Core-Collapse Supernova Archive\,\cite{garching_sne_data}. We consider radial profiles of three simulations, namely SFHo-18.6, SFHo-18.8, and SFHo-20.0, where the numbers denote the progenitor masses in $M_\odot$.
%
%
%%%%%%%%%%%%%%%%%%%%%%%%%%%%%%%%%%%%%%%%%%%%%%%%%%%%%%%%%%%%%%%%%%%%%%%%%%%
\subsection{Primakoff process}
\label{subsec:prim}

In the presence of the axion–photon coupling introduced in Eq.~\eqref{eq:L_agg}, photons can convert into axions in the electric fields of charged particles in the stellar plasma through the Primakoff process\,\cite{Raffelt:1985nk}. The Primakoff rate describing the conversion of photons into axions is\,\cite{Manzari:2024jns}
\begin{equation}\label{eq:Gam_Prim}
\begin{aligned}
    \Gamma_{\rm {Prim}} & = \dfrac{g_{a \gamma \gamma}^2 \, Z_k^2 \, \alpha_{\rm EM} \, n^{\rm eff}_k \, (1 - \zeta)}{4 \, E_a^2 \, (2 - \zeta)^3} \left[ - 2 E_a^2 (2 - \zeta) \right. \\
    & \left. + \left( \kappa_{\rm eff}^2 + 2 (2 - \zeta) E_a^2 \right) \  {\rm log} \left( \dfrac{\kappa_{\rm eff}^2 + 2 E_a^2 (2 - \zeta)}{\kappa_{\rm eff}^2} \right) \right] \, ,
\end{aligned}
\end{equation}
where $E_a$ is the axion energy, $\zeta = (\omega_p^2 + m_a^2) / E_a^2$, with $\omega_p$ denoting the photon plasma frequency, $m_a$ being the axion mass. The atomic number and effective number density of the target particle $k$ are represented by $Z_k$ and $n_k^{\rm eff}$, respectively. Protons constitute the dominant targets for the Primakoff process in the CCSN interior. Although other charged particles such as electrons, muons, and pions are also present, their contributions are subdominant. In particular, electrons are highly degenerate in the CCSN core, which suppresses their contribution to the Primakoff process. Muons and pions possess non-zero and non-degenerate abundances, which can together contribute an additional $\mathcal{O}(10\%)$ to the axion production\,\cite{Fiorillo:2025gnd}. However, to remain conservative, we include only the dominant proton contribution in our calculation of the axion flux from the Primakoff process. The effective number density is obtained by integrating the proton Fermi–Dirac distribution function $f_p$ over momentum $(q)$, and can be written as\,\cite{Payez:2014xsa}
\begin{equation}\label{eq:n_p_eff}
    n_p^{\rm eff} = 2 \int \dfrac{d^3q}{(2 \pi)^3} \, f_p \, (1 - f_p) \,,
\end{equation}
where the factor 2 accounts for the two spin states of the proton. In Eq.~\ref{eq:Gam_Prim}, $\kappa_{\rm eff}$ represents the screening scale that accounts for the finite range of the electric field of charged particles inside the supernova, and can be written as $\kappa_{\rm eff}^2 = 4 \pi \alpha_{\rm EM} (n_p^{\rm eff} / T)$, with $T$ being the temperature. Using the Primakoff production rate discussed above, the differential axion emission rate per unit volume per unit energy is\,\cite{Manzari:2024jns}
\begin{equation}\label{eq:d2na_dEa_dt_Prim}
    \left. \dfrac{d^2n_a}{dE_a \, dt} \right|_{\rm Prim} = \dfrac{E_a^2}{\pi^2} \left(1 - \frac{\omega_p^2}{E_a^2} \right) \, \left(e^{E_a/T} - 1 \right)^{-1} \, \Gamma_{\rm Prim} \, .
\end{equation}
At high nuclear densities, scalar interactions lead to a reduction in the nucleon mass. In our analysis, we use the local effective proton mass $m_p^*$ and chemical potential $\mu_p$ as provided in the supernova profiles. After being produced in the supernova, axions experience gravitational redshift as they propagate outward. We account for this effect following Ref.\,\cite{Caputo:2021rux} and incorporate the local gravitational lapse factor provided in the supernova profile data.
%
%
%%%%%%%%%%%%%%%%%%%%%%%%%%%%%%%%%%%%%%%%%%%%%%%%%%%%%%%%%%%%%%%%%%%%%%%%%%%
\subsection{Nucleon-nucleon bremsstrahlung}
\label{subsec:brem}

In addition to the Primakoff channel, axions are also efficiently produced through nucleon–nucleon bremsstrahlung in the proto-neutron star (PNS) when axion–nucleon couplings, as introduced in Eq.~\eqref{eq:L_aNN}, are taken into account\,\cite{Turner:1987by, Carena:1988kr, Brinkmann:1988vi, Carenza:2019pxu, Candon:2025fnb}. The differential axion production rate per unit volume per unit energy can be written as\,\cite{Carenza:2019pxu, Carenza:2023lci}

\begin{equation}\label{eq:d2na_dEa_dt_Brem}
\begin{split}
    \left. \dfrac{d^2n_a}{dE_a \, dt} \right|_{\rm Brem} &= \frac{n_B}{16\pi^2 f_a^{2}} \, (E_a-m_a^2)^{3/2} \\
    &\quad \times e^{-E_a/T} \, S_\sigma(E_a) \, \Theta(E_a-m_a) \, ,
\end{split}
\end{equation}
where $n_B$ is the baryon number density. The spin structure function is defined as\,\cite{Carenza:2019pxu, Carenza:2023lci}
\begin{equation}\label{eq:S_sigma}
    S_\sigma = \dfrac{\Gamma_\sigma}{E_a^2 \, + \, \Gamma^2} \, s_f \! \left(\frac{E_a}{T}\right) \, ,
\end{equation}
where the dimensionless function $s_f(E_a/T)$ encodes the structure of the matrix element together with the five dimensional phase-space integration, and its explicit form is taken from Ref.\,\cite{Carenza:2023lci}. To evaluate $s_f(E_a/T)$, we use the nucleon fractions $Y_N$, which denote the number of nucleons per baryon, together with the nucleon degeneracy parameter $\hat{\mu}_N = (\mu_N - m_N - U_N)/T$, where $\mu_N$, $m_N$, and $U_N$ are the nucleon chemical potential, mass and self-energy, respectively. The nucleon fractions are determined by assuming charge neutrality, $Y_p \approx Y_e$, together with $Y_p + Y_n= 1$\,\cite{Janka:2006fh, 2010A&A...522A..25A}, where $Y_k$ denotes the number fraction of a given particle $k$ relative to the total baryon number. Although muons can reach an abundance of $\mathcal{O}(10\%)$, we do not include them in the evaluation of these fractions. Furthermore, we neglect the presence of exotic hadronic states that may have non-zero abundance such as hyperons or thermal mesons\,\cite{Banik:2014qja, Fore:2019wib}. The width in Eq.~\eqref{eq:S_sigma} is given by $\Gamma = g_\sigma \, \Gamma_\sigma$, where $\Gamma_\sigma$ denotes the nucleon spin-fluctuation rate defined as\,\cite{Carenza:2023lci}
\begin{equation}\label{eq:Gamma_sigma}
    \Gamma_\sigma = \dfrac{4 \, \rho}{\pi^{3/2}} \, \left( \dfrac{g_A}{2 f_\pi} \right)^4 \, (m_N \, T)^{1/2} \, .
\end{equation}
To determine $g_\sigma$, we follow the normalization  prescription \cite{Carenza:2023lci,Hannestad:1997gc,Raffelt:1996di}
\begin{equation}\label{eq:g_sigma_find}
\begin{aligned}
    \int_0^\infty \frac{dE_a}{2\pi} \, & \left(1+e^{-E_a/T}\right) \, S_\sigma(E_a)\, = \\
    & \frac{1}{n_B} \sum_{N=p,n} \frac{C_{aNN}^2}{C_{app}^2Y_p + C_{ann}^2Y_n} \! \int \! \frac{2 \, d^3 q}{(2\pi)^3} \, f_N (1-f_N) \, ,
\end{aligned}
\end{equation}
where $f_N$ is the Fermi-Dirac distribution function of nucleon $N$. We solve
Eq.~\eqref{eq:g_sigma_find} for $g_\sigma$ by first computing a core-averaged value
$g_{\rm core}(t)$ over the high-density emitting region at each snapshot time $t$, and then taking a time average via $\bar g_\sigma \equiv \frac{1}{N_t}\sum_{i=1}^{N_t} g_{\rm core}(t_i)$. For the supernova profiles used in this work\,\cite{Mirizzi:2015eza, Bollig:2017lki, Bollig:2020xdr, garching_sne_data}, we obtain
\begin{equation}\label{eq:g_sigma_vals}
    \bar g_\sigma =
    \begin{cases}
        3.277 \times 10^{-3} \, ,  & \text{SFHo-18.6} \, ; \\
        3.617 \times 10^{-3} \, ,  & \text{SFHo-18.8} \, ; \\
        3.982 \times 10^{-3} \, ,  & \text{SFHo-20.0} \, .
    \end{cases}
\end{equation}
To calculate the axion flux from nucleon bremsstrahlung using Eq.~\eqref{eq:d2na_dEa_dt_Brem}, we fixed $g_\sigma=\bar{g}_\sigma$ for each profile.
%
%
%%%%%%%%%%%%%%%%%%%%%%%%%%%%%%%%%%%%%%%%%%%%%%%%%%%%%%%%%%%%%%%%%%%%%%%%%%%
\subsection{Pion conversion}
\label{subsec:pion_conv}

For axions with energies larger than the pion mass $(m_\pi)$, the pion-induced production mechanism becomes dominant within the supernova core. In particular, axions can be efficiently produced via pion–nucleon Compton-like scattering, where a pion scatters off a nucleon and converts into an axion\,\cite{Raffelt:1993ix, Keil:1996ju, Carenza:2020cis, Fischer:2021jfm, Lella:2022uwi}. The corresponding differential axion emission rate per unit volume per unit energy is\,\cite{Carenza:2023lci}
\begin{equation}\label{eq:d2na_dEa_dt_Pion}
\begin{aligned}
    \dfrac{d^2n_a}{dE_a \, dt} & \bigg|_{\rm \pi-conv} = C_a^{p \pi^-} \, \dfrac{(m_N \, T)^\frac{3}{2}}{\sqrt{8} \, \pi^5 \, f_a^2} \, \left( \dfrac{g_A}{2 f_\pi} \right)^2 \, (E_a^2 - m_a^2)^\frac{3}{2} \,  \\ 
    & \times \left( \dfrac{\Theta \left( E_a - {\rm max}[m_a, m_\pi] \right)}{e^{\frac{(E_a - m_\pi)}{T} - \hat{\mu}_\pi} - 1} \right) \, \left( \dfrac{\sqrt{E_a^2 - m_\pi^2)}}{E_a^2 + \Gamma^2} \right) \\
    & \times \int_0^\infty dy \, y^2 \, \left( \dfrac{1}{e^{y^2 - \hat{\mu}_p} + 1} \right) \, \left( \dfrac{1}{e^{-y^2 + \hat{\mu}_n} + 1} \right) \, ,
\end{aligned}
\end{equation}
where $\Gamma = \Gamma_\sigma/2$ incorporates the impact of multiple nucleon scatterings. The pion degeneracy parameter $\hat{\mu}_\pi$ is determined from the beta equilibrium condition. To calculate the axion production from pion conversion, we consider the process $\pi^- \, p \to n \, a$. Similar diagrams involving $\pi^+$ and $\pi^0$ are also possible; however, their contributions are negligible due to the much smaller abundances of $\pi^+$ and $\pi^0$ compared to $\pi^-$ in the supernova environment\,\cite{Fischer:2021jfm}. The coefficient $C_a^{p \pi^-}$ associated with the $\pi^-$ conversion process is adopted from\,\cite{Ho:2022oaw, Carenza:2023lci}, retaining only the leading-order terms in the $(1/m_N)$ expansion.
%
%
%%%%%%%%%%%%%%%%%%%%%%%%%%%%%%%%%%%%%%%%%%%%%%%%%%%%%%%%%%%%%%%%%%%%%%%%%%%
\subsection{Total Axion Emission}
\label{subsec:total_alp}

After computing the axion emission rate at a radial distance $r$ from the center of the supernova for each production channel, the total differential axion production rate per unit energy from the supernova is
\begin{equation}\label{eq:ALP_prod_rate}
\begin{aligned}
    \dfrac{d^2N_a}{dE_a \, dt} = \int_{r_{\rm min}}^{r_{\rm max}} dr \, 4 \pi r^2 \, \Bigg( & \left. \dfrac{d^2 n_a}{dE_a \, dt} \right|_{\rm Prim} \, + \, \left. \dfrac{d^2 n_a}{dE_a \, dt} \right|_{\rm Brem} \\
    & + \, \left. \dfrac{d^2 n_a}{dE_a \, dt} \right|_{\rm \pi-conv} \Bigg) \, .
\end{aligned}
\end{equation}
In our calculation, we take $r_{\rm min}$ to be the smallest radius available in the supernova profiles; for instance, for the SFHo-18.6 model, this corresponds to $\sim 117 \, {\rm m}$. While the profiles extend to much larger radii, we adopt $r_{\rm max} = 10^3 \, {\rm km}$, which remains well within the progenitor radius to ensure that axions are produced inside the stellar radius. Furthermore, contributions from larger radii are strongly suppressed due to the rapid decrease in density and temperature. The differential axion production rate per unit energy is computed using Eq.~\eqref{eq:ALP_prod_rate} for each time slice encoded the supernova profiles, which typically extend up to $\sim 10 \, {\rm s}$ after the core-bounce. The resulting rates are then integrated numerically over time to obtain the total differential axion production from a single supernova.

In Fig.~\ref{fig:alp_flux_supernova}, we present the time-integrated differential axion spectra, $dN_a/dE_a$, as a function of energy for $g_{a \gamma \gamma} = 10^{-12} \, {\rm GeV^{-1}}$. Since the axion masses considered are much smaller than the characteristic supernova temperature, the spectrum is evaluated in the massless limit. The flux contributions from Primakoff production, nucleon-nucleon bremsstrahlung, and pion conversion are displayed in the left, middle, and right panels, respectively, for the three supernova models under consideration. The Primakoff flux is identical for the KSVZ and ALP models, as it depends solely on the axion–photon coupling. For the KSVZ scenario, where UV contributions to the nucleon couplings are present, nucleon-nucleon bremsstrahlung and pion conversion dominate the production. In contrast, for the ALP case, loop-induced nucleon couplings introduce these channels such that their contributions become similar in magnitude to that of the Primakoff process. As seen in Fig.~\ref{fig:alp_flux_supernova}, the axion production flux is highest for the SFHo-20.0 profile, as its central density and temperature is higher than the other two models\,\cite{Bollig:2020xdr}.
%
%
%%%%%%%%%%%%%%%%%%%%%%%%%%%%%%%%%%%%%%%%%%%%%%%%%%%%%%%%%%%%%%%%%%%%%%%%%%%
%%%%%%%%%%%%%%%%%%%%%%%%%%%%%%%%%%%%%%%%%%%%%%%%%%%%%%%%%%%%%%%%%%%%%%%%%%%

%
%
%%%%%%%%%%%%%%%%%%%%%%%%%%%%%%%%%%%%%%%%%%%%%%%%%%%%%%%%%%%%%%%%%%%%%%%%%%%
%%%%%%%%%%%%%%%%%%%%%%%%       ALP FLUX      %%%%%%%%%%%%%%%%%%%%%%%%%%%%%%
\section{Axion-photon conversion}
\label{sec:alp_conv}

The effective axion-photon interaction Lagrangian described in Eq.~\eqref{eq:L_agg} can also be expressed as $\mathcal{L}_{a \gamma \gamma} = g_{a \gamma \gamma} \ a \ \textbf{E} \cdot \textbf{B}$, where $\textbf{E}$ and $\textbf{B}$ represent the electric and magnetic components of $F_{\mu \nu}$. With this interaction, axions can be converted into photons while propagating through external astrophysical magnetic fields. The axions produced in supernovae travel through the magnetic fields of the progenitor star, the host galaxy, the IGM, and the MW before arriving at us. In this process, a fraction of the axions can be converted into gamma rays, which can be detected by terrestrial observatories. In this paper, we calculate, \textit{for the first time}, the diffuse extra-galactic gamma-ray flux produced by axions from the cosmic supernovae population, taking into account the axion-photon conversion in all the magnetic fields encountered along the line of sight. Additionally, the host galaxy may reside within a galaxy cluster, which can introduce an additional contribution to the conversion. However, this contribution is expected to be subdominant relative to other magnetic field components.

In this section, we address the astrophysical magnetic field uncertainties and outline our calculation of the axion-photon conversion probabilities $(P_{a\to\gamma})$ using \texttt{gammaALPs}\,\cite{Meyer:2021pbp}. For the gamma-ray energies considered, absorption effects are negligible; therefore, we do not include gamma-ray attenuation in our analysis. 

%
%
%%%%%%%%%%%%%%%%%%%%%% FIGURE: ALP-Photon conversion %%%%%%%%%%%%%%%%%%%%%%
\alpPhotonConv
%%%%%%%%%%%%%%%%%%%%%%%%%%%%%%%%%%%%%%%%%%%%%%%%%%%%%%%%%%%%%%%%%%%%%%%%%%%
%
%
%%%%%%%%%%%%%%%%%%%%%%%%%%%%%%%%%%%%%%%%%%%%%%%%%%%%%%%%%%%%%%%%%%%%%%%%%%%
\subsection{Progenitor magnetic field}
\label{subsec:conv_rsg_bsg}

The axion flux from a supernova drops off quickly after $\sim 10 \, {\rm s}$ from core bounce, and it has been recently shown in Ref.\,\cite{Manzari:2024jns} that, within this timescale, the progenitor magnetic field remains essentially unchanged, providing the conditions for axion-photon conversion. We concentrate on two kinds of progenitors of CCSN, namely red supergiants (RSGs) and blue supergiants (BSGs). The magnetic field of both kinds of stars is assumed to have a dipole configuration\,\cite{Fiorillo:2025gnd}, i.e., $B_{\rm prog}(R) = B_0 \left( R_0/R \right)^3$, where $B_0$ is the magnetic field at the photospheric radius $R_0$. For RSGs, we fix the minimum surface magnetic field to be $B^{\rm min}_{0, {\rm RSG}} = 1 \, {\rm G}$, following the observed surface magnetic fields of $\mathcal{O}(1) \, {\rm G}$\,\cite{tessore2017measuring}. It is suggested that RSG magnetic fields could reach $\mathcal{O}(100) \, {\rm G}$ and even as high as $\sim 500 \, {\rm G}$\,\cite{Dorch:2004af, auriere2010magnetic}. We conservatively set the maximum value to be $B^{\rm max}_{0, {\rm RSG}} = 100 \, {\rm G}$, which is consistent with the values derived in Ref.\,\cite{Manzari:2024jns} using the simulated RSG profile of Ref.\,\cite{goldberg2022numerical}. The photospheric radius is taken to be $R^{\rm min}_{0, {\rm RSG}} = 800 \, R_\odot$ and $R^{\rm max}_{0, {\rm RSG}} = 900 \, R_\odot$, which are the photospheric radii of the RSG1L4.5 and RSG2L4.9 hydrodynamical models, respectively\,\cite{goldberg2022numerical}. Although the effective gamma-ray photosphere may be slightly smaller than these values, as discussed in Ref.\,\cite{Manzari:2024jns}. For BSGs, we explore surface magnetic fields from $B_{0, {\rm BSG}} = 100 \, {\rm G}$ to $10 \, {\rm kG}$\,\cite{donati2009magnetic} with minimum and maximum surface radius $(R_{0, {\rm BSG}})$ be $30 \, R_\odot$ and $60 \, R_\odot$, respectively\,\cite{Woosley:1988at}. In order to estimate the uncertainty band of the axion-photon conversion probability, we analyze the extreme cases with the lowest and highest magnetic field configurations for each of the progenitors.

The magnetic field for each configuration is assumed to extend from $R_0$ to $20 \, R_0$. Instead of approximating the magnetic field as a single coherent domain, we divide the propagation region into radial bins. The dipole magnetic field is first discretized on a grid of $10$ logarithmically spaced domains within the region between $R_0$ and $20 \, R_0$. For each domain, we calculate the average magnetic field strength and approximate it as a coherent magnetic field cell. This piecewise-constant approximation enables us to capture the radial dependence of the dipole magnetic field. The axion-photon conversion is calculated with \texttt{gammaALPs}\,\cite{Meyer:2021pbp}. We start the system evolution from a pure axion state at $R_0$ and then propagate it through each coherent magnetic field domain sequentially. Following each domain, we update the state vector according to the transfer matrix corresponding to that domain and repeat the process for all cells until reaching $20 \, R_0$. The obtained photon state at the outer boundary determines the axion–photon conversion probability for the given magnetic field configuration. In determining the conversion probabilities, we neglect the photon plasma frequency in the medium outside the photospheric radius, since sufficiently high densities are required for it to have a significant impact, which are generally not encountered outside stellar environments\,\cite{Manzari:2024jns}.

In contrast to BSGs, RSGs are the most common progenitors of CCSNe. To extract this quantitatively, we calculate the probability that a CCSN originates from a RSG or a BSG using the stellar initial mass function (IMF). Instead of assuming a single power-law, we use a piecewise IMF $(\psi)$ that takes the form of the Chabrier function \cite{Chabrier:2003ki} for stellar mass $M \leq 1 \, M_\odot$ and the Salpeter power-law \cite{Salpeter:1955it} for $M > 1 \, M_\odot$, with both normalized at $1\, M_\odot$ and can be expressed as\,\cite{Madau:2014bja}
\begin{equation}\label{eq:stellar_IMF}
    \psi(M) \propto
    \begin{cases}
        \dfrac{e^{-\left[\log(M/M_c)\right]^2 / 2\sigma_M^2}}{M}, & M \leq 1 \, M_\odot \\[6pt]
        M^{-2.35}, & M > 1 \, M_\odot \, ,
    \end{cases}
\end{equation}
where $M_c = 0.079 \, M_\odot$ and $\sigma_M = 0.69$. Considering the progenitor mass ranges for RSGs and BSGs as given in Ref.\,\cite{ekstrom2025stellar}, we calculate the probabilities by integrating the IMF over the corresponding mass ranges. This yields the fractions of the cosmic supernova population originating from RSGs and BSGs to be $\xi_{\rm RSG}=0.83$ and $\xi_{\rm BSG}=0.17$, respectively. Note that, these fractions are effectively determined by the Salpeter IMF alone, since the progenitor masses lie well above $1 \, M_\odot$. We describe the combined IMF for completeness, as the same prescription will be employed in the subsequent calculation of star formation rates. In Fig.~\ref{sf:patog_prog}, we plot the axion-photon conversion probability in the dipolar magnetic field of RSG (red: $P_{a\to\gamma}^{\rm RSG}$) and BSG (blue: $P_{a\to\gamma}^{\rm BSG}$). The red and blue solid curves denote the maximum and minimum conversion probabilities in each case, shown as a function of the axion mass for $E_a = 100\,{\rm MeV}$ and $g_{a\gamma\gamma}=10^{-12}\,{\rm GeV}^{-1}$. The shaded band in between the curves corresponds to the uncertainty in the conversion probability due to the variation of magnetic field configurations considered for the progenitor stars. To calculate contribution from the CCSNe population, we consider both progenitor stars by weighting the corresponding conversion probabilities according to their cosmic abundance. Consequently, the conversion probability in the progenitor magnetic field is $P_{a\to\gamma}^{\rm Prog} = \xi_{\rm RSG} \, P_{a\to\gamma}^{\rm RSG} + \xi_{\rm BSG} \, P_{a\to\gamma}^{\rm BSG}$.
%
%
%%%%%%%%%%%%%%%%%%%%%%%%%%%%%%%%%%%%%%%%%%%%%%%%%%%%%%%%%%%%%%%%%%%%%%%%%%%
\subsection{Host galaxy magnetic field}
\label{subsec:host_galaxy}

The axions surviving the progenitor magnetic field can further convert into photons as they traverse the magnetic field of the host galaxy. Since our analysis takes into account the diffuse population of CCSNe, the magnetic field structure of each host galaxy is not known in detail. To address this, we examine a wide range of typical models of the host galaxy magnetic field. For the maximum conversion case, we use the magnetic field configurations as inferred from LOFAR observations\,\cite{heesen2023nearby}, which estimate the total galactic magnetic field. The radial magnetic field profile of the host galaxy at a galactocentric distance $R_g$ is parameterized by $B_{\rm host}(R_g) = B_{g,0} \, e^{-R_g/R_{gS}}$, where $B_{g,0}$ is the strength of the magnetic field at the center of the galaxy and $R_{gS}$ is the radial scale-length. The values of the fitted parameters are given in Table~4 of Ref.\,\cite{heesen2023nearby} for 21 galaxies. We choose the galaxies for which the definite values of the fit parameters are given and derive the magnetic field profiles. Since axion-photon conversion is majorly driven by the coherent regular field component, we adopt an effective regular field $B_{\rm reg} = f_{\rm reg}\,B_{g,0}$ where $f_{\rm reg} \in (0.1\text{--}0.5)$, consistent with observations of spiral  and starburst galaxies\,\cite{Beck:2015jta, Heesen:2011kj, 2017A&A...608A..29A}. For calculating the maximum conversion probability within the host galaxy, we take $f_{\rm reg} = 0.5$. The effect of the turbulent component on the axion–photon conversion probability is expected to be subdominant\,\cite{Tavecchio:2012um, Galanti:2022chk, Galanti:2023uam} and is not considered in our analysis. For each galaxy, the region of propagation is divided into domains of width $1 \, {\rm kpc}$ up to $20 \, {\rm kpc}$. In each domain, we assume a constant field strength equal to the domain-averaged value, and the axion–photon conversion probability is calculated using the prescription given in section~\ref{subsec:conv_rsg_bsg}. For these galaxies, we adopt a fiducial electron density of $n_e = 0.05 \, {\rm cm}^{-3}$ at $z=0$, consistent with values observed in nearby spiral galaxies,\cite{Beck:2019jyi, Eckner:2022rwf}. Among the galaxies considered, we find that the conversion probabilities are very similar for NGC~5055 and NGC~5194, with slightly higher values for the later. We will use the values for NGC~5194 as the maximal host galaxy conversion. For the minimum conversion scenario, we use a MW-like host galaxy with a magnetic field strength of $\mathcal{O}(1) \, {\rm \mu G}$, a coherent propagation length of $10 \, {\rm kpc}$, and $n_e = 0.01 \, {\rm cm}^{-3}$. As the minimal case, we consider a MW-like galaxy, the choice of $f_{\rm reg}=0.5$ for the maximal conversion scenario provides a broader uncertainty band. The range of values for $f_{\rm reg}$ discussed earlier leads to conversion probabilities that lie within this wider uncertainty band.

The above discussion applies to galaxies residing in the local Universe at $z = 0$. However, to compute the cumulative contribution from cosmic CCSNe, we need to integrate over redshift, and the galactic environment evolves accordingly. To account for this, we consider the magnetic-field coherence length to scale as $L(z) = L_0 / (1+z)$, where $L_0$ denotes the coherence length at $z = 0$. The redshift evolution of both the electron density and the magnetic field is implemented following the approach described in Refs.\,\cite{Schober:2016ebm, subPeVALPs2022}, using the explicit relations provided in Ref.\,\cite{Eckner:2022rwf}. In Fig.~\ref{sf:patog_host}, we plot the host galaxy axion-photon conversion probability ($P_{a\to\gamma}^{\rm Host}$) as a function of $m_a$ for galaxies at $z = 0.01$ (red) and $z = 2$ (blue), for $E_a = 100 \, {\rm MeV}$ and $g_{a \gamma \gamma} = 10^{-12} \, {\rm GeV}^{-1}$. The shaded band corresponds to the uncertainty associated with the range of magnetic field configurations considered for host galaxies.

%
%
%%%%%%%%%%%%%%%%%%%%%%%%%%%%%%%%%%%%%%%%%%%%%%%%%%%%%%%%%%%%%%%%%%%%%%%%%%%
\subsection{Intergalactic magnetic field}
\label{subsec:igm}

The magnetic fields in the large-scale structure, i.e., the intergalactic medium (IGM), are poorly constrained, as direct measurements are scarce and the allowed parameter space remains wide. An overview of IGM magnetic field generation mechanisms and observational bounds are provided in Refs.\,\cite{Durrer:2013pga, AlvesBatista:2021sln}. Following the commonly adopted benchmark\,\cite{DeAngelis:2007dqd, sanchez2009hints, dominguez2011axion, Galanti:2015rda, Kartavtsev:2016doq, Galanti:2018nvl, Buehler:2020qsn, Galanti:2023uam}, we assume an optimistic configuration with a field strength of $1 \, {\rm nG}$ and a coherence length of $1 \, {\rm Mpc}$, which represents the maximal conversion probability in the IGM. To encompass the uncertainty, we also consider a conservative scenario with the same coherence length but a smaller magnetic field strength of $B = 0.01 \, {\rm nG}$, for which the conversion probability is reduced by approximately four orders of magnitude. Although the lowest allowed magnetic field strength for a $1 \, {\rm Mpc}$ coherence length can be as small as $\mathcal{O}(10^{-7}) \, {\rm nG}$\,\cite{HESS:2023zwb, Blunier:2025ddu}, we do not adopt this extreme value when constructing the uncertainty band. For such weak fields, the conversion probability becomes negligible, and the overall signal would instead be dominated by contributions from the host galaxy and the MW. The calculation of the conversion probability within the IGM by using \texttt{gammaALPs} is straightforward: we specify the source redshift $z$, IGM field strength and coherence length, and $n_e = 10^{-7} \, \mathrm{cm^{-3}}$\,\cite{Peebles1993} for both scenarios. However, at large redshifts the coherence length becomes much smaller than the total propagation distance, which substantially increases the computational time. We therefore average the conversion probability over $10$ random realizations of the magnetic field domains. For completeness, we include the \texttt{dominguez} model in the propagation modules, which is based on Ref.\,\cite{dominguez2011extragalactic}. Although as mentioned earlier, absorption is negligible at the gamma-ray energies considered here. Fig.~\ref{sf:patog_igm} displays the resulting averaged IGM conversion probabilities ($P_{a\to\gamma}^{\rm IGM}$) for the minimum and maximum configurations at $E_a = 100 \, \mathrm{MeV}$ and $g_{a\gamma\gamma}=10^{-12} \, \mathrm{GeV}^{-1}$. The red (blue) shaded bands represent the uncertainties associated with the allowed range of IGM field strengths for sources located at $z=0.01$ ($z=5$).
%
%
%%%%%%%%%%%%%%%%%%%%%%%%%%%%%%%%%%%%%%%%%%%%%%%%%%%%%%%%%%%%%%%%%%%%%%%%%%%
\subsection{Milky Way magnetic field}
\label{subsec:milky_way}

Axions survived through previous regions of magnetic fields may convert into gamma rays within the MW. Using \texttt{gammaALPs}, we adopt the Galactic coordinates $(\ell,b)$, and compute the probability of converting axions into photons along the line of sight. We concentrate on the high-latitude region, i.e., $|b| > 20^\circ$, to isolate the isotropic emission. This approach is consistent with the \textit{Fermi}-LAT isotropic diffuse gamma-ray background (IGRB) analysis\,\cite{Fermi-LAT:2014ryh}, which excludes the bright emission from the Galactic plane. The region of interest (ROI) averaged conversion probability can be written as
\begin{equation}\label{eq:PMW_avg}
    P_{a\to\gamma}^{\rm MW} = \frac{1}{\Delta\Omega}\int_{|b|>20^\circ} d\Omega\; P_{a\to\gamma}^{\rm GMF}(\ell,b) \, ,
\end{equation}
where $\Delta\Omega = 4\pi [ 1 - \sin (\pi/9) ]$. The averaged conversion probability can vary if the ROI corresponding to telescopes other than \textit{Fermi}-LAT is considered. A comparison of the ROI-averaged conversion probability with COMPTEL and EGRET is presented in the appendix~\ref{app:roi_comparison}. In calculating $P_{a\to\gamma}^{\rm MW}$, the dominant astrophysical uncertainty in the axion-photon conversion probability comes from the coherent Galactic magnetic field (GMF). We model the GMF using the Unger-Farrar (UF23) suite of models\,\cite{Unger:2023lob}. UF23 provides eight viable coherent field models. For each model, we compute an ROI-averaged conversion probability. We define the uncertainty on $ P_{a\to\gamma}^{\rm MW}$ as the envelope spanned by the eight models. We note that systematic differences between GMF models dominate statistical uncertainties within a single model\,\cite{Unger:2023lob}. In Fig.~\ref{sf:patog_MW}, we plot $P_{a\to\gamma}^{\rm MW}$ as a function $m_a$ for $g_{a\gamma\gamma}=10^{-12} \, \rm GeV^{-1}$ and a fixed energy $E_a=100 \, \rm MeV$. We show the probabilities for the eight models and the gray shaded area corresponds to the total uncertainty. The upper and lower edges of this uncertainty band are driven by the \texttt{nebCor} (red solid) and \texttt{twistX} (blue solid) models, respectively.  The fiducial Jansson–Farrar 2012 (JF12) model\,\cite{Jansson:2012pc} lies within the uncertainty band. However, our analysis accounts for the wider uncertainty encompassed by the UF23 envelope. It is worth noting that radio observations indicate the presence of a turbulent magnetic field in the MW with a correlation length of $\mathcal{O}(10\text{--}100) \, {\rm pc}$\,\cite{MagnetizedISM:2003Antalya, Uyaniker:2004mim}, and including such effects can modify the conversion probability by up to a factor of $\sim 2$\,\cite{Carenza:2021alz}. However, in this analysis, we focus only on the coherent magnetic field component. 
%
%
%%%%%%%%%%%%%%%%%%%%%%%%%%%%%%%%%%%%%%%%%%%%%%%%%%%%%%%%%%%%%%%%%%%%%%%%%%%
\subsection{Total conversion}
\label{subsec:total_conv}

In Fig.~\ref{fig:patog_combined}, we present the conversion probabilities as a function of $m_a$ at a fixed energy $E_a$. For the energy range relevant to this work, the conversion probabilities for BSGs can vary by up to an order of magnitude, while across all other astrophysical environments the probabilities remain within an $\mathcal{O}(1)$ factor. Next, we evaluate the overall probability of an axion produced in a supernova at redshift $z$ converts into a photon before reaching us. Since the axion propagates sequentially through multiple magnetic field regions, the effective probability includes the possibility that conversion occurs in any of these regions, and can be written as
\begin{equation}\label{eq:eff_prob}
\begin{aligned}
    \mathcal{P}& _{a\to\gamma}^{\rm eff} (z) = P_{a\to\gamma}^{\rm Prog}(z) + \big( 1 - P_{a\to\gamma}^{\rm Prog}(z) \big) P_{a\to\gamma}^{\rm Host}(z) \\
    & + \big( 1 - P_{a\to\gamma}^{\rm Prog}(z) \big) \big( 1 - P_{a\to\gamma}^{\rm Host}(z) \big) P_{a\to\gamma}^{\rm IGM}(z) \\
    & + \left[ 1 - P_{a\to\gamma}^{\rm Prog}(z) \right] \big( 1 - P_{a\to\gamma}^{\rm Host}(z) \big)  \big( 1 - P_{a\to\gamma}^{\rm IGM}(z) \big)  P_{a\to\gamma}^{\rm MW} \\
    &= 1 - \left[ \big( 1 - P_{a\to\gamma}^{\rm Prog}(z) \big) \big( 1 - P_{a\to\gamma}^{\rm Host}(z) \big) \right. \\ 
    & \left. \hspace*{3.75cm} \times \big( 1 - P_{a\to\gamma}^{\rm IGM}(z) \big) \big( 1 - P_{a\to\gamma}^{\rm MW} \big) \right] \, .
\end{aligned}
\end{equation}
For the conversion within the MW, no redshift dependence appears since the conversion occurs locally. For the progenitor magnetic fields, we assume that the field configuration does not evolve with redshift. Therefore, only the redshifted energy dependence, i.e., $P_{a\to\gamma}(z) = P_{a\to\gamma}^{\rm Prog}(E_a(1+z))$ is implemented. The explicit dependence of $\mathcal{P}_{\rm eff}(z)$ on the source redshift is particularly important for the calculation of the diffuse gamma-ray flux, where the contribution from supernovae must be integrated over cosmic redshift. The gamma-ray photons created at each environment can reconvert into axions in the subsequent magnetic field regions. To estimate this, we track the photon and axion probabilities after each step and include the possibility of reconversion in subsequent regions. This gives us a more general formula than Eq.~\eqref{eq:eff_prob}, and is presented in appendix~\ref{app:prob_wt_reconv}. We find that the effective probability can change $\sim \mathcal{O}(1)\%$, and we neglect this reconversion possibility in further analysis.
%%%%%%%%%%%%%%%%%%%%%%%%%%%%%%%%%%%%%%%%%%%%%%%%%%%%%%%%%%%%%%%%%%%%%%%%%%%
%%%%%%%%%%%%%%%%%%%%%%%%%%%%%%%%%%%%%%%%%%%%%%%%%%%%%%%%%%%%%%%%%%%%%%%%%%%
%
%

%
%
%%%%%%%%%%%%%%%%%%%%%%%%%%%%%%%%%%%%%%%%%%%%%%%%%%%%%%%%%%%%%%%%%%%%%%%%%%%
%%%%%%%%%%%%%%%%%%%%%%%%%%%%%%%%%%%%%%%%%%%%%%%%%%%%%%%%%%%%%%%%%%%%%%%%%%%
\section{Diffuse gamma-ray flux}
\label{sec:diff_gamma_ray}

Having the estimation of axion production spectrum from a single supernova and the corresponding axion–photon conversion probability along the line of sight, we integrate over the cosmic population of CCSNe to obtain the diffuse gamma-ray flux. Following the prescription outlined in\,\cite{Raffelt:2011ft, Calore:2020tjw}, the axion-induced differential $\gamma$-ray flux is given by
\begin{equation}\label{eq:dphi_dE_gam}
\begin{aligned}
    \dfrac{d \Phi_\gamma}{d E_\gamma} = \dfrac{1}{4 \pi} \int_{z_{\rm min}}^{z_{\rm max}} dz \, (1 + z) \, R_{\rm CCSN} & (z) \, \dfrac{dN_a \left( E_a (1+z) \right)}{dE_a} \\
    & \times \mathcal{P}_{a\to\gamma}^{\rm eff}(z) \ c \left| \dfrac{dt}{dz} \right| \, ,
\end{aligned}
\end{equation}
where the integration limits are taken from $z_{\rm min} = 0.01$ to $z_{\rm max} = 6$, beyond which the contribution is not significant. Moreover, the axion spectrum from a single supernova is appropriately redshifted. Note that in calculating the diffuse $\gamma$-ray spectrum, we assume that the axion spectrum from a single supernova follows one of the three benchmark models considered in our work. In principle, the spectrum may vary with the progenitor mass. Since supernova profiles across a wide range of progenitor masses are not readily available, we do not include an explicit mass dependence in the axion spectrum. To provide an estimate of the associated uncertainty, we present our results for each of the supernova models considered here. The term $\left| dt/dz \right|$ is
\begin{equation}\label{eq:dt_dz}
    \left| \dfrac{dt}{dz} \right| = \dfrac{1}{H_0 \, (1 + z) \, \sqrt{\Omega_M \, (1 + z)^3 \, + \, \Omega_\Lambda}} \, ,
\end{equation}
assuming a flat $\Lambda$CDM cosmology with parameters $\Omega_M = 0.315$, $\Omega_\Lambda = 0.685$, and $H_0 = 67.4 \, {\rm km \, s^{-1} Mpc^{-1}}$\,\cite{Planck:2018vyg}. We neglect the contribution from radiation as CCSNe occur relatively late in cosmic history, when the radiation energy density is negligible compared to matter and dark energy components.
%
%
%%%%%%%%%%%%%%%%%%%%%%%%% FIGURE: SFRD fit %%%%%%%%%%%%%%%%%%%%%%%%%%%%%%%%
\sfrdFit
%%%%%%%%%%%%%%%%%%%%%%%%%%%%%%%%%%%%%%%%%%%%%%%%%%%%%%%%%%%%%%%%%%%%%%%%%%%
%
%

In Eq.~\eqref{eq:dphi_dE_gam}, $R_{\rm CCSN}(z)$ denotes the CCSN explosion rate at redshift $z$, which is inferred from recent observational measurements of the cosmic SFRD $(\dot{\rho}_{_{\rm SFRD}})$. Since massive progenitors are short-lived, $R_{\rm CCSN}(z)$ closely traces the SFRD\,\cite{Kennicutt:1998zb, Hopkins:2006bw}. We model the redshift evolution of SFRD using the widely adopted Madau \& Dickinson functional form\,\cite{Madau:2014bja},
\begin{equation}\label{eq:sfrd_fit}
    \dot{\rho}_{_{\rm SFRD}}(z) = A \, \dfrac{(1 + z)^B}{1 + \left[ (1 + z) / C \right]^D} \, ,
\end{equation}
where $A, B, C,$ and $D$ are parameters obtained from fitting the observational data. To accurately capture the SFRD over a broad range of redshifts, we determine separate best-fit parameters for two distinct redshift regimes. At low redshifts, we fit Eq.~\eqref{eq:sfrd_fit} to the binned SFRD dataset of Ref.\,\cite{Ekanger:2023qzw}, using the inverse-variance weighted averages. For higher redshifts, we adopt the combined IR and UV measurements shown in Fig.~18 of Ref.\,\cite{Fujimoto:2023vfa}, where the UV data are taken from Ref.\,\cite{Bouwens:2016xxx}. Both these datasets are fitted using Eq.~\eqref{eq:sfrd_fit}, and the corresponding $1\sigma$ uncertainty band is obtained from the fit covariance. The two fits overlap around $z=1$; therefore, we use the fit to the data from Ref.\,\cite{Ekanger:2023qzw} for $z \leq 1$, and the fit obtained using the data from Ref.\,\cite{Fujimoto:2023vfa} for $z > 1$. In Fig.~\ref{fig:sfrd_fit}, we show the combined fit to $\dot{\rho}_{_{\rm SFRD}}$ along with the corresponding uncertainty band and the observational datasets. Since measurements at high redshift involve larger observational uncertainties, the resulting uncertainty band in this regime is significantly broader compared to the low redshift regime, where the measurements are more precise and the associated uncertainties are comparatively small. To obtain the CCSN rate, we convolve the SFRD with the stellar IMF, which can be written as\,\cite{Ekanger:2023qzw}
\begin{equation}\label{eq:R_CCSN}
    R_{\rm CCSN}(z) = \dot{\rho}_{_{\rm SFRD}}(z) \, \dfrac{\int_{8 \, M_\odot}^{40 \, M_\odot} \psi(M) \, dM}{\int_{0.1 \, M_\odot}^{100 \, M_\odot} M \, \psi(M) \, dM} \, ,
\end{equation}
where the functional form of $\psi(M)$ is given in Eq.~\eqref{eq:stellar_IMF}. To estimate the CCSN fraction, we perform the integration up to $40 \, M_\odot$ that corresponds to the upper mass limit considered earlier in the calculation of the progenitor fractions. In Fig.~\ref{fig:gamma_ray_flux}, we show the CCSN axion-induced diffuse $\gamma$-ray flux as a blue shaded region, obtained using Eq.~\eqref{eq:dphi_dE_gam} for the KSVZ model with $m_a = 10^{-12} \, {\rm eV}$ and $g_{a \gamma \gamma} = 10^{-12} \, {\rm GeV}^{-1}$, adopting the SFHo-18.6 progenitor profile. The upper (lower) edge of the band corresponds to simultaneously taking the maximal (minimal) values of both the axion–photon conversion probability and the CCSN rate, yielding optimistic (conservative) upper limits. We note that the predicted flux can vary moderately while using the SFHo-18.8 and SFHo-20.0 profiles. However, for clarity, the diffuse $\gamma$-ray flux is shown for the SFHo-18.6 case, while the corresponding limits are derived for all three profiles.
%
%
%%%%%%%%%%%%%%%%%%%%%%%%%%%%%%%%%%%%%%%%%%%%%%%%%%%%%%%%%%%%%%%%%%%%%%%%%%%
%%%%%%%%%%%%%%%%%%%%%%%%%%%%%%%%%%%%%%%%%%%%%%%%%%%%%%%%%%%%%%%%%%%%%%%%%%%
%
%
%%%%%%%%%%%%%%%%%%%%%%%%% FIGURE: Gamma ray flux %%%%%%%%%%%%%%%%%%%%%%%%%%
\gammaRayFlux
%%%%%%%%%%%%%%%%%%%%%%%%%%%%%%%%%%%%%%%%%%%%%%%%%%%%%%%%%%%%%%%%%%%%%%%%%%%
%
%
%%%%%%%%%%%%%%%%%%%%%%%%%%%%%%%%%%%%%%%%%%%%%%%%%%%%%%%%%%%%%%%%%%%%%%%%%%%
%%%%%%%%%%%%%%%%%%%%%%%%%%%%%%%%%%%%%%%%%%%%%%%%%%%%%%%%%%%%%%%%%%%%%%%%%%%
\section{Analysis and Results}
\label{sec:results}

After outlining the framework for calculating the CCSNe axion-induced diffuse gamma-ray flux in the previous sections, we now discuss the observational measurements of diffuse gamma-ray in the MeV range from different telescopes. We use the COMPTEL cosmic diffuse $\gamma$-ray background data in the interval $0.8 - 30~\mathrm{MeV}$ (see Table~VII.2 of Ref.\,\cite{1998PhDT.........3K}) which provides the observational data of high latitudes $(|b|>30^\circ)$\,\cite{1998PhDT.........3K, 2000AIPC..510..467W}. The EGRET has reported the extragalactic diffuse $\gamma$-ray background measurement in the interval $30 - 1000~\mathrm{MeV}$(see Table~3 of Ref.\,\cite{Strong:2004ry}) from observations of $10^\circ <|b|<80^\circ$. We also use the \textit{Fermi}-LAT IGRB data (Foreground Model~A, Table~3 of Ref.\,\cite{Fermi-LAT:2014ryh}) in the range $100 - 1100~\mathrm{MeV}$, obtained from observations at high Galactic latitudes $(|b| > 20^\circ)$.

We select datasets based on how each experiment defines the diffuse component relevant for our axion-induced signal. For COMPTEL, we use the standard MeV-band cosmic diffuse gamma-ray (CDG) observations\,\cite{1998PhDT.........3K}. For EGRET, we use the latest estimate of the extragalactic diffuse background presented in Ref.\,\cite{Strong:2004ry}, which supersedes Ref.\,\cite{EGRET:1997qcq}. The Galactic diffuse emission is modeled with \texttt{GALPROP}\,\cite{Strong:1998fr, Strong:2000fr} and the extragalactic component is extracted from the measured diffuse $\gamma$-ray data. For \textit{Fermi}-LAT, we use the IGRB rather than the total extragalactic gamma-ray background (EGB)\,\cite{Fermi-LAT:2014ryh}. The EGB includes the full extragalactic $\gamma$-ray emission (both resolved and unresolved), but the IGRB is the residual isotropic component. As the axion-induced signal considered in our analysis is a diffuse component, the IGRB is the more appropriate choice. We model the background in each dataset with a power-law fit of the form
\begin{equation}
\label{eq:bg_powerlaw}
    \frac{d\Phi_{\gamma}^{\rm bg}}{dE_{\gamma}} = \mathcal{C} \, \left( \dfrac{E_{\gamma}}{\rm 1\,MeV} \right)^{-\gamma_{\rm bg}} \, ,
\end{equation}
where $\mathcal{C}$ and $\gamma_{\rm bg}$ are the normalization and spectral index of the fit, respectively. The fit is performed directly at the level of the experimental observables reported by each instrument. For COMPTEL and EGRET, the data are treated as bin-averaged differential fluxes, and the corresponding energy-dependent part of the flux (upto normalization) is
\begin{equation}\label{eq:Fi_diff}
    F_{\text{diff},i} (\gamma_{\rm bg}) = \frac{1}{\Delta E_{\gamma,i}}\int_{E_{\gamma,i}^-}^{E_{\gamma,i}^+} dE_{\gamma}\,E_{\gamma}^{-\gamma_{\rm bg}} \, .
\end{equation}
For \textit{Fermi}-LAT, the IGRB values are quoted as bin-integrated intensities, for which we define the energy-dependent part of the integrated flux (upto normalization) as
\begin{equation}\label{eq:Fi_int}
    F_{\text{int},i} (\gamma_{\rm bg}) = \int_{E_{\gamma,i}^-}^{E_{\gamma,i}^+} dE_{\gamma}\,E_{\gamma}^{-\gamma_{\rm bg}} \, .
\end{equation}
%
%
%%%%%%%%%%%%%%%%%%%%%%% TABLE: Background Fits %%%%%%%%%%%%%%%%%%%%%%%%%%%%
\begin{table}[t]
    \centering
    \begin{tabular}{l|c|c|c|c}
        \hline
        \hline
        Telescope & $\mathcal{C} \, \left[ \rm MeV^{-1} \, cm^{-2} \, s^{-1} \, sr^{-1} \right]$ & $\gamma_{\rm bg}$ & $\chi^2/{\rm dof}$ & $\chi^2_{\rm red}$ \\
        \hline
        COMPTEL & $4.721\times10^{-3}$ & $2.367$ & $3.038/7$ & $0.434$ \\
        EGRET   & $5.824\times10^{-3}$ & $2.309$ & $3.797/5$ & $0.759$ \\ 
        \textit{Fermi}-LAT & $4.147\times10^{-3}$ & $2.320$ & $13.5/22$ & $0.613$ \\
        \hline
        \hline
    \end{tabular}
    \caption{\justifying Best-fit parameters of the power-law background model for the COMPTEL, EGRET, and \textit{Fermi}-LAT datasets.}
    \label{tab:bgfits}
\end{table}
%%%%%%%%%%%%%%%%%%%%%%%%%%%%%%%%%%%%%%%%%%%%%%%%%%%%%%%%%%%%%%%%%%%%%%%%%%%
%
%

The fits for COMPTEL and EGRET are performed using the quoted bin values and their associated uncertainties. For each energy bin $i$, we denote the reported diffuse-background measurement and associated uncertainty by $y_i$ and $\sigma_i$, respectively. The corresponding $\chi^2$ is
\begin{equation}\label{eq:chi2_bg}
    \chi^2_{\rm diff, bg}(\mathcal{C},\gamma_{\rm bg}) = \sum_i \frac{\left[y_{\text{diff,}i} - \mathcal{C}\,F_{\text{diff,}i} (\gamma_{\rm bg})\right]^2}{\sigma_{\text{diff,}i}^2} \, .
\end{equation}
For a given value of $\gamma_{\rm bg}$, we first compute the corresponding best-fit normalization, $\widehat{\mathcal{C}}(\gamma_{\rm bg})$, by imposing $d\chi_{\rm diff, bg}^2/d\mathcal{C}=0$, which can be expressed as
\begin{equation}\label{eq:C_hat}
    \widehat{\mathcal{C}}(\gamma_{\rm bg}) = \frac{\sum_i \sigma_{\text{diff,}i}^{-2} \, F_{\text{diff,}i} (\gamma_{\rm bg}) \, y_{\text{diff,}i}} {\sum_i \sigma_{\text{diff,}i}^{-2} \, F_{\text{diff,}i} ^2(\gamma_{\rm bg})} \, .
\end{equation}
Subsequently, we minimize the resulting $\chi^2_{\rm diff, bg}$ with respect to $\gamma_{\rm bg}$ to obtain the final best-fit parameters.

For \textit{Fermi}-LAT, we adopt the best-fit parameters of Eq.~\eqref{eq:bg_powerlaw} directly from Ref.\,\cite{Fermi-LAT:2014ryh} (Table~4, FG Model~A), which includes statistical and systematic uncertainties. Since we consider that the axion-induced diffuse gamma-ray signal extends only to $1 \, {\rm GeV}$, well below the exponential cutoff $\sim 279 \, {\rm GeV}$ for FG Model A defined in Ref.\,\cite{Fermi-LAT:2014ryh}, the pure power-law approximation is valid in our signal region. The obtained fit values for COMPTEL and EGRET, along with the adopted parameters for \textit{Fermi}-LAT, are summarised in Table~\ref{tab:bgfits}. In Fig.~\ref{fig:gamma_ray_flux}, along with the axion-induced diffuse gamma-ray flux obtained using Eq.~\eqref{eq:dphi_dE_gam}, we also show the measured diffuse gamma-ray flux from COMPTEL, EGRET, and \textit{Fermi}-LAT, represented by purple, red, and teal markers, respectively. The corresponding best-fit background power-law spectra, with parameters listed in Table~\ref{tab:bgfits}, are overlaid as solid lines in the same respective colors for each telescope.
%
%
%%%%%%%%%%%%%%%%%%%%%%%%% FIGURE: chi^2 limits %%%%%%%%%%%%%%%%%%%%%%%%%%%%
\chisqLimits
%%%%%%%%%%%%%%%%%%%%%%%%%%%%%%%%%%%%%%%%%%%%%%%%%%%%%%%%%%%%%%%%%%%%%%%%%%%
%
%
%%%%%%%%%%%%%%%%%%%%%%%%%%%%%%%%%%%%%%%%%%%%%%%%%%%%%%%%%%%%%%%%%%%%%%%%%%%
\subsection{$\chi^2$ analysis with current data}
\label{subsec:chi2_analysis}

We derive exclusion limits on the axion–photon coupling by performing a $\chi^2$ analysis using the measured binned $\gamma$-ray data, the predicted diffuse axion-induced $\gamma$-ray flux, and the fitted background model. We obtain the signal contribution to energy bin $i$, denoted as $\mathcal{S}_i(m_a, g_{a\gamma\gamma})$, by integrating the differential flux $d\Phi_\gamma/dE_\gamma$ over the energy range $[E_i^-, E_i^+]$. The signal definition depends on the data representation adopted by each instrument. For COMPTEL and EGRET, which provide bin-averaged differential measurements, the signal is expressed as
\begin{equation}\label{eq:signal_diff}
    \mathcal{S}_{{\rm diff},i} (m_a,g_{a\gamma\gamma}) = \frac{1}{\Delta E_{\gamma,i}} \int_{E_{\gamma,i}^-}^{E_{\gamma,i}^+} \frac{d\Phi_\gamma}{dE_\gamma}(m_a,g_{a\gamma\gamma}) \, dE_\gamma \, .
\end{equation}
For \textit{Fermi}-LAT, which reports energy bin-integrated measurements, we use
\begin{equation}\label{eq:signal_intg}
    \mathcal{S}_{{\rm int}, i} (m_a,g_{a\gamma\gamma}) = \int_{E_{\gamma,i}^-}^{E_{\gamma,i}^+} \frac{d\Phi_\gamma}{dE_\gamma}(m_a,g_{a\gamma\gamma}) \, dE_\gamma \, .
\end{equation}
The background contribution in bin $i$ is constructed from the best-fit power-law model of Eq.~\eqref{eq:bg_powerlaw} with the fitted parameters as mentioned in Table~\ref{tab:bgfits}. For COMPTEL and EGRET, the $\chi^2$ is defined as
\begin{equation}\label{eq:chi2_diff}
\begin{aligned}
    & \chi_{\rm diff}^2  (m_a,g_{a\gamma\gamma}) = \\
    &\sum_i \left[y_{{\rm diff}, i} -\mathcal{S}_{{\rm diff}, i} (m_a,g_{a\gamma\gamma}) -\widehat{\mathcal{C}}\,F_{{\rm diff}, i} (\widehat{\gamma}_{\rm bg})\right]^2 \bigg/ \sigma_{{\rm diff}, i}^2\, ,
\end{aligned}
\end{equation}
while for \textit{Fermi}-LAT, it takes the form
\begin{equation}\label{eq:chi2_intg}
\begin{aligned}
    & \chi_{\rm int}^2 (m_a,g_{a\gamma\gamma}) = \\
    & \sum_i \left[y_{{\rm int}, i} -\mathcal{S}_{{\rm int}, i} (m_a,g_{a\gamma\gamma}) -\widehat{\mathcal{C}}\,F_{{\rm int}, i} (\widehat{\gamma}_{\rm bg})\right]^2 \bigg/ \sigma_{{\rm int}, i}^2 \, .
\end{aligned}
\end{equation}
In this analysis, we keep the background model fixed while deriving the upper limits on the axion-photon coupling. In principle, the background parameters can also be varied in the $\chi^2$ analysis; however, this is not included here and is instead considered in the forecast analysis. At fixed $m_a$, we define
\begin{equation}
    \Delta\chi^2(m_a,g_{a\gamma\gamma}) = \chi^2(m_a,g_{a\gamma\gamma})-\chi^2(m_a,0) \, ,
\end{equation}
and extract the upper limit with $95\%$ confidence level (CL) in $g_{a\gamma\gamma}$ considering $\Delta\chi^2=2.71$\,\cite{Wilks:1938dza}. The limits are evaluated separately for COMPTEL, EGRET, and \textit{Fermi}-LAT based on their respective definition of $\chi^2$.

The resulting exclusion limits with $95\%$\,CL for the SFHo-18.6 CCSN model are shown in Fig.~\ref{fig:chisq_limits} for the KSVZ (left panel) and ALP (right panel) models. In each panel, the purple, red, and teal bands show the exclusion limits from COMPTEL, EGRET, and \textit{Fermi}-LAT, respectively. The width of the bands arises from uncertainties in the CCSN rate and conversion probabilities in different astrophysical environments. The shape of exclusion limits is dictated by the mass dependence of the conversion probabilities. The low mass regime ($\lesssim 10^{-8}$~eV) is dominated by host galaxy, extragalactic and MW conversion, and the progenitor contribution becomes increasingly important at higher masses ($\gtrsim 10^{-8}$~eV). We note that across the mass range, EGRET provides the strongest limits among the three datasets for both the KSVZ and ALP models. This can be attributed to the smaller uncertainties of EGRET in the diffuse measurements compared to other telescopes. We consider the \textit{Fermi}-LAT results to be more robust and conservative. For the KSVZ model (optimistic case), the \textit{Fermi}-LAT exclusion limit exhibits three flat plateaus separated by two rising transitions. The first plateau lies at $g_{a\gamma\gamma} \simeq (1.86 - 2.07) \times 10^{-13} \, \mathrm{GeV}^{-1}$ for $m_a \sim (10^{-12} - 10^{-10}) \, \mathrm{eV}$, followed by a steep rise into a second, shallower plateau at $(4.35 - 4.64)\times 10^{-13} \, \mathrm{GeV}^{-1}$ across $m_a \sim (10^{-10} - 10^{-9}) \, \mathrm{eV}$. A second rise then leads to the third plateau at $\simeq 3.76 \times 10^{-12} \, \mathrm{GeV}^{-1}$ spanning $m_a \sim (10^{-7} - 10^{-5}) \, \mathrm{eV}$, beyond which the limit weakens to $2.85 \times 10^{-12} \, \mathrm{GeV}^{-1}$ toward $m_a \sim 10^{-3}\,\mathrm{eV}$. For the ALP model (optimistic case), two plateaus are present. The first spans $m_a \sim (10^{-12} - 10^{-9}) \, \mathrm{eV}$ at $(2.06 - 2.49)\times 10^{-11} \, \mathrm{GeV}^{-1}$, with values nearly constant below $m_a \sim 10^{-10} \, \mathrm{eV}$ and rising slowly thereafter. Further rising intermediate region across $m_a \sim (10^{-9} - 10^{-7}) \, \mathrm{eV}$ leads to a second, extremely flat plateau at $\simeq 1.33\times 10^{-10} \, \mathrm{GeV}^{-1}$ 
from $m_a \sim (10^{-7} - 10^{-5}) \, \mathrm{eV}$, beyond which the limit rises to $\sim 1.10\times 10^{-9} \, \mathrm{GeV}^{-1}$ at $m_a \sim 10^{-3} \, \mathrm{eV}$. The limits obtained for the KSVZ model are stronger than those for the ALP, as in the former the axion–nucleon coupling is present, whereas in the latter it arises through loop corrections. This leads to orders of magnitude variation in the axion production flux, as shown in Fig.~\ref{fig:alp_flux_supernova}. Our result remains complementary to existing astrophysical\,\cite{Wouters:2013hua, HESS:2013udx, Marsh:2017yvc, Kohri:2017ljt, Reynolds:2019uqt, Buen-Abad:2020zbd, Calore:2020tjw, Dessert:2020lil, Li:2020pcn, Meyer:2020vzy, Xiao:2020pra, Calore:2021hhn, Chan:2021gjl, Dessert:2021bkv, Li:2021gxs, Keller:2021zbl, Reynes:2021bpe, Dessert:2022yqq, Dolan:2022kul, Hoof:2022xbe, Jacobsen:2022swa, Foster:2022fxn, Noordhuis:2022ljw, Escudero:2023vgv, Battye:2023oac, Li:2024zst, Cyr:2024sbd, Manzari:2024jns, Ning:2024eky, Ruz:2024gkl, MAGIC:2024arq, Benabou:2025jcv}, laboratory searches including haloscope experiments\,\cite{DePanfilis:1987dk, Wuensch:1989sa, Hagmann:1990tj, Hagmann:1996qd, 2010PhRvL.104d1301A, Brubaker:2016ktl, McAllister:2017lkb, Ouellet:2018beu, HAYSTAC:2018rwy, ADMX:2018gho, ADMX:2018ogs, ADMX:2019uok, Alesini:2019ajt, Lee:2020cfj, CAPP:2020utb, Alesini:2020vny, HAYSTAC:2020kwv, CAST:2020rlf, Gramolin:2020ict, Jeong:2020cwz, Devlin:2021fpq, Salemi:2021gck, Grenet:2021vbb, Thomson:2021zvq, ADMX:2021nhd, Adair:2022rtw, Kim:2022hmg, Yoon:2022gzp, Lee:2022mnc, Alesini:2022lnp, Quiskamp:2022pks, TASEH:2022vvu, Kim:2023vpo, Yang:2023yry, Thomson:2023moc, Quiskamp:2023ehr, QUAX:2023gop, HAYSTAC:2023cam, Heinze:2023nfb, Bae:2024kmy, Ahyoune:2024klt, QUAX:2024fut, Quiskamp:2024oet, HAYSTAC:2024jch, MADMAX:2024sxs, ADMX:2024xbv, Pandey:2024dcd, GigaBREAD:2025lzq, ADMX:2025vom} and helioscope experiments\,\cite{CAST:2007jps, Ehret:2010mh, Betz:2013dza, OSQAR:2015qdv, CAST:2017uph, CAST:2024eil}. However, haloscope and some of the astrophysical constraints  are obtained assuming axions to be dark matter, while our limits are more general. The limits for other CCSN profiles in particular SFHo-18.8 and SFHo-20.0 are outlined in appendix~\ref{app:other_chi2_results}.
%
%
%%%%%%%%%%%%%%%%%%%%%%%%% FIGURE: chi^2 limits %%%%%%%%%%%%%%%%%%%%%%%%%%%%
\fisherLimits
%%%%%%%%%%%%%%%%%%%%%%%%%%%%%%%%%%%%%%%%%%%%%%%%%%%%%%%%%%%%%%%%%%%%%%%%%%%
%
%
%%%%%%%%%%%%%%%%%%%%%%%%%%%%%%%%%%%%%%%%%%%%%%%%%%%%%%%%%%%%%%%%%%%%%%%%%%%
\subsection{Fisher forecast}
\label{subsec:fisher_forecast}

\textit{Fermi}-LAT constraints derived above are the most robust among all current instruments. Currently, \textit{Fermi}-LAT is dominated by systematic uncertainties associated with Galactic FG modeling. Future MeV telescopes with wider energy coverage and larger effective areas have an optimistic outlook. These telescopes will resolve numerous point sources, consequently reducing the residual isotropic background. Additionally, the cosmic-ray propagation model can also decrease the FG systematics. We estimate the projected sensitivities to $g_{a\gamma\gamma}$ for different future telescopes using a Fisher forecast matrix formalism, including correlated background systematics\,\cite{Edwards:2017mnf}. The Fisher information matrix is given by\,\cite{Tegmark:1996bz, ODonnell:2024aaw}
\begin{equation}\label{eq:F_ij}
    \mathcal{F}_{ij} = \sum_{m,n} \frac{\partial \lambda_m}{\partial \theta_i} (\mathcal{D}^{-1})_{mn} \frac{\partial \lambda_n}{\partial \theta_j} \, ,
\end{equation}
where $\boldsymbol{\theta} = (\eta,\, \mathcal{C},\, \gamma_{\rm bg})$, with $\eta \equiv \ln g_{a\gamma\gamma}$, and $\mathcal{D}$ is the total covariance matrix. For a fixed mass $m_a$, $\lambda_k$ and $\lambda'_k$ denote the expected total (signal + background) and background-only photon counts in the $k^{\rm th}$ energy bin:
\begin{equation}\label{eq:lambda_k}
    \begin{split}
        \lambda_k = \frac{d\Phi_\gamma}{dE_\gamma}(E_k;m_a,\eta) \, A_{\rm eff}(E_k)\, T_{\rm obs}\, \Delta\Omega\, \Delta E_k+ \lambda_k' \, ,
    \end{split}
\end{equation}
with
\begin{equation}\label{eq:lambda_k_bg}
    \begin{split}
        \lambda'_k = \frac{d\Phi_\gamma^{\rm bg}}{dE_\gamma}(E_k;m_a,\mathcal{C}, \gamma_{\rm bg}) \, A_{\rm eff}(E_k) \, T_{\rm obs} \, \Delta\Omega \, \Delta E_k \, .
    \end{split}
\end{equation}
Here, $d\Phi_\gamma/dE_\gamma$ and $d\Phi_\gamma^{\rm bg}/dE_\gamma$ are axion-induced and background diffuse gamma-ray fluxes defined through Eqs.\,\eqref{eq:dphi_dE_gam} and \eqref{eq:bg_powerlaw}, respectively. To robustly model the background, following the approach introduced in \cite{Edwards:2017mnf}, the total covariance matrix is constructed as $\mathcal{D}_{mn} = \lambda_m \delta_{mn} + \lambda'_m \lambda'_n \Sigma^{\rm sys}_{mn}$. The first term represents the standard Poisson statistical variance and the second term encapsulates correlated systematic uncertainties in the background shape. By modeling these fractional shape uncertainties as a continuous Gaussian random field, we define the dimensionless systematic covariance matrix as
\begin{equation}\label{eq:sys_cov}
    \Sigma^{\rm sys}_{mn} = \sigma_{\rm sys}^2 \exp\left[ -\frac{1}{2} \left( \frac{\log_{10}(E_m/E_n)}{\ell_c} \right)^2 \right] \, ,
\end{equation}
where $\sigma_{\rm sys}$ represents the fractional amplitude of the background uncertainty and $\ell_c$ is the correlation length across energy bins (in decades). Here, $E_m$ and $E_n$ denote the central energies of bins $m$ and $n$ respectively, 
so that $\log_{10}(E_m/E_n)$ measures their separation in decades of energy. For our sensitivity projections, we adopt a benchmark of $10\%$ background systematic uncertainty ($\sigma_{\rm sys} = 0.1$) with a correlation length of $\ell_c = 0.1$. These values are chosen to reflect realistic systematic uncertainties in the MeV diffuse gamma-ray background.

The terms $A_{\rm eff}$ and $\Delta E_k$ denote the effective area of the telescope and width of the $k^{\rm th}$ energy bin, respectively. The energy binning for each instrument is generated adaptively using $E_{k+1} = E_k (1 + \frac{\Delta E}{E}|_{_{E_k}}),$ where $\Delta E / E$ is the instrument's energy resolution evaluated at $k^{\rm th}$ energy bin. This makes sure that each bin is approximately one energy-resolution element wide. For each instrument, we assume the benchmark observation time of $T_{\rm obs} = 10^7 \, {\rm s}$ (about 3.9 months). The observed solid angle $\Delta\Omega$ is taken to be $4\pi\,[1 - {\rm sin}(\pi/9)]$, corresponding to high-latitude sky observation with the Galactic plane region $|b|<20^\circ$ excluded, similar to the \textit{Fermi}-LAT IGRB measurements.

The fiducial background parameters $(\mathcal{C}^*,\,\gamma_{\rm bg}^*)$ entering the Fisher matrix are obtained from power-law fits to the COMPTEL 
($0.8 - 30$~MeV), EGRET ($30 - 100$~MeV), and \textit{Fermi}-LAT ($100 - 1000$~MeV) diffuse $\gamma$-ray data, as described in section~\ref{sec:results} and summarized in Table~\ref{tab:bgfits}. For each instrument, we define the geometric mean energy $E_{\rm mid} = \sqrt{E_{\rm min} E_{\rm max}}$ of its operating range and select the fiducial parameters based on the observational dataset (COMPTEL/EGRET/\textit{Fermi}-LAT) that covers this energy. For example, MAST operating in the $10.6 – 1000 \, {\rm MeV}$ range ($E_{\rm mid} \approx 102.9$ MeV) is assigned the \textit{Fermi}-LAT fiducial parameters. Note that, MAST and some other telescopes can operate up to much higher energies, but we consider energies only up to $1000 \, {\rm MeV}$, which is our energy range of interest.

We treat the background parameters $(\mathcal{C},\,\gamma_{\rm bg})$ as nuisance parameters and marginalise over them via the Schur complement\,\cite{schur_complement_wikipedia, zhang2005schur}, obtaining the profiled Fisher information $\mathcal{F}_{\rm prof}(\eta)$ that depends only on $\eta$ (see appendix~\ref{app:schur_comp}). No fiducial signal value is assumed \emph{a priori}; instead, the fiducial point $\eta^*\equiv\ln g^*_{a\gamma\gamma}$ is determined self-consistently as the solution to $\mathcal{F}_{\rm prof}(\eta^*)=2.71$, which corresponds to the one-sided $95\%$\,CL upper limit on $\eta$. In this treatment, the $m_a$ is fixed and later scanned over the considered axion mass range. Additionally, to illustrate the degeneracy between the signal and background parameters, we show the Fisher ellipses in appendix~\ref{app:fisher_ellipse}.

We perform this analysis with instrument-specific $A_{\rm eff}$ and operating energy ranges for various future telescopes: AMEGO-X\,\cite{2021SPIE11444E..31K}, e-ASTROGAM\,\cite{e-ASTROGAM:2017pxr}, MAST\,\cite{Dzhatdoev:2019kay}, GRAMS-satellite\,\cite{GRAMS:2021tax,Aramaki:2019bpi}, AdEPT\,\cite{Hunter:2013wla}, PANGU\,\cite{Wu:2014tya}, APT\,\cite{APT:2021lhj} and VLAST\,\cite{Pan:2024adp}. In Fig.~\ref{fig:fisher_limits}, we show the projected sensitivity on $g_{a\gamma\gamma}$ with $95\%$\,CL for the SFHo-18.6 CCSN model. The sensitivities obtained with other CCSN models are shown in appendix~\ref{app:other_fisher_results}. In the left (right) panel of Fig.~\ref{fig:fisher_limits}, we show the case for KSVZ (ALP) model. The previously existing limits are represented with a grey shaded region\,\cite{Wouters:2013hua, HESS:2013udx, Marsh:2017yvc, Kohri:2017ljt, Reynolds:2019uqt, Buen-Abad:2020zbd, Calore:2020tjw, Dessert:2020lil, Li:2020pcn, Meyer:2020vzy, Xiao:2020pra, Calore:2021hhn, Chan:2021gjl, Dessert:2021bkv, Li:2021gxs, Keller:2021zbl,Reynes:2021bpe, Dessert:2022yqq, Dolan:2022kul, Hoof:2022xbe, Jacobsen:2022swa, Foster:2022fxn,Noordhuis:2022ljw, Escudero:2023vgv,Battye:2023oac, Li:2024zst, Cyr:2024sbd, Manzari:2024jns, Ning:2024eky, Ruz:2024gkl, MAGIC:2024arq, Benabou:2025jcv, DePanfilis:1987dk, Wuensch:1989sa, Hagmann:1990tj, Hagmann:1996qd, 2010PhRvL.104d1301A, Brubaker:2016ktl, McAllister:2017lkb, Ouellet:2018beu, HAYSTAC:2018rwy, ADMX:2018gho, ADMX:2018ogs, ADMX:2019uok, Alesini:2019ajt, Lee:2020cfj, CAPP:2020utb, Alesini:2020vny, HAYSTAC:2020kwv, CAST:2020rlf, Gramolin:2020ict, Jeong:2020cwz, Devlin:2021fpq, Salemi:2021gck, Grenet:2021vbb, Thomson:2021zvq, ADMX:2021nhd, Adair:2022rtw, Kim:2022hmg, Yoon:2022gzp, Lee:2022mnc, Alesini:2022lnp, Quiskamp:2022pks, TASEH:2022vvu, Kim:2023vpo, Yang:2023yry, Thomson:2023moc, Quiskamp:2023ehr, QUAX:2023gop, HAYSTAC:2023cam, Heinze:2023nfb, Bae:2024kmy, Ahyoune:2024klt, QUAX:2024fut, Quiskamp:2024oet, HAYSTAC:2024jch, MADMAX:2024sxs, ADMX:2024xbv, Pandey:2024dcd, GigaBREAD:2025lzq, ADMX:2025vom,CAST:2007jps,Ehret:2010mh,Betz:2013dza, OSQAR:2015qdv, CAST:2017uph,CAST:2024eil}. The colored bands show the projected sensitivities of the future MeV gamma-ray 
telescopes: AMEGO-X (blue), e-ASTROGAM (yellow), APT (green), GRAMS Satellite (orange), MAST (dark red), AdEPT (light blue),  PANGU (pink), and VLAST (dark blue). With the APT telescope (optimistic case with KSVZ model), we can probe couplings as low as $g_{a\gamma\gamma} \sim (6.9 \times 10^{-14} - 1.5 \times 10^{-11}) \, \mathrm{GeV}^{-1}$ over the considered mass range. After accounting for systematic uncertainties, most instruments provide limits within the same ballpark, reflecting their broadly comparable energy coverage, effective areas, and more importantly similar treatment of systematics. As discussed earlier, the width of the bands reflects the uncertainty associated with the astrophysical magnetic fields modeling and SFRD fits. This shows the potential of near-future MeV telescopes to probe the unconstrained parameter space across a wide mass range or to potentially discover axions. 
%
%
%%%%%%%%%%%%%%%%%%%%%%%%%%%%%%%%%%%%%%%%%%%%%%%%%%%%%%%%%%%%%%%%%%%%%%%%%%%
%%%%%%%%%%%%%%%%%%%%%%%%%%%%%%%%%%%%%%%%%%%%%%%%%%%%%%%%%%%%%%%%%%%%%%%%%%%

%
%
%%%%%%%%%%%%%%%%%%%%%%%%%%%%%%%%%%%%%%%%%%%%%%%%%%%%%%%%%%%%%%%%%%%%%%%%%%%
%%%%%%%%%%%%%%%%%%%%%%%%%%%%%%%%%%%%%%%%%%%%%%%%%%%%%%%%%%%%%%%%%%%%%%%%%%%
\section{Conclusion}
\label{sec:conclusion}

We construct a comprehensive, detailed framework to calculate the diffuse gamma-ray signal originating from axions produced in the vast cosmic population of CCSNe and their subsequent conversion into photons in the presence of astrophysical magnetic fields. \textit{For the first time}, we calculate axion–photon conversion by accounting for all magnetic field environments traversed by axions en route to Earth, including the progenitor star, the host galaxy, the IGM, and the MW, incorporating the uncertainties in each component. We study two axion models, the KSVZ and the ALP model along with three supernova progenitor profiles from the Garching archive. \textit{In contrast to previous works}, which typically employ single analytic fit to the SFRD, we construct a data-driven, piecewise SFRD using recent low and high redshift observations. We propagate the associated fit uncertainties and the conversion probability uncertainties arising from different magnetic field environments to estimate the diffuse gamma-ray flux for each CCSN profile independently. 

Exploiting existing measurements of the diffuse gamma-ray from COMPTEL, EGRET, and \textit{Fermi}-LAT, we derive $95\%$\,CL upper limits on the axion-photon coupling $(g_{a\gamma\gamma})$ over the axion mass range $m_a = 10^{-12}-10^{-3} \, {\rm eV}$. We find that the EGRET data provide the strongest constraints on axion-photon coupling. However, the \textit{Fermi}-LAT results are considerably more robust and conservative. For low mass regime, $m_a \sim (10^{-12} - 10^{-10}) \, {\rm eV}$, our optimistic limit from \textit{Fermi}-LAT reaches $g_{a\gamma\gamma} \sim 2 \times 10^{-13} \, {\rm GeV}^{-1}$ for the KSVZ model (with SFHo-18.6 profile). For both the KSVZ and the ALP models, our limits are competitive and complementary to existing limits. The width of the exclusion band reflects the uncertainties in the CCSNe rate and magnetic field estimates. 

Further, we perform a detailed Fisher analysis to estimate the projected upper limits on the axion-photon coupling for future MeV telescopes. Adopting a benchmark observation time of $10^7 \, {\rm s}$ and sky coverage of $|b| > 20^\circ$, we estimate the $95\%$\,CL exclusion limits on the axion-photon coupling for AMEGO-X, e-ASTROGAM, APT, AdEPT, GRAMS Satellite, PANGU, MAST and VLAST telescopes. Our forecasts use power-law background templates fitted to existing COMPTEL, EGRET, and \textit{Fermi}-LAT data as fiducial models, and marginalize over the background nuisance parameters $(\mathcal{C},\,\gamma_{\rm bg})$  via the Schur complement to obtain the profiled Fisher sensitivity on $g_{a\gamma\gamma}$. Adopting realistic foreground systematic uncertainties ($\sigma_{\rm sys}=10\%$, $\ell_c=0.1$), most instruments achieve comparable projected sensitivities across the low-mass regime, $m_a \sim (10^{-12} - 10^{-10}) \, {\rm eV}$, reaching $g_{a\gamma\gamma} \sim (7 - 9) \times 10^{-14} \, \mathrm{GeV}^{-1}$ for the KSVZ model with SFHo-18.6 profile (optimistic scenario). These projections represent improvements over current constraints in parts of the parameter space, demonstrating the potential of next-generation MeV gamma-ray observatories to reveal axion interactions via the diffuse gamma-ray signal.

Looking ahead, extending the gamma-ray flux calculation to include the cluster magnetic field environments of host galaxies would further improve the accuracy of the 
signal prediction. Notably, the axion flux calculation for a variety of CCSN mass profiles, weighted by a realistic distribution, would additionally strengthen the robustness of the diffuse flux calculation. Finally, a joint analysis combining multiple datasets and cross-correlating with the diffuse supernova neutrino background could open a new multi-messenger window onto the axion parameter space.
%
%
%%%%%%%%%%%%%%%%%%%%%%%%%%%%%%%%%%%%%%%%%%%%%%%%%%%%%%%%%%%%%%%%%%%%%%%%%%%
%%%%%%%%%%%%%%%%%%%%%%%%%%%%%%%%%%%%%%%%%%%%%%%%%%%%%%%%%%%%%%%%%%%%%%%%%%%
%
%

%
%
%%%%%%%%%%%%%%%%%%%%%%%%%%%%%%%%%%%%%%%%%%%%%%%%%%%%%%%%%%%%%%%%%%%
\section*{Acknowledgments}

We are especially grateful to Hans-Thomas Janka and Daniel Kresse for providing access to the Garching Core-Collapse Supernova archive. We sincerely thank Pierluca Carenza, Shiuli Chatterjee, Deep Jyoti Das, Sougata Ganguly, Durba Ghosh, Shunsaku Horiuchi, Rajeev Kumar Jain, Tanmoy Kumar, Ranjini Mondol, Nirmal Raj, Georg G. Raffelt, Sanjay Reddy, Akash Kumar Saha, Abhishek Tiwari and Ujjwal Kumar Upadhyay for their helpful discussions and insightful suggestions. The previous limits presented in this work are obtained from Ref.\,\cite{AxionLimits}. We acknowledge the use of high-performance computational facilities at the PTG Cluster at the Department of Physics, Indian Institute of Science, Bengaluru, India. D.B. acknowledges the Council of Scientific and Industrial Research (CSIR), Government of India, for supporting his research under the Research Associateship program through grant no.\,\,09/0079(24106)/2025-EMR-I, and support from the ISRO-IISc STC for the grant no.\,\,ISTC/PHY/RL/499.  S.B. acknowledges the Council of Scientific and Industrial Research (CSIR), Government of India, for supporting his research under the CSIR Junior/Senior Research Fellowship program through grant no.\,\,09/0079(15488)/2022-EMR-I. R.L.\,\,acknowledges financial support from the institute start-up funds, ISRO-IISc STC for the grant no.\,\,ISTC/PHY/RL/499, and ANRF for the grant no.\,\,ANRF/ARG/2025/005140/PS.
%%%%%%%%%%%%%%%%%%%%%%%%%%%%%%%%%%%%%%%%%%%%%%%%%%%%%%%%%%%%%%%%%%%
%
%
%%%%%%%%%%%%%%%%%%%%%%%%%%%%% Appendix %%%%%%%%%%%%%%%%%%%%%%%%%%%%
\appendix
%
%
%%%%%%%%%%%%%%%%%%%%%%%%%%%%%  Appendix %%%%%%%%%%%%%%%%%%%%%%%%%%%%%%%%%%%
%%%%%%%%%%%%%%%%%%%%%%%%%%%%%%%%%%%%%%%%%%%%%%%%%%%%%%%%%%%%%%%%%%%%%%%%%%%
%
%
%%%%%%%%%%%%%%%%%%%%%%%%%%%%%%%%%%%%%%%%%%%%%%%%%%%%%%%%%%%%%%%%%%%%%%%%%%%
\section{ROI Dependence of the Milky Way Conversion Probability}
\label{app:roi_comparison}

\patogMWROIavg
Space-based gamma-ray telescopes measure the isotropic diffuse flux over instrument-specific ROI, defined by latitude cuts that mask the bright Galactic plane. Since the MW axion-photon conversion probability $P_{a\to\gamma}^{\rm MW}$ is computed as a directional average over the line of sight, it inherits a mild dependence on the chosen ROI. For each instrument, we define the averaged probability, analogous to Eq.~\eqref{eq:PMW_avg}, with different ROIs, and is expressed as
\begin{equation}\label{eq:PMW_avg_gen}
    P_{a\to\gamma}^{\rm MW} = \frac{1}{\Delta\Omega}\int_{\rm ROI} d\Omega\; P_{a\to\gamma}^{\rm GMF}(\ell,b)\,,
\end{equation}
where $\Delta\Omega$ is the solid angle corresponding to the ROI, and $P_{a\to\gamma}^{\rm GMF}(\ell,b)$ is computed using  \texttt{gammaALPs}~\cite{Meyer:2021pbp}. The three telescopes use the following angular windows: \textit{Fermi}-LAT with $|b|>20^\circ$\,\cite{Fermi-LAT:2014ryh}, EGRET with $10^\circ < |b| < 80^\circ$\,\cite{Strong:2004ry}, and COMPTEL with $|b|>30^\circ$\,\cite{1998PhDT.........3K, 2000AIPC..510..467W}.
%
%
%%%%%%%%%%%%%%%%%%%%%%%%% FIG: ROI avg MW Conversion  %%%%%%%%%%%%%%%%%%%%%%%%%%%%
%%%%%%%%%%%%%%%%%%%%%%%%%%%%%%%%%%%%%%%%%%%%%%%%%%%%%%%%%%%%%%%%%%%%%%%%%%%
%
%
Fig.~\ref{fig:roi_comparison} compares $P_{a\to\gamma}^{\rm MW}$ evaluated over these three ROIs for two representative axion masses, using the \texttt{nebCor} (maximal) and \texttt{twistX} (minimal) variants of the UF23 model\,\cite{Unger:2023lob}. The ROI variation is subdominant relative to the magnetic field uncertainty across all three instruments. The ratio of conversion probabilities between \texttt{nebCor} and \texttt{twistX} is a factor of $3$–$4$, while the inter-instrument ratio at fixed field model can reach up to $\sim 1.25$. This confirms that the choice of ROI does not significantly affect the derived constraints, and the dominant theoretical uncertainty in $P_{a\to\gamma}^{\rm MW}$ stems from the GMF model rather than the observation window. In our analysis, we adopt the \textit{Fermi}-LAT ROI for all instruments.
\section{Total probability with reconversion}
\label{app:prob_wt_reconv}

In section~\ref{sec:alp_conv}, we outline the axion–photon conversion probability in each of the astrophysical magnetic field environments considered in our framework and present the effective probability for an axion produced in an extra-galactic supernova to convert into a gamma-ray photon before reaching the detector. However, an axion that converts into a photon in one magnetic field region can subsequently reconvert into an axion in later regions. By accounting for such reconversion effects at each step, we construct a more general expression than Eq.~\eqref{eq:eff_prob} by tracking the axion and photon fractions through successive magnetic field domains, which can be written as
\begin{widetext}
\begin{equation}\label{eq:eff_prob_full}
\begin{aligned}
    \mathcal{P}^{\rm eff}_{a\to\gamma}(z)\big|_{\rm reconv} 
    ={}& P^{\rm MW}_{a\to\gamma} \Bigg\{ P^{\rm IGM}_{a\to\gamma}(z) \bigg[ \Big(1 - P^{\rm Prog}_{a\to\gamma}(z)\Big) P^{\rm Host}_{a\to\gamma}(z)
    + \Big(1 - P^{\rm Host}_{a\to\gamma}(z)\Big) P^{\rm Prog}_{a\to\gamma}(z) \bigg] + \Big(1 - P^{\rm IGM}_{a\to\gamma}(z) \Big) \\
    & \times \bigg[ \Big(1 - P^{\rm Prog}_{a\to\gamma}(z) \Big) \Big(1 - P^{\rm Host}_{a\to\gamma}(z) \Big) + P^{\rm Prog}_{a\to\gamma}(z) P^{\rm Host}_{a\to\gamma}(z) \bigg] \Bigg\} + \Big(1 - P^{\rm MW}_{a\to\gamma} \Big) \Bigg\{ \Big(1 - P^{\rm IGM}_{a\to\gamma}(z) \Big) \\
    & \times \bigg[ \Big(1 - P^{\rm Prog}_{a\to\gamma}(z) \Big) \times P^{\rm Host}_{a\to\gamma}(z) + \Big(1 - P^{\rm Host}_{a\to\gamma}(z) \Big) P^{\rm Prog}_{a\to\gamma}(z) \bigg] + P^{\rm IGM}_{a\to\gamma}(z) \bigg[ \Big(1 - P^{\rm Prog}_{a\to\gamma}(z) \Big) \\ 
    & \times \Big(1 - P^{\rm Host}_{a\to\gamma}(z) \Big) + P^{\rm Prog}_{a\to\gamma}(z) P^{\rm Host}_{a\to\gamma}(z) \bigg] \Bigg\} \\
    ={}& \frac{1}{2}\Bigg[1 - \Big(1 - 2P^{\rm Prog}_{a\to\gamma}(z) \Big) \Big(1 - 2P^{\rm Host}_{a\to\gamma}(z) \Big) \Big(1 - 2P^{\rm IGM}_{a\to\gamma}(z) \Big) \Big(1 - 2P^{\rm MW}_{a\to\gamma} \Big) \Bigg].
\end{aligned}
\end{equation}
\end{widetext}
%
%
%%%%%%%%%%%%%%%%%%%%%%%%% FIGURE: chi^2 limits (Other) %%%%%%%%%%%%%%%%%%%%%%%%%%%%
%%%%%%%%%%%%%%%%%%%%%%%%%%%%%%%%%%%%%%%%%%%%%%%%%%%%%%%%%%%%%%%%%%%%%%%%%%%
%
%
To assess the impact of including reconversion effects encoded in Eq.~\eqref{eq:eff_prob_full}, compared to Eq.~\eqref{eq:eff_prob}, we compute the differential diffuse $\gamma$-ray flux using Eq.~\eqref{eq:dphi_dE_gam} for representative parameter values $m_a = 10^{-12} \, {\rm eV}$, $g_{a\gamma\gamma} = 10^{-12} \, {\rm GeV}^{-1}$, and $E_a = 100 \, {\rm MeV}$, adopting the SFHo-18.6 profile. The resulting fluxes are $2.750 \times 10^{-5} \, {\rm MeV}^{-1} \, {\rm cm}^{-2} \, {\rm s}^{-1} \, {\rm sr}^{-1}$ and $2.737 \times 10^{-5} \, {\rm MeV}^{-1} \, {\rm cm}^{-2} \, {\rm s}^{-1} \, {\rm sr}^{-1}$, respectively, for the KSVZ model (optimistic case), corresponding to the two prescriptions for the conversion probability given in Eqs.~\eqref{eq:dphi_dE_gam} and \eqref{eq:eff_prob_full}, respectively. The relative difference between these results is at the level of $\mathcal{O}(1\%)$, indicating a mild suppression of the total $\gamma$-ray flux when reconversion effects are included. Given the minor effect compared to other uncertainties in the analysis, we consider it negligible and therefore do not include it in calculating our results.
%%%%%%%%%%%%%%%%%%%%%%%%%%%%%%%%%%%%%%%%%%%%%%%%%%%%%%%%%%%%%%%%%%%%%%%%%%%
%
%
\chisqLimitsOther
%
%
%%%%%%%%%%%%%%%%%%%%%%%%% FIGURE: Fisher limits (Other) %%%%%%%%%%%%%%%%%%%%%%%%%%%%

%%%%%%%%%%%%%%%%%%%%%%%%%%%%%%%%%%%%%%%%%%%%%%%%%%%%%%%%%%%%%%%%%%%%%%%%%%%
%
%
%%%%%%%%%%%%%%%%%%%%%%%%%%%%%%%%%%%%%%%%%%%%%%%%%%%%%%%%%%%%%%%%%%%%%%%%%%%%%%%%%%%%%%%%%%%%%%%
\section{$\chi^2$ results for other CCSN models}
\label{app:other_chi2_results}

We redo the $\chi^2$ analysis for other alternative progenitor models, i.e., SFHo-18.8 and SFHo-20.0, and plotted the $95\%$ CL exclusion limits in Fig.~\ref{fig:chisq_limits_other_models}. The three models are ordered by their total axion production rate as $\text{SFHo-}20.0 > \text{SFHo-}18.6 > \text{SFHo-}18.8$, so a higher production rate yields a larger photon signal for a given coupling, allowing the same observed background to constrain a smaller $g_{a\gamma\gamma}$. Thus, the exclusion limits follow the same ordering, with SFHo-20.0 yielding the strongest constraints and SFHo-18.8 the weakest, as shown in Fig.~\ref{fig:chisq_limits_other_models}.

For the KSVZ model (optimistic case) with \textit{Fermi}-LAT, the limits from SFHo-20.0 are, on average, $\sim 36\%$ stronger than those from SFHo-18.6, while those from SFHo-18.8 are $\sim 34\%$ weaker. Both deviations are most pronounced at low masses ($m_a \lesssim 10^{-10}\,\mathrm{eV}$), where the deviations reach $\sim 44\%$, and diminish to $\sim 33\%$ at high masses ($m_a \gtrsim 10^{-5}\,\mathrm{eV}$). For the ALP model (optimistic case) with \textit{Fermi}-LAT, the overall variation is larger, with SFHo-20.0 and SFHo-18.8 differing from the SFHo-18.6 by $\sim 39\%$ and $\sim 47\%$, respectively, on average. The low-mass plateau ($m_a \lesssim 10^{-9}\,\mathrm{eV}$) shows the largest model sensitivity, with SFHo-20.0 and SFHo-18.8 deviating by $\sim 47\%$ and $\sim 68\%$, respectively, while the high-mass regime ($m_a \gtrsim 10^{-7}\,\mathrm{eV}$) converges to within $\sim 30\%$ for both models. For all CCSN models, the shape of the exclusion band is unchanged, confirming that our main conclusions are robust to reasonable variations in the profiles.
%%%%%%%%%%%%%%%%%%%%%%%%%%%%%%%%%%%%%%%%%%%%%%%%%%%%%%%%%%%%%%%%%%%%%%%%%%%%%%%%%%%%%%%%%%%%%%%
%
%
\FisherLimitsOther
\section{Forecasts for other CCSN models}
\label{app:other_fisher_results}

We also perform the CCSN profile comparison for results obtained via Fisher forecast. We find that APT gives us the most stringent projected limit in each case. The sensitivity ordering across CCSN models is the same as in the $\chi^2$ analysis shown in the previous appendix~\ref{app:other_chi2_results}. For the KSVZ model (optimistic case) with APT, the SFHo-20.0 projected limit is $\sim 30\%$ more stringent than SFHo-18.6, while SFHo-18.8 is $\sim 26\%$ less stringent. The deviations are largest in the low mass plateau ($\sim 31\%$ and $\sim 24\%$, respectively) and diminish slightly at high masses ($\sim 29\%$ and $\sim 27\%$, respectively). For the ALP model (optimistic case) with APT, the overall deviations are larger, with SFHo-20.0 and SFHo-18.8 differing from SFHo-18.6 by $\sim 36\%$ and $\sim 40\%$, respectively. The low mass plateau ($m_a \lesssim 10^{-9}\,\mathrm{eV}$) again shows the largest model sensitivity, with deviations reaching $\sim 44\%$ and $\sim 56\%$ for SFHo-20.0 and SFHo-18.8, respectively, while the high mass regime ($m_a \gtrsim 10^{-7}\,\mathrm{eV}$) converges to within $\sim 30\%$ for both models. Similar to $\chi^2$ results, in all cases the shape of the projected exclusion band is unchanged, confirming that the Fisher forecast conclusions remain insensitive under reasonable variations in the progenitor model. The forecast results for SFHo-18.8 and SFHo-20.0 are shown in Fig.~\ref{fig:fisher_limits_other_models}.
%
%
%%%%%%%%%%%%%%%%%%%%%%%%% FIGURE: Fisher Corner %%%%%%%%%%%%%%%%%%%%%%%%%%%%
\FisherCorner
%%%%%%%%%%%%%%%%%%%%%%%%%%%%%%%%%%%%%%%%%%%%%%%%%%%%%%%%%%%%%%%%%%%%%%%%%%%
%
%%%%%%%%%%%%%%%%%%%%%%%%%%%%%%%%%%%%%%%%%%%%%%%%%%%%%%%%%%%%%%%%%%%%%%%%%%%
%
%
\section{Schur complement}
\label{app:schur_comp}

In our analysis, as discussed in section\,\ref{subsec:fisher_forecast}, the background parameters ($\mathcal{C},\gamma_{\rm bg}$) are not perfectly known. To obtain Fisher information with respect to $\eta$, we marginalize over 
these background nuisance parameters using the Schur complement 
technique\,\cite{Wilks:1938dza, schur_complement_wikipedia}. With the parameter vector $\boldsymbol{\theta}=(\eta,\,\mathcal{C},\,\gamma_{\rm bg})$, the $3\times3$
Fisher matrix takes the block form
\begin{equation}
    \mathcal{F} =
    \begin{pmatrix}
        \mathcal{F}_{\eta\eta} & \mathbf{f}^{T} \\[4pt]
        \mathbf{f}             & \mathcal{F}_{bb}
    \end{pmatrix},
\end{equation}
where $\mathcal{F}_{\eta\eta}$ is the signal--signal information,
$\mathcal{F}_{bb}$ is the $2\times2$ background--background block spanned by
$(\mathcal{C},\,\gamma_{\rm bg})$, and
$\mathbf{f}=(\mathcal{F}_{\eta\mathcal{C}},\,\mathcal{F}_{\eta\gamma_{\rm bg}})^{T}$
encodes the signal--background cross-correlations.

Since $\lambda_k = S_k + \lambda_k'$ and $\lambda_k'$ does not depend on $\eta$,
the signal--signal block reduces to
\begin{equation}
    \mathcal{F}_{\eta\eta}
    = \sum_{m,n}\frac{\partial S_m}{\partial\eta}
      \left(\mathcal{D_{\rm bg}}^{-1}\right)_{mn}
      \frac{\partial S_n}{\partial\eta}\,,
      \label{eq:}
\end{equation}
where $S_k\equiv\lambda_k - \lambda_k'$ is the signal count at $k^{\rm th}$ bin. The covariance matrix $\mathcal{D}$ in Eq.~\eqref{eq:F_ij} is evaluated under the background-only hypothesis, such that
$(\mathcal{D_{\rm bg}})_{mn}=\lambda_m'\,\delta_{mn}+\lambda_m'\lambda_n'\,\Sigma^{\rm sys}_{mn}$, with $\Sigma^{\rm sys}_{mn}$ defined in Eq.~\eqref{eq:sys_cov}.

Marginalizing over $(\mathcal{C},\,\gamma_{\rm bg})$ via the Schur complement of
$\mathcal{F}_{bb}$ yields the profiled Fisher information on $\eta$ alone,
\begin{equation}\label{eq:F_prof}
    \mathcal{F}_{\rm prof}(\eta)
    = \mathcal{F}_{\eta\eta}
      - \mathbf{f}^{T}\mathcal{F}_{bb}^{-1}\mathbf{f}\,.
\end{equation}
The correction term $\mathbf{f}^{\,T}\mathcal{F}_{bb}^{-1}\mathbf{f}$ quantifies the loss of information due to degeneracy. Next, we obtain the 95\% CL upper limits on $g_{a\gamma\gamma}$ by solving $\mathcal{F}_{\rm prof}\!\left(\ln g_{a\gamma\gamma}^*\right) = 2.71$. The signal fiducial $\eta^*=\ln g_{a\gamma\gamma}^*$ is determined 
self-consistently as the solution to this equation rather than fixed in advance, which avoids the case $F_{\eta\eta}(g=0) = 0$ that arises when the 
Fisher matrix is evaluated at vanishing signal. Given the instrument's effective area, observation time, and sky coverage, 
$g_{a\gamma\gamma}^*$ is the minimum coupling at which the axion signal is statistically distinguishable from background fluctuations at 95\% CL.
%
%%%%%%%%%%%%%%%%%%%%%%%%%%%%%%%%%%%%%%%%%%%%%%%%%%%%%%%%%%%%%%%%%%%%%%%%%%%
\section{Fisher ellipses}
\label{app:fisher_ellipse}

The covariance matrix $\boldsymbol{\Sigma} = \mathcal{F}^{-1}$, which is estimated at $\boldsymbol{\theta}^*=(\eta^*,\,\mathcal{C}^*,\,\gamma_{\rm bg}^*)$ contains the joint uncertainties and correlations among all three considered parameters. We reparametrise the signal axis from $\eta$ to $\log_{10}(g_{a\gamma\gamma})$ for optimal visualization. To see the degeneracy among them, we construct confidence ellipses. For each pair $(\theta_i, \theta_j)$, we extract the $2\times2$ sub-block of $\boldsymbol{\Sigma}$ and diagonalize it to obtain eigenvalues $\nu_1 \geq \nu_2$ and the corresponding eigenvectors. Next, we draw ellipses with semi-major and semi-minor axes as
\begin{equation}
    a = \sqrt{\nu_1\,\Delta\chi^2}\,, \qquad b = \sqrt{\nu_2\,\Delta\chi^2}\,,
\end{equation}
oriented along the eigenvectors. We draw joint $1\sigma$ and $2\sigma$ confidence regions using $\Delta\chi^2 = 2.30$ and $6.18$, respectively. The degree of degeneracy between each pair is expressed by the Pearson correlation coefficient
\begin{equation}
    \rho(\theta_i,\theta_j) = \frac{\Sigma_{\theta_i\theta_j}}{\sqrt{\Sigma_{\theta_i\theta_i}\,\Sigma_{\theta_j\theta_j}}}\,.
\end{equation}

In Fig.\,\ref{fig:fisher_corner_ksvz} we show the forecast corner plot for APT with $m_a=10^{-12}\,\mathrm{eV}$ for the KSVZ  model with SFHo-$18.6$ CCSN model. Each panel shows the joint 
$1\sigma$ (inner, darker) and $2\sigma$ (outer, lighter) confidence regions from the 
$2\times 2$ sub-block of $\boldsymbol{\Sigma}$. The mean Pearson correlation 
coefficient $\langle \rho \rangle$ is given in each panel, averaged over the two 
flux realizations. We find a negative tilt in the 
$(\log_{10}g_{a\gamma\gamma},\,\mathcal{C})$ plane, indicating a moderate 
anti-correlation: a larger background amplitude can partially mimic the signal, so 
the fit compensates by requiring a larger coupling, thereby weakening the constraint on $g_{a\gamma\gamma}$. The $(\log_{10} g_{a\gamma\gamma},\,\gamma_{\rm bg})$ ellipse  shows a moderate anti correlation with $\langle\rho(g_{a\gamma\gamma},\gamma_{\rm bg})\rangle \approx -0.23$. The 
$(\mathcal{C},\,\gamma_{\rm bg})$ ellipse has a strong correlation ($\langle\rho(\mathcal{C},\gamma_{\rm bg})\rangle \approx 0.95$). This near-degeneracy between the two nuisance parameters is the dominant source of background uncertainty and directly justifies their marginalization via the Schur complement of the Fisher matrix.
%%%%%%%%%%%%%%%%%%%%%%%%%%%%%%%%%%%%%%%%%%%%%%%%%%%%%%%%%%%%%%%%%%%%%%%%%%%%%%%%%%
%%%%%%%%%%%%%%%%%%%%%%%%%%%%%%%%%%%%%%%%%%%%%%%%%%%%%%%%%%%%%%%%%%%
%
%

\bibliographystyle{JHEP}
\bibliography{refs}

\end{document}